\documentclass[fleqn,usenatbib]{mnras}
\usepackage{lineno}
\usepackage[T1]{fontenc}
\usepackage{ae,aecompl}
\usepackage{graphicx}	
\usepackage{amsmath}	
\usepackage{amssymb}	
\usepackage{dirtytalk}
\usepackage{siunitx}
\usepackage{tablefootnote}

\usepackage{xspace}
\usepackage{newtxtext,newtxmath}
\usepackage{xcolor}
\usepackage{multirow,longtable,supertabular}

\def \msun{{\rm ~M_{\odot}}}

\newcommand{\thisgrb}{GRB~140102A\xspace}
\newcommand{\fermi}{{\em Fermi}\xspace}
\newcommand{\kw}{{\em Konus}-Wind\xspace}
\newcommand{\kwT}{{T$_{\rm kw,0}$}\xspace}
\newcommand{\swiftT}{{T$_{\rm 0}$}\xspace}
\newcommand{\fermiT}{{T$_{\rm 0}$}\xspace}
\newcommand{\keV}{{\rm keV}\xspace}

\newcommand{\swift}{{\em Swift}\xspace}
\newcommand{\tninty}{{$T_{\rm 90}$}\xspace}
\newcommand{\mvts}{{$t_{\rm mvts}$}\xspace}

\newcommand{\Ep}{$E_{\rm p}$\xspace}
\newcommand{\sw}[1]{\texttt{#1}}
\graphicspath{{./}{figures/}}

\title[\thisgrb]{\thisgrb : Insight into Prompt Spectral Evolution and Early Optical Afterglow Emission}

\author[Rahul Gupta et al.]
{Rahul Gupta,$^{1, 2}$\thanks{E-mail: rahulbhu.c157@gmail.com, rahul@aries.res.in}
S. R. Oates,$^{3}$ 
S. B. Pandey,$^{1}$\textsuperscript{\thanks{shashi@aries.res.in}} 
A. J. Castro-Tirado,$^{4,5}$ 
Jagdish C. Joshi,$^{6,7}$ 
\newauthor Y.-D. Hu,$^{4,8}$ 
A. F. Valeev,$^{9,10}$
B. B. Zhang,$^{6,7}$
Z. Zhang,$^{6}$
Amit Kumar,$^{1,11}$ 
A. Aryan,$^{1,2}$
\newauthor A. Lien,$^{12,13}$
B. Kumar,$^{1}$
Ch. Cui,$^{14}$
Ch. Wang,$^{15}$
Dimple,$^{1, 2}$
D. Bhattacharya,$^{16}$
E. Sonbas,$^{17}$
\newauthor J. Bai,$^{15}$
J. C. Tello,$^{4}$
J. Gorosabel,\dag$^{4,18,19}$
J. M. Castro Cer\'on,$^{20}$
J. R. F. Porto,$^{21}$
K. Misra,$^{1}$
\newauthor M. De Pasquale,$^{22,23}$
M. D. Caballero-Garc\'ia,$^{4}$
M. Jelínek,$^{4,24}$
P. Kub\'anek,$^{25}$
P. Yu. Minaev,$^{26}$
\newauthor R. Cunniffe,$^{4}$
R. S\'anchez-Ram\'irez,$^{27}$
S. Guziy,$^{28,29}$
S. Jeong,$^{4}$
S. N. Tiwari,$^{2}$
S. Razzaque,$^{30}$
\newauthor V. Bhalerao,$^{31}$
V. C. Pintado,$^{32}$
V. V. Sokolov,$^{9}$
X. Zhao,$^{15}$
Y. Fan,$^{15}$
and Y. Xin$^{15}$ 
}
  
\date{Accepted 2021 May 26. Received 2021 May 22; in original form 2021 February 19}
\pubyear{2021}

\begin{document}
\label{firstpage}
\maketitle

\begin{abstract} 
We present and perform a detailed analysis of multi-wavelength observations of \thisgrb, an optical bright GRB with an observed reverse shock (RS) signature. Observations of this GRB were acquired with the BOOTES-4 robotic telescope, the \fermi, and the \swift missions. Time-resolved spectroscopy of the prompt emission shows that changes to the peak energy (\Ep) tracks intensity and the low-energy spectral index seems to follow the intensity for the first episode, whereas this tracking behavior is less clear during the second episode. The fit to the afterglow light curves shows that the early optical afterglow can be described with RS emission and is consistent with the thin shell scenario of the constant ambient medium. The late time afterglow decay is also consistent with the prediction of the external forward shock (FS) model. We determine the properties of the shocks, Lorentz factor, magnetization parameters, and ambient density of \thisgrb, and compare these parameters with another 12 GRBs, consistent with having RS produced by thin shells in an ISM-like medium. The value of the magnetization parameter ($R_{\rm B} \approx 18$) indicates a moderately magnetized baryonic dominant jet composition for \thisgrb. We also report the host galaxy photometric observations of \thisgrb obtained with 10.4m GTC, 3.5m CAHA, and 3.6m DOT telescopes and find the host (photo $z$ = $2.8^{+0.7}_{-0.9}$) to be a high mass, star-forming galaxy with a star formation rate of $20 \pm 10 \msun$ $\rm yr^{-1}$.

\end{abstract}


\section{Introduction}
\label{intro}

Gamma-ray Burst (GRB) emission can be divided into two phases: a short-lived and highly variable, initial prompt $\gamma$-ray emission phase, and a long-lived multi-wavelength afterglow emission phase \citep{2004RvMP...76.1143P, 2013FrPhy...8..661G, 2015PhR...561....1K}. Based on the observed duration of the prompt emission, GRBs are classified into two broad classes \citep{kou93}: long GRBs (LGRBs: \tninty\footnote{\tninty is the duration during which 5 percent to 95 percent of the $\gamma$-ray/hard X-ray fluence is received.} $>$ 2 sec) and short GRBs (SGRBs: \tninty $\leq$ 2 sec). It implies a distinction between their progenitors \citep{2016SSRv..202...33L}. The LGRBs are associated with the deaths of massive stars \citep{1999ApJ...524..262M, 2006ARA&A..44..507W, 2007AIPC..906...69D, 2020MNRAS.492.4613O} and SGRBs with the mergers of compact binary objects such as two neutron stars or a neutron star and black hole \citep{2002ApJ...570..252P, 2002ApJ...571..394B, Abbott_2017, 2017ApJ...848L..14G}. There is, however, a significant overlap between the bi-modal duration distribution of GRBs \citep{kou93, 2013MNRAS.430..163Q, 2017ApJ...848L..14G, 2016ApJS..227....7L, 2017AstL...43....1M}. A doubtful third class with intermediate duration is also proposed \citep{1998ApJ...508..314M, 2000ApJ...538..165H, 2007ApJ...667.1017C, 2008A&A...489L...1H, 2010AstBu..65..326M, 2016Ap&SS.361..155H}. 
 
\par
The prompt emission in GRBs is believed to be produced in a relativistic jet via energy dissipation in internal shocks when fast-moving shells catch-up with slower shells or due to catastrophic reconfiguration of the magnetic fields in a Poynting flux dominated outflow \citep{2015AdAst2015E..22P, 2001MNRAS.321..177L}. The radiation physics of the prompt emission is still under debate \citep{2015PhR...561....1K}. The prompt emission may be explained as synchrotron emission from a cooling population of particles \citep{2019A&A...628A..59O, 2020NatAs...4..174B, 2020NatAs...4..210Z}. However, photospheric models also can equally well explain the data \citep{Vianello:2017, Ahlgren:2019ApJ, Zeynep:2020arXiv}. 
 
\par
Examining the prompt emission spectral evolution is a powerful tool to investigate the radiation mechanisms of GRBs \citep{2015AdAst2015E..22P}. The evolution of peak energy has been observed to have three types of patterns: (i) a hard-to-soft pattern, where {\bf \Ep} decreases continuously \citep{1986ApJ...301..213N, 1994ApJ...426..604B, 1997ApJ...486..928B};
(ii) an intensity-tracking pattern, where {\bf \Ep} increases/decreases as the intensity increases/decreases  \citep{1983Natur.306..451G, 1999ApJ...512..693R}; (iii) a soft-to-hard pattern or disordered pattern, where \Ep increases continuously or does not show any correlation with flux \citep{1985ApJ...290..728L, 1994ApJ...422..260K}. \cite{2018ApJ...869..100U} reproduced the hard-to-soft and intensity-tracking patterns using a synchrotron radiation model in a bulk-accelerating emission region. Furthermore, they suggested a direct connection between these two patterns (hard-to-soft and intensity-tracking) and spectral lags. They predicted that only the positive spectral lag could be observed for the hard-to-soft pattern, but both positive and negative spectral lags are possible in the case of intensity-tracking. \cite{2014AstL...40..235M} suggested that individual pulses of GRBs demonstrate hard-to-soft spectral evolution (positive spectral lags), depending on the pulse duration. A complicated spectral evolution behavior could be connected with superposition effects. The low energy spectral index ($\it \alpha_{\rm pt}$) evolves with time but does not exhibit any strong typical pattern \citep{1997ApJ...479L..39C}. Recently, \cite{2019ApJ...884..109L} found that both \Ep and $\it \alpha_{\rm pt}$ show an intensity-tracking pattern (`double-tracking') for GRB 131231A.

\par
The relativistically moving blastwave inevitably crashes into the circumburst medium and results in external shocks. The afterglow phase, in general, is well explained by the emission originating in these external shocks, and any deviations from this model can generally be explained (e.g., see \citealt{2015PhR...561....1K, Meszaros:2019MmSAI} for a review). According to the standard afterglow model, the external shocks can be divided into two forms: a long-lived forward shock that propagates into the circumburst medium and produces a broadband afterglow, and a short-lived RS that propagates into the ejecta and produces a short-lived optical flash and a radio flare \citep{2003ApJ...595..950Z, 2004MNRAS.353..647N, 2015AdAst2015E..13G}. For most GRBs, the FS component can generally explain the observed afterglow. Investigations of the afterglow using the FS model offer detailed information about the late time afterglow emission, jet geometry, circumburst medium, and total energy \citep{2007MNRAS.379..331P, Wang_2015, 2019arXiv191101558J}. On the other hand, the short-lived RS emission is useful in probing the nature of magnetization and composition of GRB ejecta from the central engine \citep{2015AdAst2015E..13G}. RS emission has mainly two types of evolution \citep{2000ApJ...542..819K}: In the first situation, RS is relativistic enough to decelerate the shell (thick shell case, \tninty $>$ $T_{\rm dec}$  where $T_{\rm dec}$ is the deceleration time defined as $T_{\rm dec}$ =(3$E_{\rm k} (1+z)^3/32 \pi n_0 $ $\rm m_{\rm p}$ $\Gamma_0^8$ $\rm c^5)^{1/3}$, for blast-wave kinetic energy ($E_{\rm k}$), initial Lorentz factor ($\Gamma_0$), traversing into a constant density circumbusrt medium with density ($n_0$) \citep{zhang03}. The blast-wave radius evolves with time and it is defined as $R_{\rm dec}(t = T_{\rm dec}) \approx 2$$\rm c$$T_{\rm dec}\Gamma_0^2/(1+z)$.  On the other hand, for the thin shell case (\tninty $<$ $T_{\rm dec}$), the RS could be sub-relativistic and too weak to decelerate the shell \citep{2015ApJ...810..160G}. It has been found that most of the GRBs with early optical RS signatures could be explained well within the thin shell case \citep{2014ApJ...785...84J, 2015ApJ...810..160G, 2020ApJ...895...94Y}.  

\par
Fast optical follow-up of GRBs is vital for detecting and studying the relatively early and short-lived RS component. GRB 990123, the first burst with a simultaneous optical flash as the signature of RS was detected using the Robotic Optical Transient Search Experiment (ROTSE)-I telescope \citep{1999Natur.398..400A, 1999MNRAS.306L..39M}. The RS emission has been observed for only a few GRBs, even after rapid follow-up observations by the \swift Ultra-Violet and Optical Telescope \citep[UVOT;][]{2005SSRv..120...95R} and other ground-based robotic telescopes network \citep{2003ApJ...595..950Z, oates09}. The lack of observed RS could be due to strongly magnetized outflows such that the RS component of the external shock is suppressed, either the RS emission component peaks at frequencies lower than optical frequencies and is thus generally missed, and/or the RS emission originating in external shock is masked by the prompt emission originating in an internal shock \citep{2005ApJ...628..315Z, 2016ApJ...825...48R, 2013ApJ...772...73K}. 

\par
In a comprehensive sample (118 GRBs with known redshift) by \cite{2014ApJ...785...84J}, 10 bursts\footnote{GRBs 990123, 021004, 021211, 060908, 061126, 080319B, 081007, 090102, 090424, and
130427A} showed dominant RS signatures originated from external shock with most of these having an ISM-like external medium (constant density). At late times ($>$ 10 ks), the RS emission showed magnetization parameters\footnote{It is the ratio between $\epsilon_{\rm B,r}$ (fraction of reverse shock energy into the magnetic field) and $\epsilon_{\rm B,f}$ (fraction of forward shock energy into the magnetic field), respectively.} ($R_{\rm B}$ $\sim$ 2 to $10^4$) and was fainter than average optical FS emission \citep{2014ApJ...785...84J}. \cite{2015ApJ...810..160G} identified 15 GRBs with RS signatures based on a morphological analysis of the early optical afterglow light curves of 63 GRBs and an estimated $R_{\rm B}$ $\sim$ 100. More recently, \cite{2020ApJ...895...94Y} studied the early optical afterglows of 11 GRBs\footnote{GRBs 990123, 041219A, 060607A, 061007, 081007, 081008, 090102, 110205A, 130427A, 140512A, and 161023A} with RS emission signatures. They found that the external medium density w.r.t. blast-wave radius follows a power-law type behavior with index ($k$) ranges in between 0 to 1.5. Among other shock parameters, the densities of the external medium for thin shell RS dominated bursts are compared in this work (see \S~\ref{sample comparison}) and shown to vary in a range 0.1-500 $\rm cm^{-3}$. 

\par
In this paper, we present multiwavelength data and analysis of \thisgrb including our early optical afterglow observations using Burst Observer and Optical Transient Exploring System (BOOTES)-4 robotic telescope starting $\sim$ 29 sec after the \fermi ~Gamma-Ray Burst Monitor (GBM; also the \swift ~Burst alert telescope (BAT) at the same time) trigger (\fermiT). The very early detection of optical emission from \thisgrb along with GeV Large Area Telescope (LAT) detection inspired us to study this burst in detail and compare it with other similar events. This paper is organized as follows. In \S~\ref{multiwaveength observations and data analyisis}, we discuss the multiwavelength observations and data analysis. The main results are presented in \S~\ref{results}. This is followed by discussion in \S~\ref{discussion} and finally summary and conclusion in \S~\ref{conclusions}, respectively. All the uncertainties are quoted at 1 $\sigma$ throughout this paper unless otherwise mentioned. The temporal ($\alpha$) and spectral indices ($\beta$) for the afterglow are given by the expression $\rm F (t,\nu)\propto t^{-\alpha}\nu^{-\beta}$. We consider the Hubble parameter $\rm H_{0}$ = 70 km $\rm sec^{-1}$ $\rm Mpc^{-1}$, density parameters $\rm \Omega_{\Lambda}= 0.73$, and $\rm \Omega_m= 0.27$.

\begin{figure*}
\centering
\includegraphics[scale=0.35]{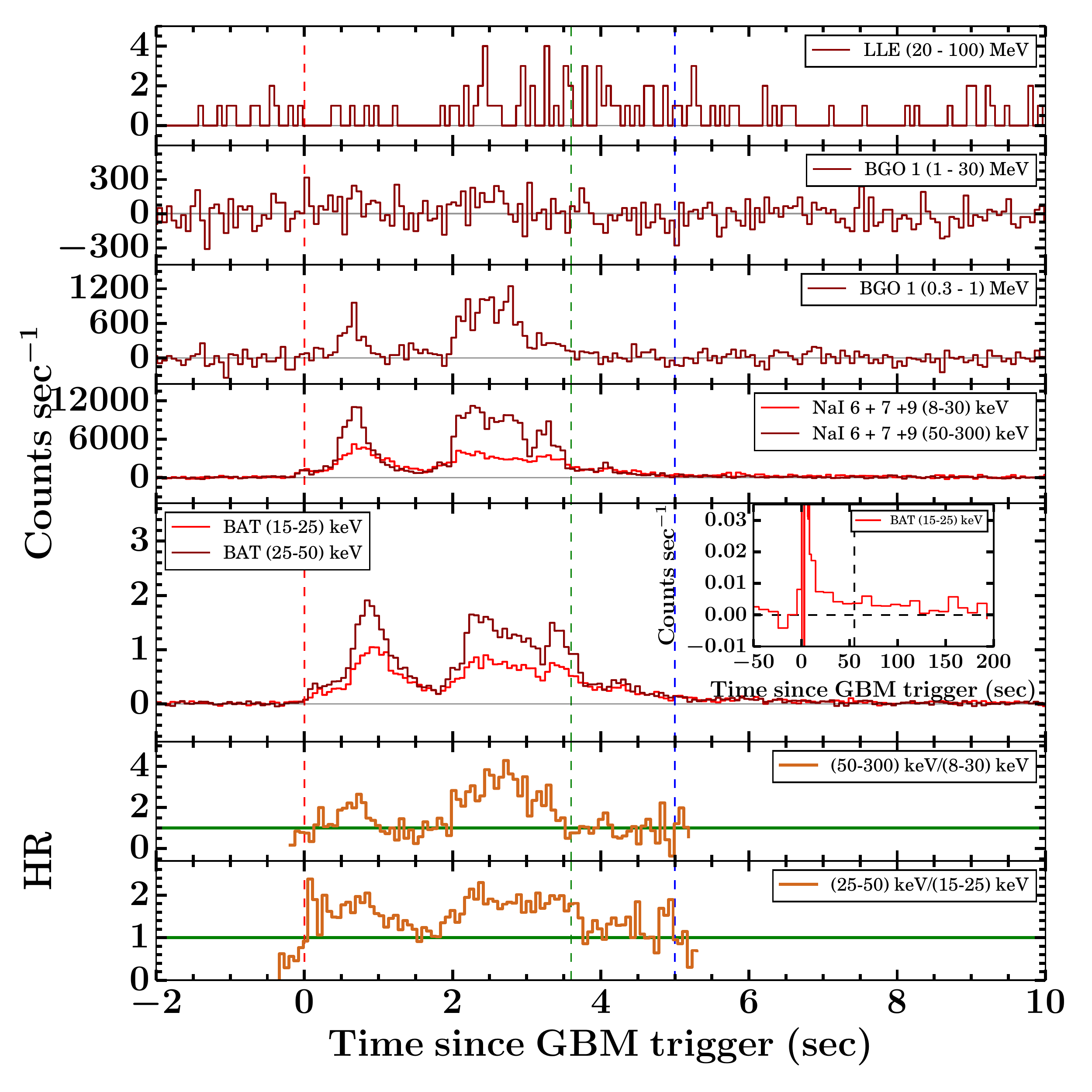}
\caption{ {\bf Multi-channel prompt $\gamma$-ray/ hard X-ray light curve:} The top five panels show the background-subtracted light curve of \thisgrb in multiple energy channels of \fermi LLE, \fermi GBM, and \swift BAT detectors with 64 ms temporal bins. The red and green vertical dashed lines indicate the \fermi trigger time and end of \tninty duration for \fermi GBM detector in 50-300 \keV energy range, respectively. The blue vertical dashed line indicates the end time of the time-averaged spectral analysis, with the start time taken as \fermiT. The horizontal grey solid lines differentiate between signal and background (at count rate equal to zero). Inset shows the BAT light curve in the soft (15-25 \keV) energy band with SN = 5 or 10 sec temporal binning. It indicates the presence of soft tail emission in the \swift BAT light curve. The vertical black dashed line represents the \tninty end time for \swift BAT detector in 15-350 \keV energy range. {\bf Evolution of hardness ratio (HR) :} The last two  panels show the evolution of HR in 50-300 \keV (hard) to 8-30 \keV (soft) energy channels of \fermi sodium iodide (NaI 6 + 7 + 9) and in 25-50 \keV (hard) to 15-25 \keV (soft) energy channels of \swift BAT detectors, respectively. The horizontal green solid line corresponds to HR equal to one.}
\label{promptlc}
\end{figure*}

\section{MULTI-WAVELENGTH OBSERVATIONS \& DATA ANALYSIS}
\label{multiwaveength observations and data analyisis}

\subsection{ \bf Gamma-ray/ Hard X-ray observations}

\thisgrb triggered the BAT \citep{2005SSRv..120..143B} on board the {\em Neil Gehrels Swift} observatory \citep{2004ApJ...611.1005G} on January 02, 2014 at 21:17:37 UT. The best on-ground location was found to be RA, Dec = 211.902, +1.331 degrees (J2000) with 3' uncertainty with 90 \% containment \citep{2014GCN.15653....1H}. 
\begin{table}
\caption{Prompt emission properties of \thisgrb. Redshift has been obtained using the SED modeling at 700 - 1000 sec (see \S~\ref{SED}). The peak flux is calculated in the 1-10,000 keV energy range in the source frame.}
\label{tab:prompt_properties}
\begin{center}
\begin{tabular}{|c|c|c|}
\hline
\bf {Prompt Properties} & \bf {\thisgrb }& \bf {Detector} \\
\hline 
\hline
\tninty (sec) & 3.58 $ \pm $ 0.01 & GBM \\ \hline 
\mvts (sec) 
&  $ \sim $ 0.20 & GBM\\ \hline 
HR  &   1.05 $\pm$ 0.02 & GBM\\ \hline
$E_{\rm p}$ ($\keV$) &$186.57_{-4.51}^{+4.16}$   & BAT+GBM+LLE+LAT\\ \hline
$F_{\rm p}$  & $8.90_{-5.0}^{+11.0}$ & GBM\\ \hline
$E_{\rm \gamma, iso}$ ($\rm erg$) & $  7.14^{+0.36}_{-0.31} \times 10^{53}$ &-\\ \hline
$L_{\rm \gamma, iso}$ ($\rm erg ~sec^{-1}$) & $2.78^{+3.44}_{-1.56}$ $ \times 10^{53}$ & -\\
\hline
Redshift $z$ & $2.02^{+0.05}_{-0.05}$ & XRT + UVOT + BOOTES \\
\hline
\end{tabular}
\end{center}
$T_{90}$: Duration from GBM data in 50-300 keV; \mvts : minimum variability time scale in 8-900 keV; HR: ratio of the counts in 50-300 keV to the counts in 10-50 keV; $E_{\rm p}$: Time-integrated peak energy calculated using joint \swift BAT, \fermi GBM and LAT data; $F_{\rm p}$: peak flux in $\rm 10^{-6} erg ~cm^{-2}$;
$E_{\rm \gamma, iso}$: Isotropic $\gamma$-ray energy in the source frame; $L_{\rm \gamma, iso}$: Isotropic $\gamma$-ray peak luminosity in the source frame.  All the results presented in this table are obtained using different analyses methods discussed in \S~\ref{multiwaveength observations and data analyisis} and \S~\ref{results}.
\end{table}
The GBM \citep{2009ApJ...702..791M} on board \fermi Gamma-Ray Space Telescope triggered at 21:17:37.81 UT (\fermiT). 
The \fermi GBM light curve shows two bright overlapping peaks (see Figure \ref{promptlc}) with a \tninty duration of $3.6\pm0.1$ sec in the 50 - 300 \keV energy range \citep{Bhat:2016ApJS..223...28N}. For a time interval from \fermiT + 0.4 to \fermiT + 4 sec, the time-averaged spectrum is best described with the \sw{Band} model \citep{Band:1993} with $\it \alpha_{\rm pt}$ = -0.71 $\pm$ 0.02, high energy spectral index $\beta_{\rm pt}$ = -2.49 $\pm$ 0.07, and \Ep = 186 $\pm$ 5 \keV.  
In this time interval, an energy fluence of 1.78 $\pm$ 0.03 $\times$ 10$^{-5}$ $\rm erg$ $\rm cm^{-2}$ is calculated in the 10-10000 \keV energy band \citep{2014GCN.15669....1Z}. This fluence value is among the top 12 percent most bright GRBs observed by the \fermi-GBM, making this burst suitable for detailed analysis. The LAT \citep{2009ApJ...697.1071A} on board \fermi triggered at 21:17:37.64 UT and detected high energy emission from \thisgrb. 
The best \fermi-LAT on-ground location (RA, DEC = 211.88, 1.36 (J2000)) was at 47$^{\circ}$ from the LAT boresight angle at \fermiT and the highest-energy photon with a energy of 8 GeV is detected 520 sec after \fermiT \citep{2014GCN.15659....1S}. 
 
\par
Other gamma-ray/hard X-ray space telescopes such as MAXI/GSC \citep[at 21:19:54 UT,][]{2014GCN.15663....1K} and \kw \citep[at 21:17:36.245 UT; \kwT;][]{2014GCN.15667....1G} also detected \thisgrb. 
For a time interval from \kwT to \kwT + 10.496 sec, the time-averaged KW spectrum is best fitted with \sw{Band} function \citep{Band:1993} with parameters $\it \alpha_{\rm pt}$ = -1.05 $\pm$ 0.14, $\beta_{\rm pt}$ = -2.68 $\pm$ 0.30, and \Ep = 185 $\pm$ 19 \keV \citep{2014GCN.15667....1G}.

\subsubsection{ \bf \fermi-LAT analysis}

\begin{figure}
\centering
\includegraphics[scale=0.33]{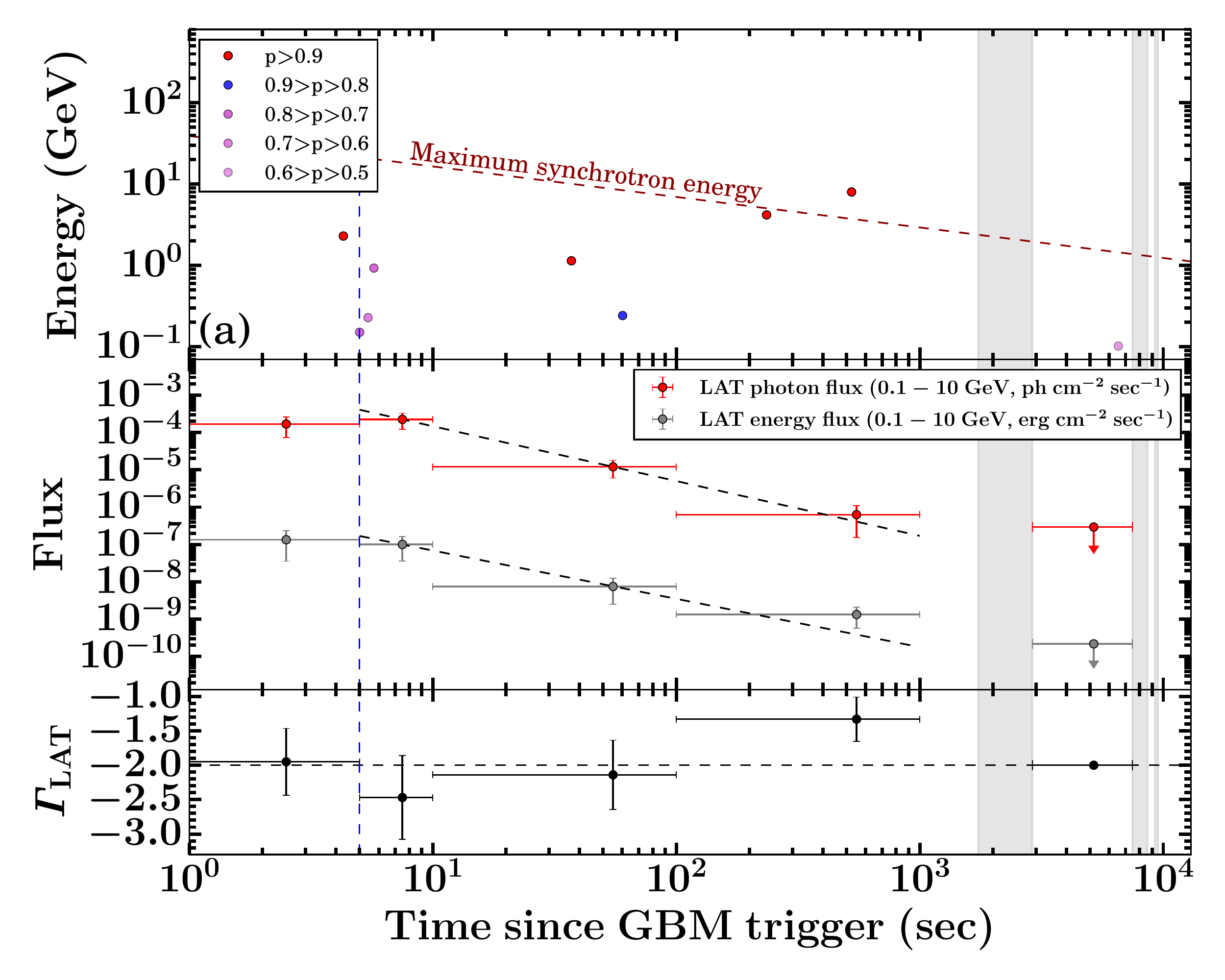}
\includegraphics[scale=0.33]{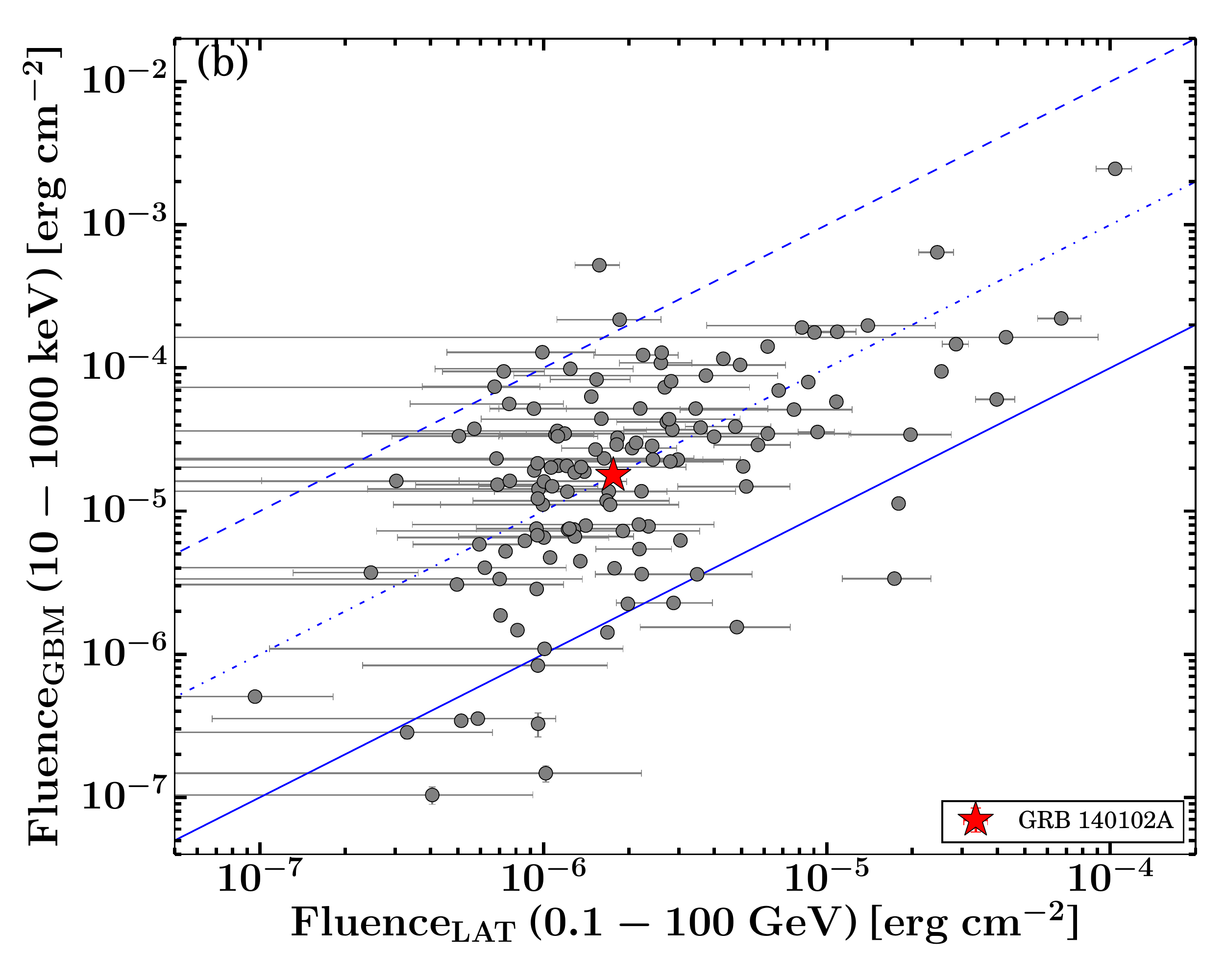}
\caption{{ \bf (a) Extended \fermi-LAT emission:} \textit{Top panel}: Temporal distribution of high energy LAT photons with energies $> 100$ MeV. The different colors of the photons represent their associated probabilities of originating from \thisgrb. \textit{Middle panel}: Evolution of \fermi-LAT energy (grey) and photon (red) fluxes in 100-10000 MeV range. In the last time bin, the LAT photon index was fixed to $-2$ to get an upper limit on the fluxes. The black dashed lines provide a power-law fit to the temporally extended photon flux and the energy flux light curves (for the photons detected 5 sec after \fermiT).
\textit{Bottom panel (b)}: Temporal evolution of \fermi-LAT photon index in the 0.1-300 GeV range. The vertical blue dashed line corresponds similarly to Figure \ref{promptlc}. The grey shaded region represents the source off-axis angle $ > 65^\circ$. (b) Comparison of GBM and LAT fluence values for \thisgrb along with LAT detected GRBs \citep{2019ApJ...878...52A}. The solid line represents equal fluence, and dashed-dotted, dashed lines denote the fluence shifted by factors of 10 and 100, respectively.}
\label{fig:LAT_LCs}
\end{figure}

We obtained the \fermi-LAT \citep{2009ApJ...697.1071A} data within a temporal window extending up to 10000 sec after \fermiT from \fermi-LAT data server\footnote{\url{https://fermi.gsfc.nasa.gov/cgi-bin/ssc/LAT/LATDataQuery.cgi}} using \sw{gtburst}\footnote{\url{https://fermi.gsfc.nasa.gov/ssc/data/analysis/scitools/gtburst.html}} version 02-03-00p33 software. We performed unbinned likelihood analysis and selected a region of interest (ROI) of $\rm 12^{\circ}$ around the enhanced \swift XRT position \citep{2014GCN.15657....1G}. Further, we filtered the high energy LAT emission by cutting on photons with energies in the range of 100 MeV-300 GeV. We also applied an angular cut of $\rm 100^{\circ}$ between the source and zenith of the satellite in order to reduce the contamination of photons coming from the Earth limb. For the time-integrated duration, we used the \sw{P8R3\_SOURCE\_V2 response}, which is appropriate for long durations ($\rm \sim 10^{3} ~sec$), and for the time-resolved bins, we used \sw{ P8R2\_TRANSIENT020E\_V6} response, which is appropriate for small durations \citep{2018arXiv181011394B}. We included a point source (for GRB) at the location of the burst, considering a power-law spectrum, an isotropic component (to show the extragalactic background emission) \sw{iso\_P8R3\_SOURCE\_V2} and a Galactic diffuse component (to represent the Galactic diffuse emission) \sw{gll\_iem\_v07} \footnote{\url{https://fermi.gsfc.nasa.gov/ssc/data/access/lat/BackgroundModels.html}}. The probability of association of the photons with \thisgrb is calculated using the \sw{gtsrcprob} tool. For the total duration of 0-10000 sec, the energy and photon flux in 100-10000 MeV energy range are $(5.96 \pm 2.37) ~\times 10^{-10}$ $\rm ~erg ~cm^{-2} ~ sec^{-1}$ and $(5.01 \pm 2.35) ~ \times 10^{-7}$ $\rm ~ph. ~cm^{-2} ~ sec^{-1}$, respectively, with a test-statistic (TS)\footnote{ It is define as TS = -2ln$L_{\rm max,0}$/$L_{\rm max,1}$, where $L_{\rm max,0}$ and $L_{\rm max,1}$ are the maximum likelihood value for a model without and with an additional source, respectively.} of detection 46. The temporal distribution of \fermi-LAT photons with energies $> 100$ MeV along with the photon flux and the energy flux light curves is shown in Figure \ref{fig:LAT_LCs} (a). During the GBM time window, we obtained the LAT fluence value equal to $(0.18\pm0.10) ~\times$ $10^{-5}$ erg $\rm cm^{-2}$ in 0.1-100 GeV energy range. We compared this value with the GBM fluence value for \thisgrb along with other LAT detected GRBs \citep[ the second GRB LAT catalogue (2FLGC);][]{2019ApJ...878...52A, 2011ApJ...726...22A}. \thisgrb lies on the line for which GBM fluence is 10 times brighter than LAT fluence for all the LAT detected sample (see Figure \ref{fig:LAT_LCs} (b)). Further, we also perform (see \S~\ref{lat_TRS}) the time-resolved spectral analysis to investigate the origin of the high energy LAT photons (see Table A1 in appendix A). 

\subsubsection{\bf \fermi-GBM analysis}
The \fermi-GBM \citep{2009ApJ...702..791M} time-tagged event (TTE) data is downloaded from the GBM trigger data archive\footnote{\url{https://heasarc.gsfc.nasa.gov/FTP/fermi/data/gbm/triggers/}}. We investigated the temporal and spectral prompt emission properties of \thisgrb using the three brightest sodium iodide (NaI 6, 7, and 9) and one brightest bismuth germanate (BGO 1) detectors. These detectors have source observing angles : NaI 9 - $\rm 31^\circ$, NaI 7 - $\rm 33^\circ$, NaI 6 - $\rm 35^\circ$, and BGO 1 - $\rm 42^\circ$. We obtained the multi-channel (energy-resolved) prompt emission light curve using \sw{RMFIT} version 4.3.2 software\footnote{\url{https://fermi.gsfc.nasa.gov/ssc/data/analysis/rmfit/}}. The background-subtracted multi-channel prompt emission $\gamma$-ray/ hard X-ray light curves along with the hardness ratio are shown in Figure \ref{promptlc}. The light curve consists of two major bright overlapping peaks. The HR shows that the first episode is softer than the subsequent episodes, which is also visible from the very low signal for the first episode in LLE data\footnote{\url{https://heasarc.gsfc.nasa.gov/W3Browse/fermi/fermille.html}} \citep[][see appendix]{2010arXiv1002.2617P, 2014ApJ...789...20A}.

\par
We created time-averaged spectra, using the \sw{gtburst} software from the \emph{Fermi Science Tools}, using the TTE mode data from the same detectors used for the temporal analysis. The main GRB emission duration (\fermiT to \fermiT+ 5 sec) is selected for the time-averaged spectral analysis. The background is fitted by selecting two intervals, one prior to the burst and another after the burst. The spectral analyses are performed using X-Ray Spectral Fitting Package \citep[\sw{XSPEC};][]{1996ASPC..101...17A} version 12.10.1 of \sw{heasoft-6.25}. The statistics \sw{pgstat} is used for optimization and testing the various models as \fermi-GBM data is Poisson in nature with Gaussian background. We modelled the data with traditional \sw{Band} function 
For the time-averaged spectrum obtained using GBM data only, we calculated  $\it \alpha_{\rm pt}$ = -0.75$^{+0.04}_{-0.04}$, $\beta_{\rm pt}$ = -2.57$^{+0.14}_{-0.18}$, and \Ep = 185.42$^{+9.54}_{-9.67}$ \keV and these parameters are consistent with those determined by \cite{2014GCN.15669....1Z}.

\subsubsection{\bf {\swift-BAT analysis}}
We retrieved the \swift-BAT \citep{2005SSRv..120..143B} data from the Swift Archive Download Portal hosted by the UK Swift Science Data Centre\footnote{\url{https://www.swift.ac.uk/swift_portal/}}. For the BAT data analysis, we used \sw{HEASOFT} software version-6.25 with latest calibration database\footnote{\url{https://heasarc.gsfc.nasa.gov/FTP/caldb/}}. We began the reduction of BAT data from the raw files (Observation Id: 00582760000). We produced detector plane image (DPI) using \sw{batbinevt} and identify the hot pixels with \sw{bathotpix}. We applied mask-weighting (background subtraction) in the event file using the \sw{batmaskwtevt} pipeline. The light curves 
is extracted using \sw{batbinevt} tool. The energy resolved \swift BAT mask-weighted light curve is presented in Figure \ref{promptlc} along with the evolution of HR in 25-50 \keV to 15-25 \keV energy ranges. The mask-weighted light curve consists of two peaks starting at \fermiT-0.2 sec, peaking at \fermiT+0.7 sec and ending at \fermiT+5 sec with a long tail out to \fermiT+200 sec with \tninty duration (15-350 \keV) of 55 $\pm$ 15 sec \citep{2014GCN.15662....1B, 2016ApJ...829....7L}. 

\par
We extracted the \swift BAT spectrum corresponding to the times of our selection for \fermi GBM time-averaged spectral analysis. The detailed method used for the reduction of the spectrum is as described in \swift-BAT software guide\footnote{\url{https://swift.gsfc.nasa.gov/analysis/bat_swguide_v6_3.pdf}}. We applied energy correction using \sw{bateconvert} to ensures that the PHA to PI energy conversion is quadratic. We used \sw{batbinevt} to create the BAT spectrum after producing DPI, correcting for hot pixels, and mask-weighting. Further, \sw{batphasyserr} and \sw{batupdatephakw} are used for compensating the residuals in the response matrix and to ensure the location of \thisgrb in instrument coordinates. In order to model the BAT spectrum, we built a detector response matrix (DRM) using tool \sw{batdrmgen}.
We modelled the BAT spectrum using the \sw{XSPEC} software. We used \sw{$\chi^2$} statistic to model the spectrum and to find the best-fit model of several different models. The time-averaged spectrum from \swiftT-0.148 sec to \swiftT+131.22 sec is best fitted by a simple power-law model with photon index ($\it \Gamma_{\rm BAT}$) of 1.37 $\pm$ 0.04 \citep{2014GCN.15662....1B,2016ApJ...829....7L}. 
 
\subsubsection{\bf {Joint \fermi and \swift spectral analysis}}
We performed a joint spectral analysis of \fermi and \swift BAT data using the Multi-Mission Maximum Likelihood framework \citep[\sw{3ML}\footnote{\url{https://threeml.readthedocs.io/en/latest/}}]{2015arXiv150708343V} version 2.3.1 software to constrain the properties of the emission mechanism of \thisgrb. We extracted the time-averaged \fermi-GBM, LAT, and \swift BAT spectrum for a duration of \fermiT to \fermiT+ 5 sec for the joint spectral analysis. We modelled the time-averaged joint spectrum with \sw{Band} function and explored various other models such as \sw{Black Body} along with \sw{Band} function to search for thermal component in the burst; a power-law with two breaks (\sw{bkn2pow}), and a smoothly broken power-law (\sw{SBPL}) or their combinations based upon residuals of the data. We calculated Bayesian information criteria  \citep[BIC;][]{Kass:1995}, Akaike information criterion (AIC), and Log (likelihood) for each model to find the best-fit model. We consider GBM spectrum over the energy range of 8-900 \keV (NaI detectors) and 200-30000 \keV (BGO detectors) for spectral analysis. However, we ignore the K-edge (33–37 \keV) energy range for the analysis of NaI data. We consider 100 MeV-300 GeV energy channels for the \fermi LAT observations.

The best-fit spectral parameters of the joint analysis are presented in appendix A. We found that of all the six models used, the \sw{Band} model with a \sw{Black Body} component has the lowest BIC value. Therefore, we conclude that the time-averaged spectrum of \thisgrb is best described with \sw{Band} + {\sw{Black Body}} function. Furthermore, time-resolved spectroscopy (see \S~\ref{trs_section}) suggests that the presence of apparent BB is due to spectral evolution.

\begin{figure}
\centering
\includegraphics[scale=0.32]{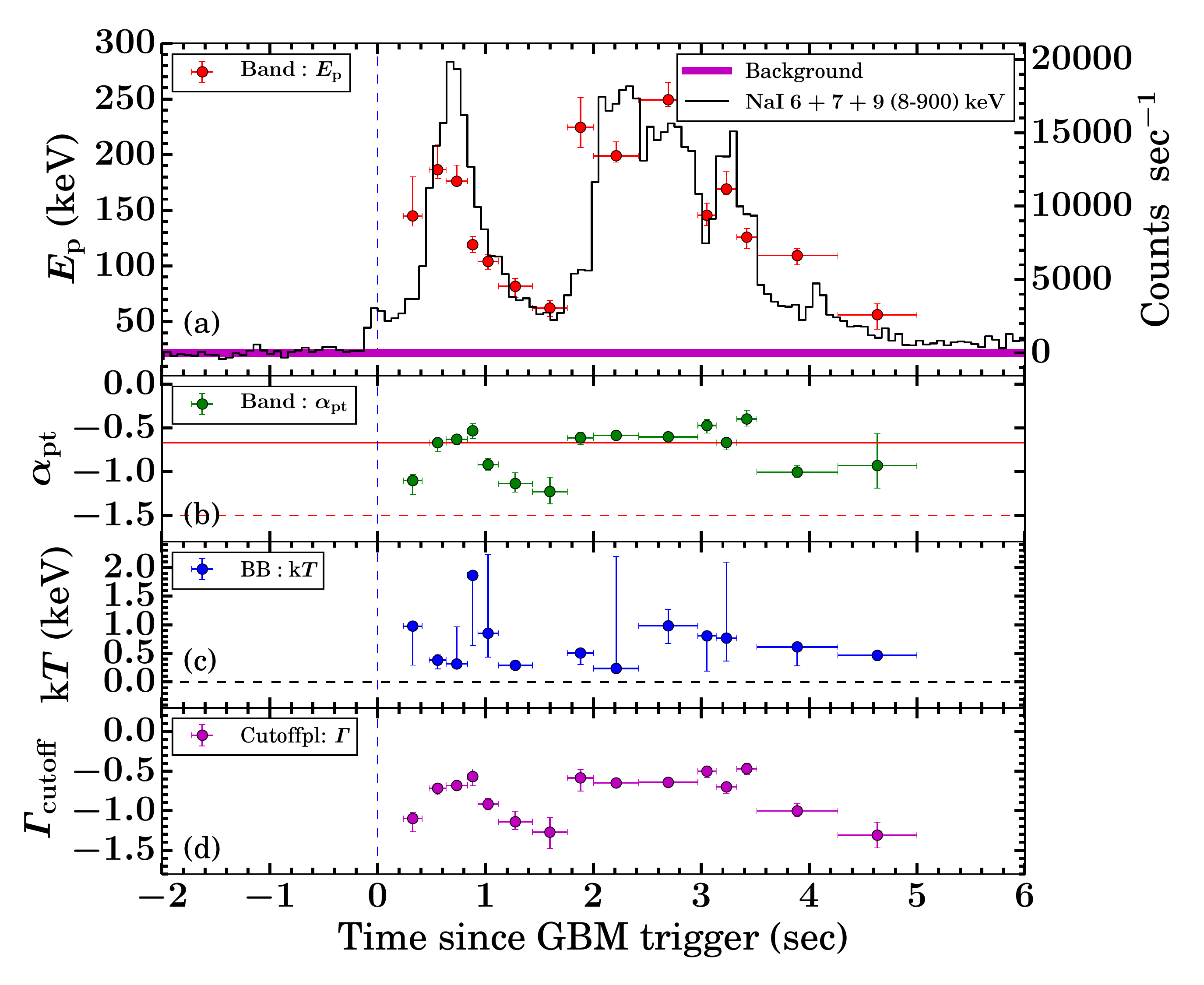}
\caption{{\bf Evolution of spectral parameters obtained for \thisgrb}: (a) The evolution of the peak energy (red circles) is overlaid on the GBM light curve (solid black line). The horizontal magenta solid line shows the background level. (b) The evolution of the low energy spectral index using the joint \fermi GBM and \swift BAT data. The two horizontal lines are lines of death for synchrotron fast cooling ($\it \alpha_{\rm pt} = -3/2$, dotted red line), and synchrotron slow cooling ($\it \alpha_{\rm pt} = -2/3$, red solid line). (c) The evolution of k$T$ (\keV) obtained from the \sw{Black Body} component. The horizontal black dashed line shows k$T$ = 0 \keV. (d) Evolution of photon indices for the \sw{Cutoff-power law} model. Binning has been performed based on the Bayesian Block algorithm.}
\label{TRS}
\end{figure}

\subsubsection{\bf {Time-resolved spectroscopy of \thisgrb}}
\label{trs_section}

Time-resolved spectroscopy is a promising tool to investigate the radiation mechanisms (non-thermal synchrotron or thermal photospheric model) and to study the correlations among the spectral parameters of GRBs, which are still an unsolved problem of the prompt emission \citep{2011CRPhy..12..206Z}. Initially, we resolved the main emission interval so that each extracted spectrum had a signal-to-noise ratio (SNR) equal to 25 using \fermi GBM data and model each spectrum with \sw{Band} function only. We notice a strong correlation of $\it \alpha_{\rm pt}$ and \Ep with flux. Further, we resolved the total emission interval of \thisgrb based on the Bayesian Block algorithm integrated over the 8-900 \keV in the detector with a maximum count rate (NaI 9) and jointly analyzed \fermi GBM and \swift BAT data as the Bayesian Block algorithm is the most suitable method to identify the intrinsic intensity change in the prompt emission light curve \citep{2013arXiv1304.2818S, 2014MNRAS.445.2589B}. This resulted in 17 spectra, however, two bins do not have sufficient counts to be modeled. We adopted the various models (\sw{Band}, \sw{Black Body}, and \sw{Cutoff-power law} or their combinations) for time-resolved spectral analysis. We find that most of the temporal bins are well described with the \sw{Band} function only. The best-fit parameters and their associated errors are listed in appendix A. The evolution of spectral parameters along with the light curve is shown in Figure \ref{TRS}. As can be seen from Figure \ref{TRS}, the value of \Ep is changing throughout the burst resulting in spectral evolution. The \Ep evolution shows an intensity-tracking trend throughout the emission. Similarly, the evolution of $\it \alpha_{\rm pt}$ also follows an intensity-tracking trend for the first episode, whereas this tracking behavior is less clear during the second peak, and it exceeds the line of death for synchrotron slow cooling \citep{2013ApJS..208...21G, 2018ApJ...862..154C} in most of the bins of the second episode, therefore the emission of the second episode may have a non-synchrotron origin. The light curve of this burst has a complicated structure and consists of two emission episodes, each of them consist of several overlapped pulses: the first episode consists of at least one pulse, and the second episode consists of at least three pulses (see Figure \ref{TRS} (a)). The superposition effect, suggested in \cite{2014AstL...40..235M}, could lead to intensity tracking behavior of the spectral evolution.

\subsection{\swift XRT observations}
\begin{table*}
  \caption{The best fit models to the X-ray, combined optical light curves, and individual optical filters.}
  \begin{tabular}{|c|c|c|c|c|c|c|}
  \hline
 \bf Wavelength & \bf Model &$\bf \alpha_1$ &\bf  $\bf \alpha_2 $ or $\bf \alpha_{\rm x1}$ &\bf  $\rm \bf {t_{break}}$ or $\rm \bf t_{\rm bx}$ (sec) & $ \bf \alpha_3$ or $\bf \alpha_{\rm \bf x2}$ & $\bf \chi^2/dof$ \\
  \hline
  \hline
 X-ray             & broken power-law              & ---                   & $-1.09\pm0.01$  & $1298^{+108}_{-74}$ & $-1.50\pm0.02$           &  575/543\\
 Optical/IR         & power + broken power           & $-1.72\pm0.04$         & $-0.47\pm0.03$  & $6160^{+740}_{-240}$   & -$1.11\pm0.15$ & 109/33 \\ 

 \hline                
  BOOTES Clear      & 2x power-law                  & $-1.72\pm0.04$        & $-0.41\pm0.04$  & ---                & ---                      & 199/104 \\
 UVOT $white$        & 2x power-law                  & $-2.19^{+0.27}_{-0.41}$ & $-0.86\pm0.07$  & ---                & ---                      & 41/48   \\
  BOOTES Clear      & Broken power-law              & ---                   & $-1.49\pm0.01$  & $475\pm15$         & $ -0.70\pm0.01$          & 315/106 \\
 UVOT $white$  & Broken power-law              & ---                   & $-1.49\pm0.05$  & $294\pm58$         & $ -0.88\pm0.07$          & 43/48   \\
 \hline
  \end{tabular}

  \label{lcfits}
\end{table*}

The X-ray Telescope \citep[XRT;][]{2005SSRv..120..165B} began observing the field of \thisgrb on January 02, 2014 at 21:18:34.3 UT, 56.8 sec after the BAT trigger. The XRT detected a bright, uncatalogued X-ray source located RA, and DEC = 211.9190, 1.3333 degrees (J2000) with an uncertainty of 4.8$\rm "$ (radius, 90 $\%$ containment). This location is 61$\rm "$ from the BAT onboard position but within the BAT error circle \citep{2014GCN.15653....1H}. In total, there are 9.5 ks of XRT data for \thisgrb, from 47 sec to 205 ks after the BAT trigger. The data comprise $\sim$ 2.2 ks in Windowed Timing (WT) mode (the first $\sim$ 8 sec were taken while \swift was slewing) with the remainder in Photon Counting (PC) mode. In this paper, we used X-ray data products (both light curve and spectrum) available on the \swift online repository\footnote{\url{https://www.swift.ac.uk/}} hosted by the University of Leicester \citep{eva07,eva09}.

\par
We modelled the X-ray light curve using a simple power-law function, and  broken power-law model (BPL). We find that the XRT light curve could be best described with a broken power-law model (see \S~\ref{xray_optical lc fitting}). We calculated $\alpha_{\rm x1}$ = $-1.09^{+0.01}_{-0.01}$, $\alpha_{\rm x2}$ = $-1.50^{+0.02}_{-0.02}$, and $t_{\rm bx}$ = $1298^{+108}_{-74}$ sec, where $\alpha_{\rm x1}$ is the temporal index before break ($t_{\rm bx}$), and $ \alpha_{\rm x2}$ corresponds to the temporal index after break. All the temporal parameters are also listed in Table \ref{lcfits}.

\par
We analyzed the \swift XRT spectra using the XSPEC package using an absorbed power-law model in 0.3-10 \keV energy channels.
The absorption includes photoelectric absorption from our Galaxy and the host galaxy of the GRB using the XSPEC components \sw{phabs} and \sw{zphabs} together with the source spectral model. We considered the Galactic hydrogen column density fixed at $\rm NH_{\rm Gal}= 3.04 \times 10^{20}{\rm cm}^{-2}$ \citep{2013MNRAS.431..394W}, and a free intrinsic hydrogen column density located at the host redshift {\bf ($\rm NH_{\rm z}$)}. The XRT spectra were grouped with a minimum of 1 count per bin unless mentioned otherwise. The statistics \sw{C-Stat} is used for optimization and testing the various models. The redshift of the second absorption component was fixed at 2.02 as discussed in \S~\ref{photoz}. We also search for an additional thermal (\sw{Black Body}) and other possible components in the early time WT mode spectra (63-200 sec), but we did not find any signature of thermal component based on the BIC value comparison obtained for simple absorption power-law model. All the spectral parameters for the absorbed power-law model have been listed in Table \ref{xspectra}.

\par
\begin{figure*}
\centering
 \includegraphics[height=9cm,width=8.82cm,angle=0]{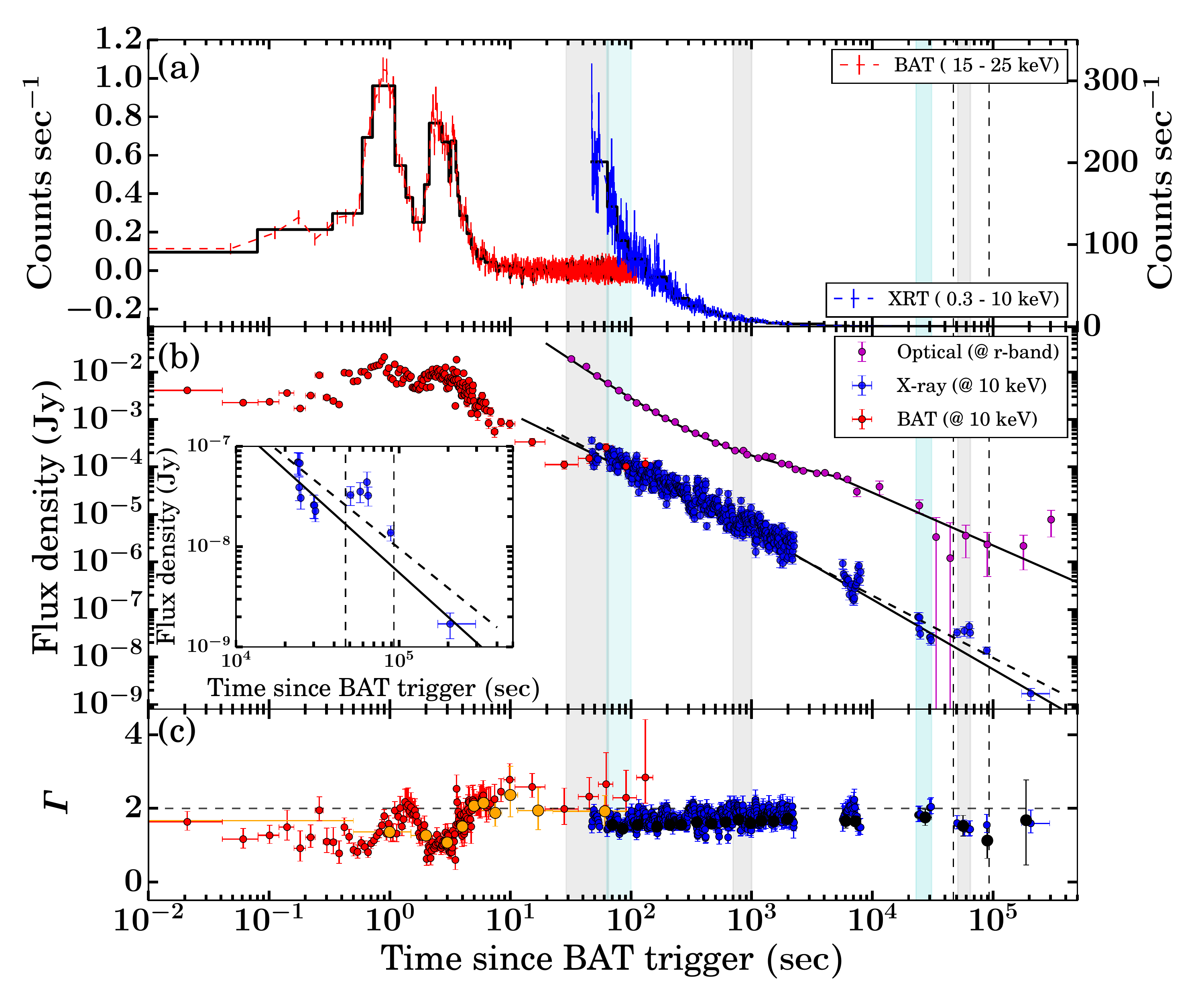}
\includegraphics[height=9cm,width=8.82cm,angle=0]{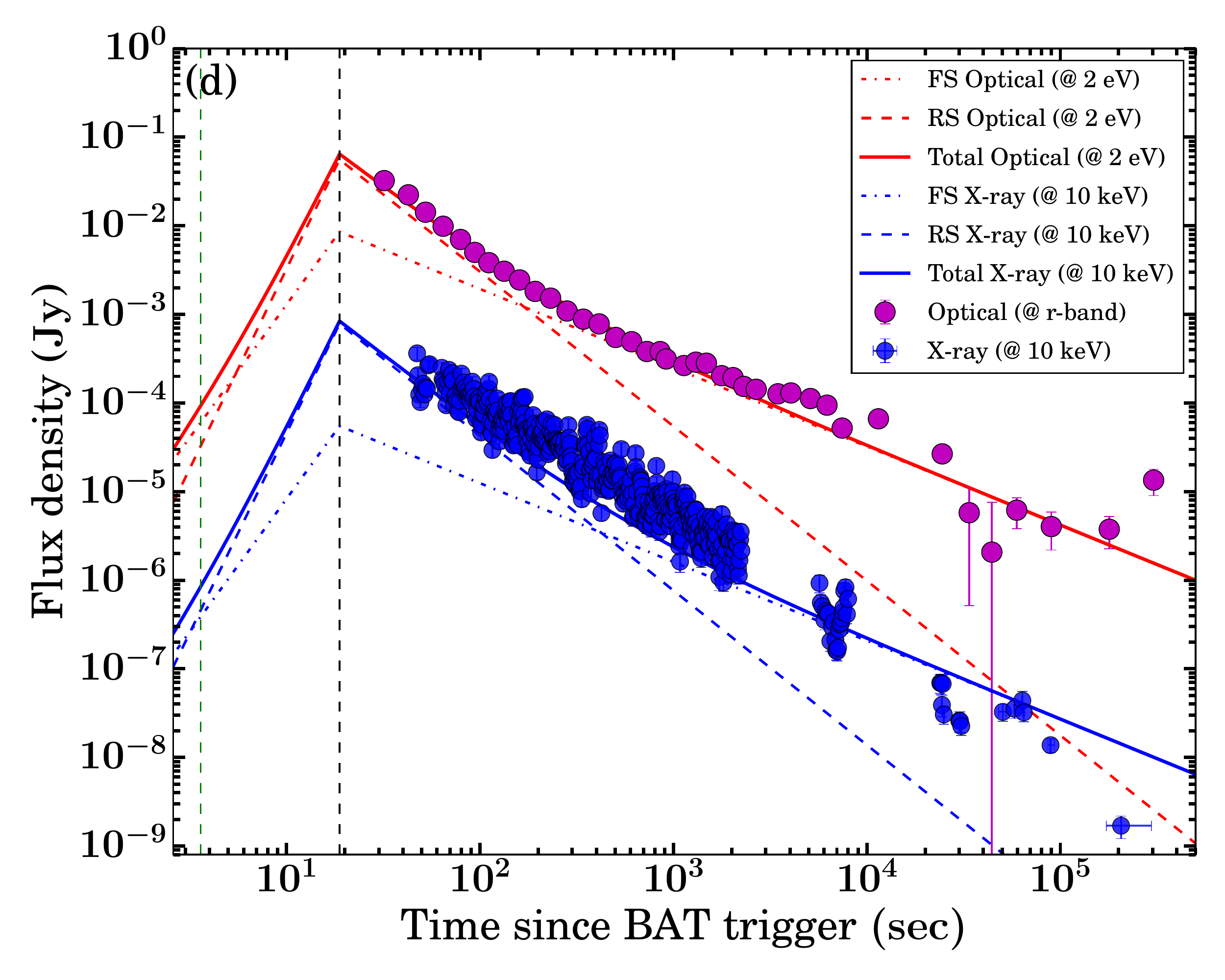}
\caption[]{{\bf Multiwavelength light curve of \thisgrb :} (a) Count rate BAT and XRT light curves overlaid with the Bayesian Block analysis. (b) The X-ray and optical/UV afterglow light curves of \thisgrb overlaid with the best-fit models: a broken power-law (X-ray, solid black line) or simple power-law (X-ray,  black dashed line) or its combinations (optical, black solid line). The inset plot shows the late re-brightening activity in the X-ray light curve. (c) Evolution of photon indices within the BAT (red) and XRT (blue) window\protect\footnotemark. The orange and black circles indicate the evolution of BAT and XRT photon indices obtained using our spectral analysis of Bayesian Blocks bins. The shaded vertical grey and cyan color bars represent the epochs used to create the spectral energy distributions (SEDs) of \thisgrb afterglow. The vertical dashed lines indicate the late re-brightening activity in the X-ray light curve (at 10 keV). (d) A combined forward and reverse shock model is used for the interpretation of the optical emission from \thisgrb. The peak flux is obtained at the crossing time $t_{\rm x}$ = 18.79 sec and for $t> t_{\rm x}$ power-law behavior of the afterglow model is followed. The optical flux is explained using the sum of reverse and forward shock components. The same set of parameters produce a lower amount of early X-ray flux in this model. The vertical green and black dashed lines indicate the end time of \tninty duration and deceleration time, respectively. The model parameters are given in Table \ref{sample_Modelling}.}
\label{Xray_optical_afterglow}

\end{figure*}

Initially, we created two time-sliced spectra (before and after the break time) using the UK Swift Science Data Centre webpages\footnote{\url{https://www.swift.ac.uk/xrt_spectra/addspec.php?targ=00582760}} and found spectral decay index $\beta_{\rm x1}$ = $0.56^{+0.02}_{-0.02}$ (before break) and  $\beta_{\rm x2}$ = $0.67^{+0.06}_{-0.06}$ (after break). Further, we created three more time-sliced spectra, using data between 63-200 sec, 200-2200 sec, and 5600- $2 \times 10^5$ sec to observed the evolution of decay slope in the X-ray band ($\beta_{\rm x}$). We notice, $\beta_{\rm x}$ evolves and increases continuously in each temporal bins.  However, the photon index obtained from the Swift Burst Analyser webpage \footnote{\url{https://www.swift.ac.uk/burst_analyser/00582760/}} could be fitted with a constant (\sw{CONS}) model\footnote{\url{https://www.ira.inaf.it/Computing/manuals/xanadu/xronos/node93.html}}, and at very late times the photon index is different and seems to have dropped (see Figure \ref{Xray_optical_afterglow} (c)). We observed a late time ($\sim$ 4.7 $\times$ $10^{4}$ to $\sim$ 9.3 $\times$ $10^{4}$ sec) re-brightening activity in the unabsorbed X-ray flux light curve (at 10 \keV). 

\begin{figure}
\includegraphics[scale=0.33]{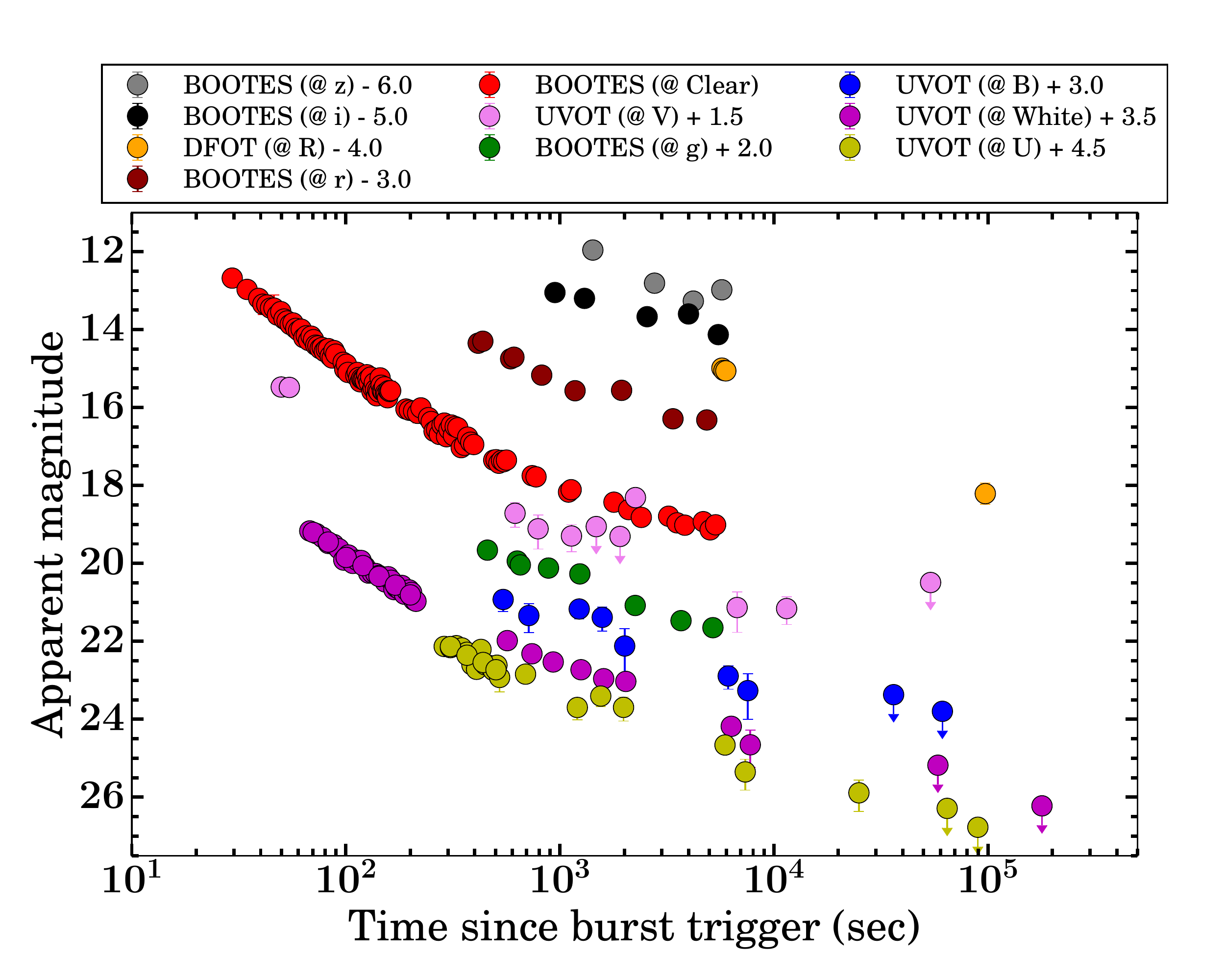}
\caption{{\bf Multi-band optical light curve:} Multi-band light curves of the afterglow of \thisgrb using UVOT, BOOTES, and {\bf Devasthal fast optical telescope (DFOT)} data as a part of the present analysis as tabulated in appendix A. Magnitudes are not corrected for the extinctions.}
\label{optical_multiband}
\end{figure}

\begin{table}
\centering
\caption{The best-fit values of Photon indices from the X-ray afterglow spectral modelling for different temporal segments. We have frozen the host hydrogen column density, obtained from SED 3 (see \S~\ref{SED}).}

\begin{tabular}{|c|c|c|c|c}
\hline
\bf Time (sec) &$\rm \bf Photon ~index$ &  \bf Mode \\
\hline
\hline
63-65196 & $ 1.59 \pm 0.09$ &  WT+PC \\
63-200   & $ 1.51 \pm 0.03$ & WT\\
200-2200   & $ 1.61 \pm 0.02$ & WT \\
5600-$2 \times 10^5$& $ 1.63 \pm 0.09$ & PC\\
\hline
\end{tabular}
\label{xspectra}
\end{table}

\footnotetext{\url{https://www.swift.ac.uk/burst_analyser/00582760/}}
    
\subsection{UV/Optical Observations}

 Ultra-violet and optical observations were carried out using \swift-UVOT, BOOTES robotic, and 1.3m DFOT telescopes as a part of the present work. Details of these observations are given below in respective sub-sections. Multi-band light curves obtained from these observations are presented in Figure \ref{optical_multiband}.

\subsubsection{\bf {\swift Ultra-Violet and Optical Telescope}}

The \swift UVOT began observing the field of \thisgrb 65 sec after the BAT trigger \citep{2014GCN.15653....1H}. The afterglow was detected in 5 UVOT filters. UVOT observations were obtained in both image and event modes. Before extracting count rates from the event lists, the astrometric corrections were refined following the methodology described in \cite{oates09}. For both the event and image mode data, the source counts were extracted using a region of 3" radius. In order to be consistent with the \swift UVOT calibration, the count rates were then corrected to 5" using the curve of growth contained in the calibration files using standard methods. Background counts were estimated using a circular region of radius 20" from a blank area of sky situated near the source position. The count rates were obtained from the event and image lists using the \swift standard tools \sw{uvotevtlc} and \sw{uvotsource}, respectively\footnote{\url{https://www.swift.ac.uk/analysis/uvot/}}. Later, these counts were converted to magnitudes using the UVOT photometric zero points \citep{bre11}. The UVOT data for this burst is provided in appendix A. As with data from all the following telescopes, the resulting afterglow photometry is given in AB magnitudes and not corrected for Galactic reddening of E(B-V)= 0.03 \citep{sch11}.

\subsubsection{\bf {BOOTES Robotic Telescope Network}}
The BOOTES-4/MET robotic telescope in Lijiang, China \citep{cast12} automatically responded to the BAT trigger \citep[][]{2014GCN.15653....1H} with observations starting at 21:18:05 UT on January 02, 2014, 28 sec after the trigger. For the first 5 minutes, observations were taken with a series of 0.5 sec clear filter exposures, after which observations were taken with a systematic increase in exposure. Starting from $\sim 6$
minutes after the trigger, observations were also performed in rotation with the $g$, $r$, $i$, $Z$, and $Y$ filters. The images were dark-subtracted and flat-fielded using custom IRAF routines. The aperture photometry was extracted using the standard IRAF software and field calibration was conducted using the SDSS DR12 catalogue \cite{ala15}. Color transforms of \cite{hew06} were used when calibrating the $Z$ and $Y$ filters. The BOOTES data
is provided in appendix A.

\subsubsection{\bf {1.3m ARIES Telescope}}
The 1.3m DFOT at ARIES, Nainital started observing the field of \thisgrb $\sim$ 1.5 hours after the trigger. Several frames in V, $\rm R_c$ and $\rm I_c$ pass-bands were obtained in clear sky conditions. Images were dark-subtracted, but not flat fielded since there were no available flats taken on the same night and flat fielding using those taken on a different night made the science images worse. Aperture photometry was extracted using the standard IRAF software, and field calibration
was conducted using the SDSS DR12 catalogue \cite{ala15} and the color transforms of Robert Lupton in the SDSS online documentation\footnote{\url{http://www.sdss.org/dr12/algoritms/sdssUBVRITransform/##Lupton2005}}. The optical data using the 1.3m DFOT telescope along with other data sets are provided in appendix A.

\subsection{Combined Optical Light curve}
\label{combLC} 
In order to maximize the SNR of the optical light curves, we followed the methodology of \cite{oates09} to
combine the individual filter light curves into one single filter \citep{2010MNRAS.401.2773S, kan10, oates12, 2017ApJS..228...13R, 2019A&A...632A.100H}. First, the ground-based photometry was converted from
magnitudes to count rate using an arbitrary zero-point. The light curves from the different filters were then normalized to the r filter. The normalization was determined by fitting a power-law to each of the light curves in a given time range simultaneously. The power-law indices were constrained to be the same for all the filters, and the normalizations were allowed to vary between the filters. For the light curve of \thisgrb, the power-law was fitted between 600-6000 sec since this is the epoch which maximized the number of filters, and the behavior in each filter appeared to be similar. However, the normalization led to a slight offset between the UVOT $white$ and $V$ band data in comparison to the BOOTES-clear filter. We will discuss the consequences on model fitting in \S~\ref{optical_afterglow_modelling}. After the light curves were normalized, they were binned by taking the weighted average of the normalized count rates in time bins of $\Delta T/T = 0.2$. The combined optical light curve is shown in Figure \ref{Xray_optical_afterglow}.

\subsection{Spectral Energy Distributions}
\label{SED}

\begin{figure}
\includegraphics[scale=0.28]{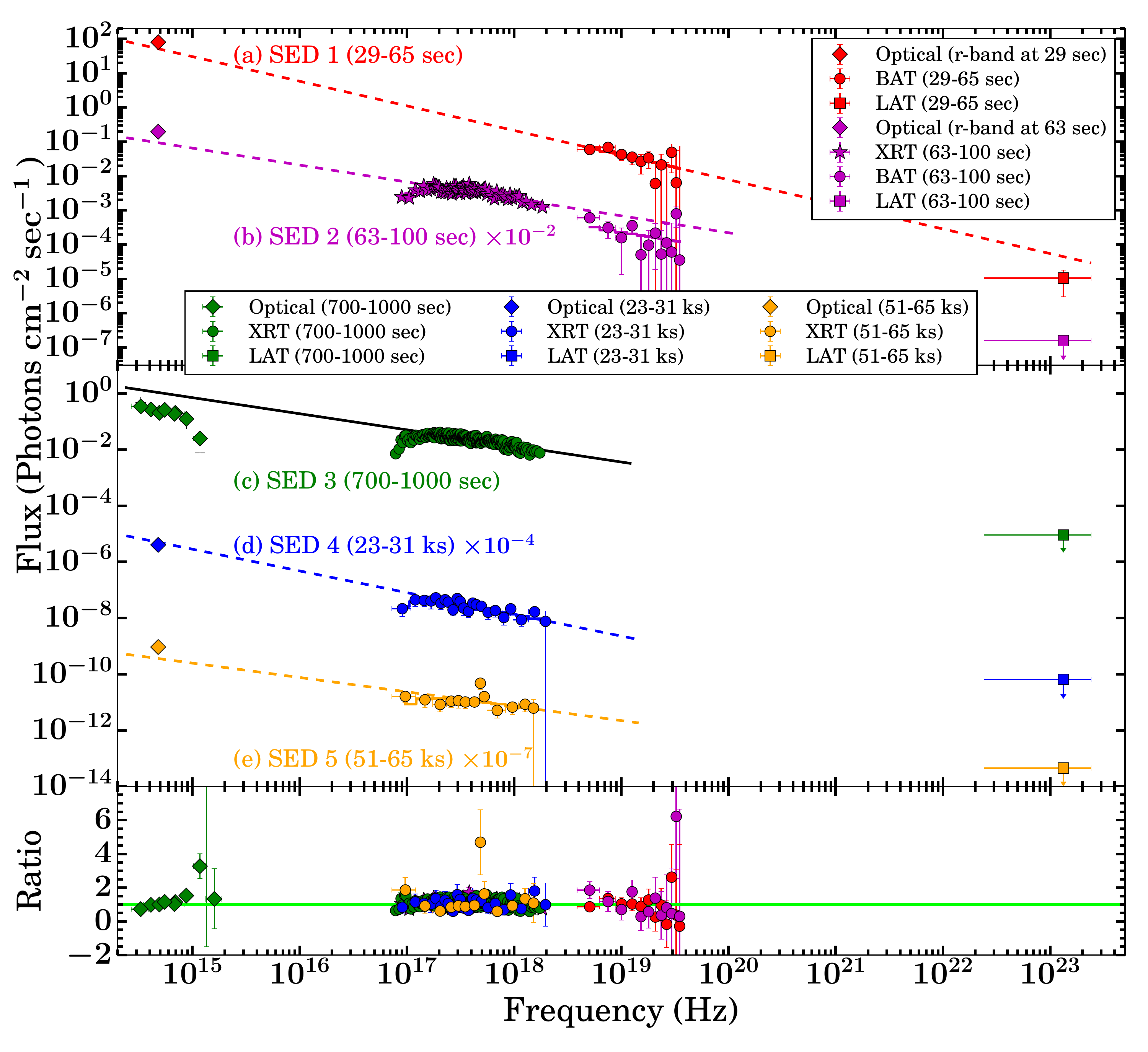}
\caption{{\bf Spectral energy distributions:} (a) SED 1 at 29-65 sec, the black dashed lines show the extrapolations of the power-law fit to the BAT spectra to the optical and LAT frequencies. The observed LAT flux is lower than the extrapolated value, and it indicates a presence of possible spectral break between BAT and LAT frequencies. (b) SED 2 at 63-100 sec, the black dashed lines show the extrapolations of the unabsorbed power-law model to the optical frequencies. (c) SED 3 at 700-1000 sec, used to constrain the redshift of the burst. The solid black line is the best-fit by a simple power-law from the joint X-ray and Optical spectral fit. (d) SED 4 23-31 ks. (e) SED 51-65 ks. The bottom panel shows the ratio of data to the model. The horizontal lime green solid line corresponds to the ratio equal to 1.}
\label{SED_fig}
\end{figure}

SEDs were produced at five epochs (29-65 sec, 63-100 sec, 700-1000 sec, 23-31 ks, and 51-65 ks) following the prescription within \cite{depas07}, which is based on the methodology of \cite{schady07}. We used XSPEC \citep{1996ASPC..101...17A} to fit the optical and X-ray data. We adopted two models, a simple power-law model and a broken-power law model. The difference between the indices of the broken power-law model was fixed at 0.5, consistent with the expectation of a synchrotron cooling break \citep[see \S~\ref{optical_afterglow_modelling},][]{zhang04}. In each model, we include a Galactic and intrinsic absorber using the XSPEC models \sw{phabs} and \sw{zphabs}. The Galactic absorption is fixed to $\rm NH_{Gal} = 3.04 \times 10^{20} {\rm cm^{-2}}$ \citep{2013MNRAS.431..394W}. We also include Galactic and intrinsic dust components using the XSPEC model
\sw{zdust}, one at redshift $z$= 0, and the other is free to vary. The Galactic reddening was fixed at E(B-V)= 0.0295 according to the map of \cite{sch11}. For the extinction at the redshift of the burst, we test Milky Way, Large and Small Magellanic Clouds (MW, LMC and SMC) extinction laws \citep{pei92}. All the five SEDs are shown in Figure \ref{SED_fig}. We extrapolated the best-fit model towards the LAT frequencies to search for a possible spectral break between X-ray and GeV. All the results of SEDs are listed in Table \ref{SED_table}. 

\subsubsection{\bf {Photometric Redshift}}
\label{photoz}
A spectroscopic redshift for the GRB was not reported and could not be determined from GTC spectroscopic observations of the host galaxy (see \S~\ref{Host Galaxy}). To determine a photometric redshift, we, therefore, created a SED using the afterglow data (X-ray and Optical) between 700 sec and 1000 sec. During this time, there is no break in the X-ray light curve, also,
no spectral evolution is observed. The evolution of the X-ray photon index ($\it \Gamma_{\rm XRT}$) measured during this time window is consistent with not changing (see Figure \ref{Xray_optical_afterglow}). Using the methodology outlined in \S~\ref{SED}, we fit the SED with the simplest model, a power-law, which we find to be statistically unacceptable with \sw{$\chi^2$} for the  LMC and SMC 282 and 316 respectively for 213 degrees of freedom. However, for the MW model, we find a better \sw{$\chi^2$} (223) for the same number of degrees of freedom. Further, we fit the SED with the broken power-law model. For the MW, LMC, and SMC, all the fits are statistically acceptable with \sw{$\chi^2$} of 169, 178, 185, respectively for 212 degrees of freedom.  We notice that for this SED (700-1000 sec), the MW model both for power-law and broken power-law fit show the smallest reduced \sw{$\chi^2$}, however, the reduced \sw{$\chi^2$} is less than 1 for broken power-law fit indicating that the model is "over-fitting". In the case of power-law model, the reduced \sw{$\chi^2$} is close to one. It confirms that the power-law model is the best fit. The redshift values for
all three extinction laws are similar and consistent within 3 $\sigma$. Taking the redshift from the MW model, we find a photometric redshift of $2.02^{+0.05}_{-0.05}$ for \thisgrb. We adopted this value for all our subsequent analyses.

\begin{table}
\begin{center}
\caption{The best fit results of optical and X-ray spectral indices at different epoch SEDs and their best describe spectral regime. $p$ is the mean value of the electron distribution indices calculated from observed value of $\alpha_{\rm opt}$/$\alpha_{\rm x-ray}$ and $\beta_{\rm opt}$/$\beta_{\rm x-ray}$ of best describe spectral regime.}
\label{SED_table}
\begin{scriptsize}

\begin{tabular}{|c|c|c|c|c|}
\hline
\textbf{SED} & \textbf{Time interval (sec)} & \textbf{$\bf \beta_{\rm \bf X-ray/opt}$}  & \textbf{\begin{tabular}[c]{@{}c@{}} $p$\\ (Spectral regime)\end{tabular}} & \textbf{$\bf \chi_r^{2}$} \\ \hline
1 & 29-65 & 0.72$^{+0.45}_{-0.42}$ & \begin{tabular}[c]{@{}c@{}}2.33 $\pm$ 0.21 \\ ($\nu_{\rm opt}$ $<$ $\nu_{\rm x-ray}$ $<$ $\nu_{\rm c}$)  \end{tabular} & 0.88 \\ \hline
2 & 63-100 &  0.49$^{+0.05}_{-0.05}$  &  \begin{tabular}[c]{@{}c@{}}2.10 $\pm$ 0.21\\ ($\nu_{\rm opt}$ $<$ $\nu_{\rm x-ray}$ $<$ $\nu_{\rm c}$)  \end{tabular}  &  1.01\\ \hline
3 & 700-1000 & 0.57$^{+0.02}_{-0.02}$ & \begin{tabular}[c]{@{}c@{}}2.09 $\pm$ 0.29\\ ($\nu_{\rm opt}$ $<$ $\nu_{\rm x-ray}$ $<$ $\nu_{\rm c}$) \end{tabular}  & 1.05 \\ \hline
4 & 23000-31000 & 0.77$^{+0.22}_{-0.22}$ &  \begin{tabular}[c]{@{}c@{}}2.64 $\pm$ 0.21\\ ($\nu_{\rm opt}$ $<$ $\nu_{\rm x-ray}$ $<$ $\nu_{\rm c}$) \end{tabular}  &  0.78 \\ \hline
5 & 51000-65000 &  0.51$^{+0.32}_{-0.32}$ &  \begin{tabular}[c]{@{}c@{}}2.38 $\pm$ 0.40 \\ ($\nu_{\rm opt}$ $<$ $\nu_{\rm x-ray}$ $<$ $\nu_{\rm c}$) \end{tabular}  & 1.27 \\ \hline

\end{tabular}
\end{scriptsize}

\end{center}

We calculated $\rm NH_{\rm z}  (0.61^{+0.11}_{-0.10} \times 10^{22}{\rm cm}^{-2}$), host extinction ($0.21^{+0.02}_{-0.02}$ mag) and redshift ($2.02_{-0.05}^{+0.05}$) using the SED 3. $\chi_r^{2}$ notify the reduced chi-square values for SEDs at different epochs.  Uncertainty in the calculation of $p$ is obtained with a confidence level of 95 \%.
\end{table}
\subsection{Host Galaxy Observations}
\label{Host Galaxy}

\subsubsection{\bf {10.4m GTC telescope}}
The 10.4m Gran Telescopio Canarias (GTC; Canary Island, Spain) obtained photometry for the host galaxy on the 18$^{th}$ July 2017. Images in the g,r,i,z filters were obtained using the Optical System for Imaging and Low-Intermediate Resolution Imaging Spectroscopy (OSIRIS) instrument \citep{cep00}. The images were dark-subtracted and flat-fielded using custom IRAF routines. Aperture photometry was performed using standard IRAF software, and field calibration was conducted using the SDSS DR12 catalogue \cite{ala15}. The host photometry from GTC and CAHA 3.5m telescope, described next, is given in AB magnitudes and not corrected for Galactic reddening of E(B-V)= 0.03 \citep{sch11}. These values are given in appendix A.

Optical spectroscopy of \thisgrb was obtained using the 10.4m telescope on Feb 26, 2014, and Feb 26, 2015. In both cases, the grism, which covers the 5400--10000 \AA{} wave range was used. Unfortunately, the S/N on the extracted spectra was very low, and we did not detect any emission features. At the photometric redshift, we would not expect any host galaxy emissions in this wavelength range.

\subsubsection{\bf {3.5m CAHA telescope}}
\begin{figure}
\centering
\includegraphics[height=4cm,width=8.0cm,angle=0]{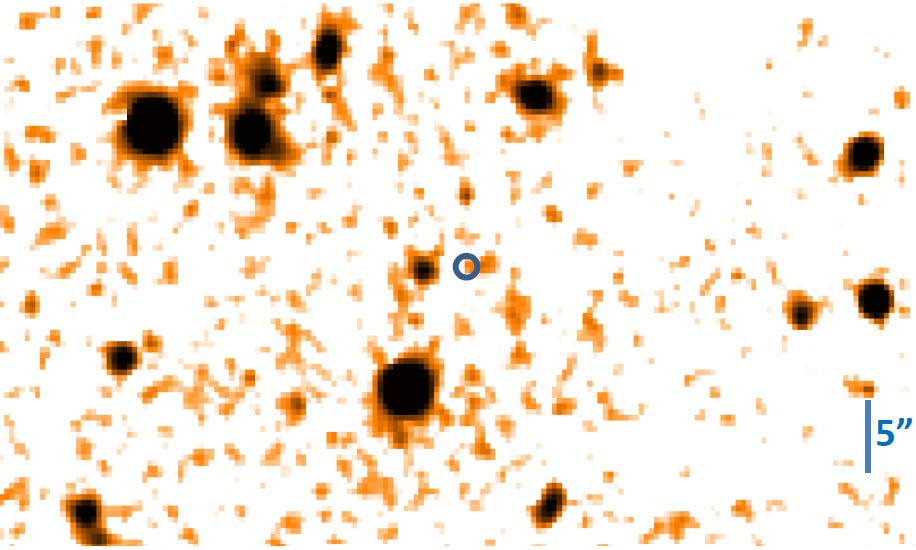}
\caption{H-band observations of the host galaxy of \thisgrb using 3.5m CAHA telescope. The blue circle marks the afterglow position. North is up and East to the left.}
\label{host_caha}
\end{figure}

The 3.5m telescope at Calar Alto Astronomical Observatory (CAHA) observed the field of \thisgrb on the 13$^{th}$ May 2014, 4 months after the trigger. A series of 59 H band images were taken, each with 65 sec exposure and, after dark-subtraction and flat-fielding using custom IRAF routines, were median combined to create a single image (see Figure \ref{host_caha}). Aperture photometry was performed using standard IRAF software, and field calibration was conducted using the 2MASS catalogue \citep{skr06}.

\subsubsection{\bf {3.6m DOT telescope}}

We observed the host galaxy of \thisgrb using the 4Kx4K CCD Imager \citep{2017arXiv171105422P} mounted at the axial port of the 3.6m Devasthal Optical Telescope (DOT) of ARIES Nainital on 16$^{th}$ January 2021 $\sim$ seven years after GRB detection. Observations were carried out in the $R$ filter with a total exposure time of 45 minutes (3 x 300 sec and 2 x 900 sec). We performed the reduction of data using IRAF packages. After pre-processing, we stacked the images to create a single image and perform the aperture photometry. More details about the reduction method can be found in \cite{Kumar2021}. We did not detect the host galaxy in the $R$-band. We calculated the 3 sigma limiting magnitude value equal to 24.10 mag (AB), calibrated with the nearby USNO-B1.0 field.

\section{Results}
\label{results}
Based on the multi-band observations obtained from various space and ground-based facilities of \thisgrb, we discuss the results based on the present analysis of both prompt emission and afterglows, including the host galaxy observations.

\subsection{Prompt emission}
 In the present subsection, we show the results of the prompt emission properties of \thisgrb and their comparison with other well-studied GRBs samples.

\subsubsection {\bf Spectral Hardness, Variability Time scale and Spectral lag}

\begin{figure}
\centering
\includegraphics[scale=0.32]{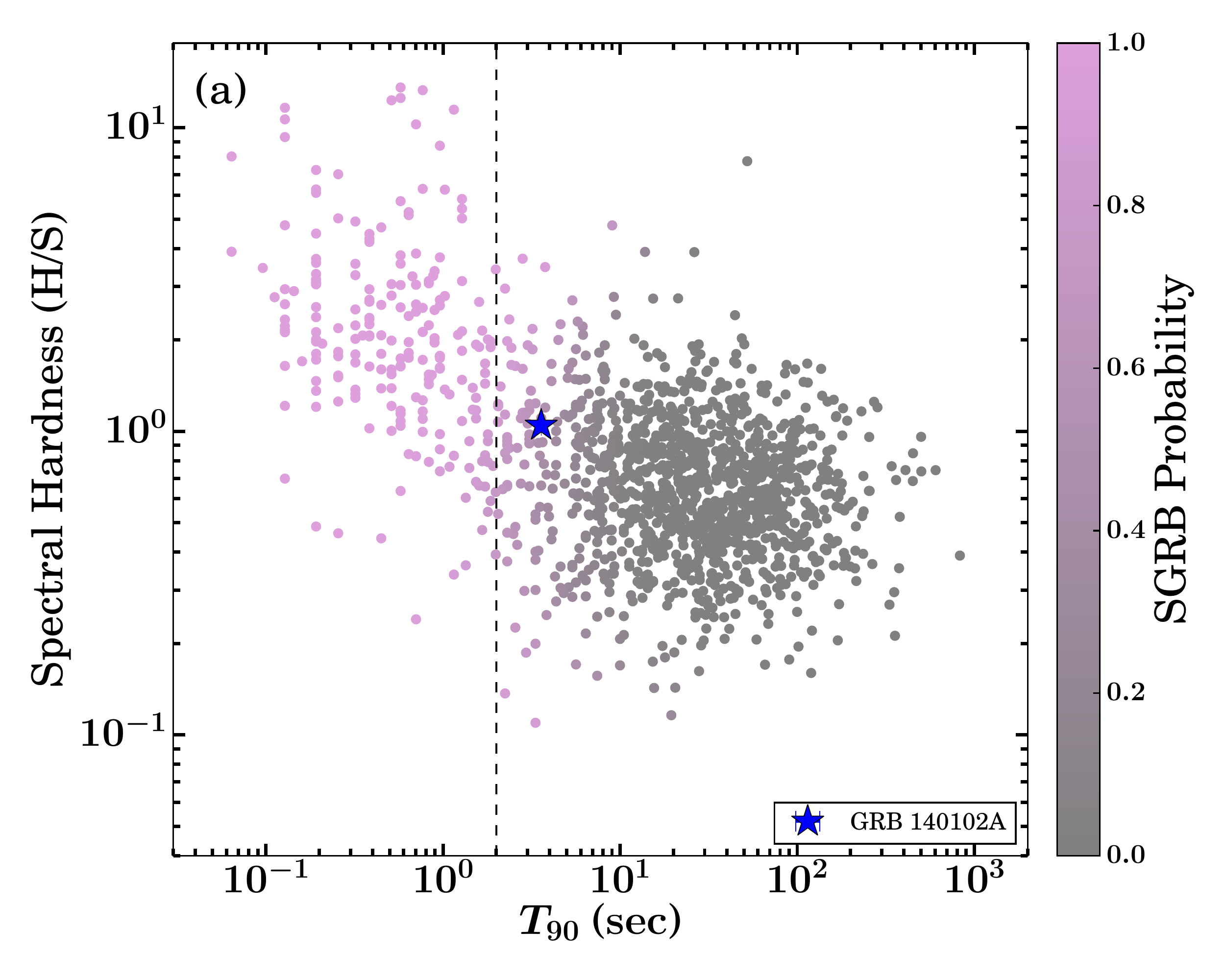}
\includegraphics[scale=0.32]{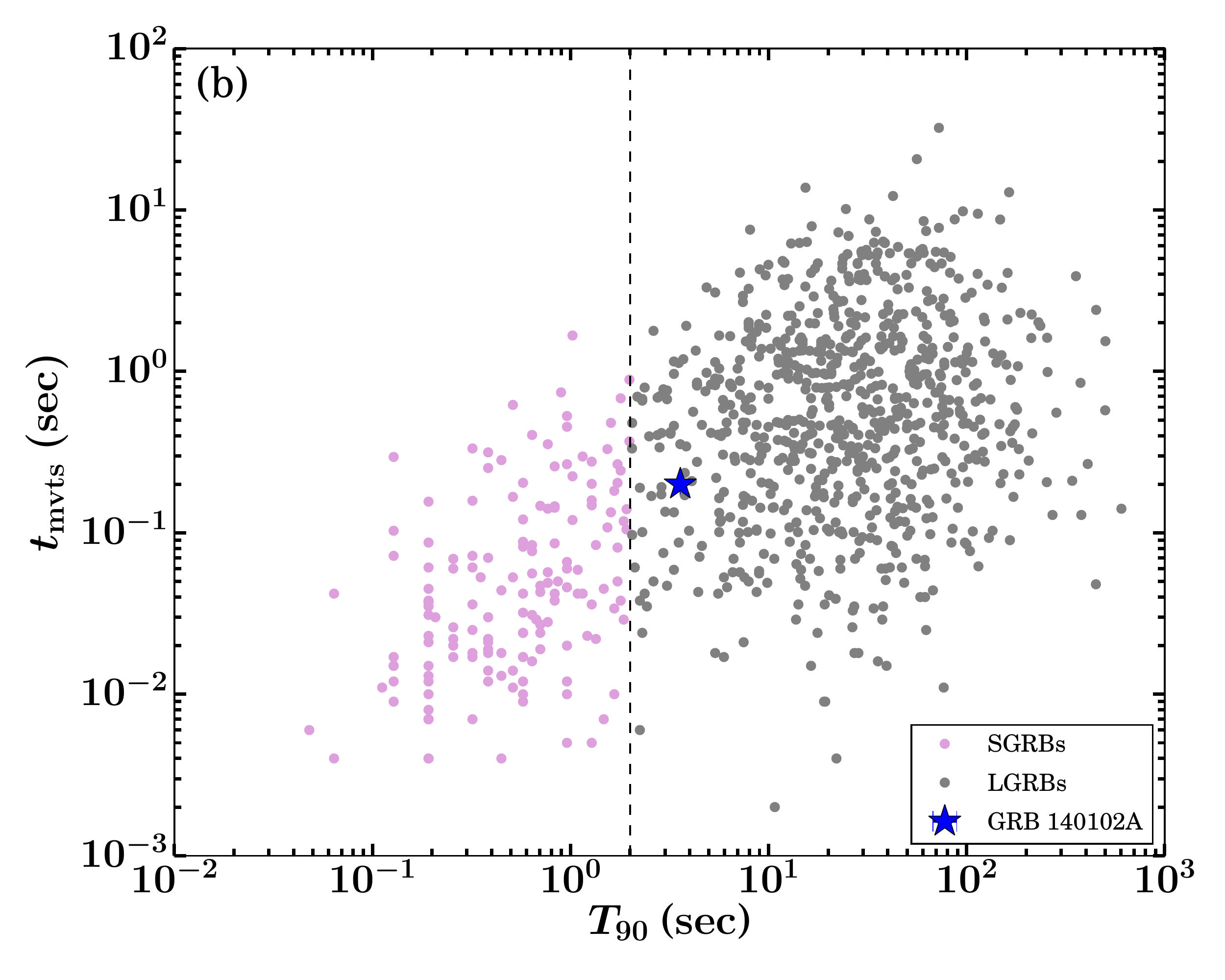}
\includegraphics[scale=0.32]{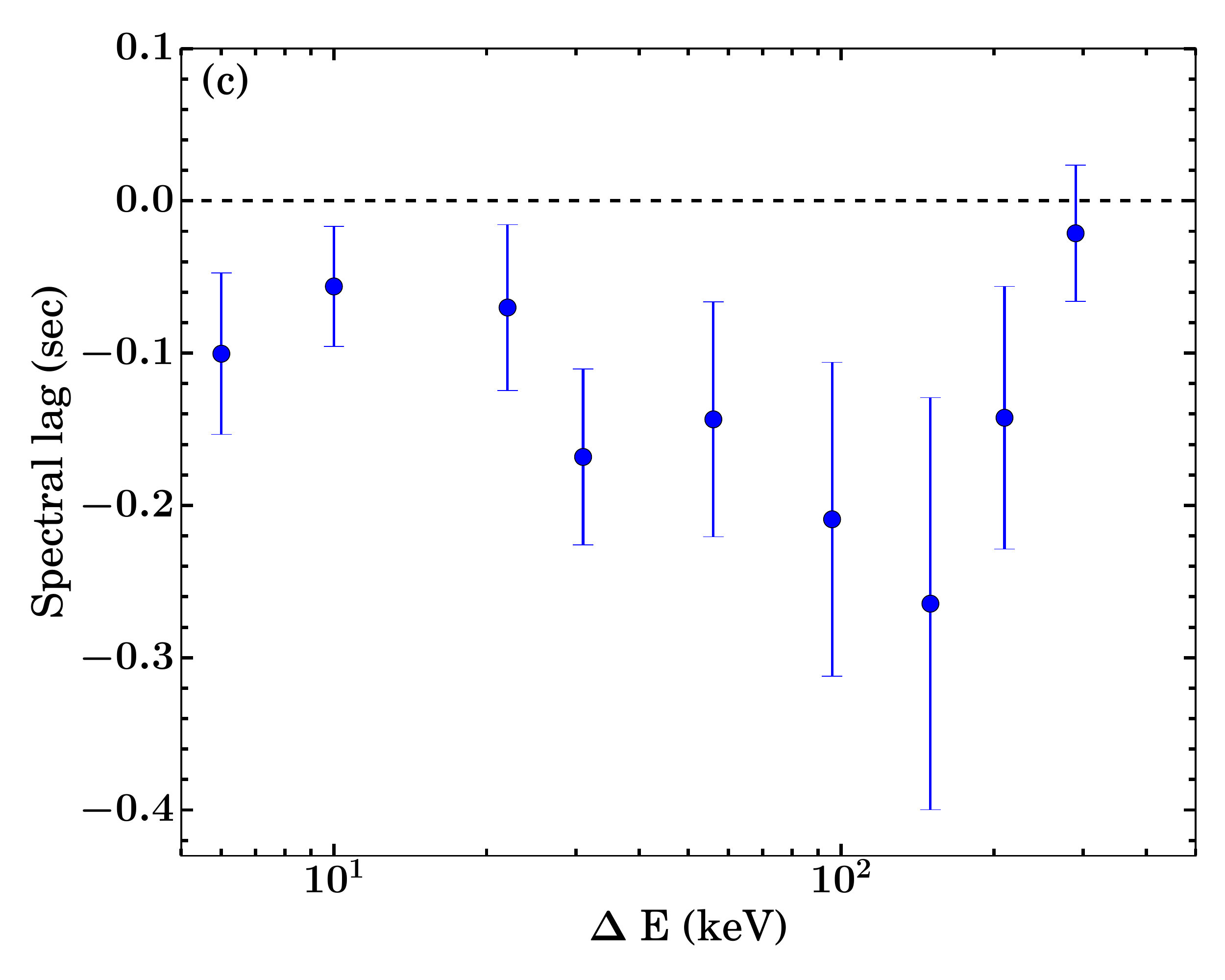}
\caption{\label{hr_t90} {\bf Prompt emission properties of \thisgrb} (a) The spectral hardness and \tninty for \thisgrb (shown with a blue star) along with the data points for short and long GRBs used in \protect\cite{2017ApJ...848L..14G}. 
The left side color scale shows the probability of a GRB belonging to SGRBs class. The vertical dashed lines show the boundary between SGRBs and LGRBs. (b) \thisgrb (shown with a blue star) in \tninty vs. minimum variability time scale (\mvts) plane along with the LGRBs and SGRBs sample studied by \protect\cite{2015ApJ...811...93G}. (c) Spectral lag as a function of energy for \thisgrb using the GBM multi-channel light curves. The negative lag value shows that higher energy photons arrived later than the lower energy photons. The black dashed line shows the zero lag.}
\end{figure}

We calculated the time-integrated HR by dividing the counts in 10 - 50 \keV and 50 - 300, \keV energy bands for the 6$^{th}$ + 7$^{th}$ + 9$^{th}$  NaI  detectors (see Table \ref{tab:prompt_properties}) to make a comparison with other \fermi GRBs published in \cite{2017ApJ...848L..14G}. We calculated \tninty durations of \thisgrb considering the total emission episodes using GBM data. Our result for the \tninty calculation (see Table \ref{tab:prompt_properties}) is consistent with \cite{2014GCN.15669....1Z} and its value lies in the overlapping interval of the bi-modal duration distribution of GRBs. The error in \tninty is calculated by simulating 10,000 light curves by adding a Poissonian noise with the mean values at observed errors \citep{2014AstL...40..235M, Bhat:2016ApJS..223...28N}. In  Figure \ref{hr_t90} (a), we show the HR-\tninty diagram of \thisgrb. The probabilities of a GRB classified as a short or long GRB from the Gaussian mixture model in the logarithmic scale are also shown in the background (taken from \citealt{2017ApJ...848L..14G}).


Minimum variability time scale \citep[][\mvts]{2013MNRAS.432..857M} is important to constrain the source emission radius ($R_{\rm c}$) of the source \citep{2015ApJ...811...93G}. We calculated the \mvts of \thisgrb using continuous wavelet transforms\footnote{\url{https://github.com/giacomov/mvts}} discussed in \cite{2018ApJ...864..163V} and found \mvts $\sim$ 0.2 sec for this burst. In Figure \ref{hr_t90} (b), we show \thisgrb in \mvts-\tninty distribution plane along with the sample of SGRBs and LGRBs studied by \citep{2015ApJ...811...93G}. Further, we calculated  $R_{\rm c}$ (see equation \ref{minimum_source}) using the following relation obtained from \cite{2015ApJ...811...93G}.


\begin{equation}
R_{\rm c} \simeq 7.3 {\times} 10^{13} \, \left (\frac{L_{\rm \gamma, iso}}{10^{51} \, \rm erg/sec} \right )^{2/5} \left (\frac{t_{\rm mvts} / 0.1 \, \rm sec }{1+z} \right )^{3/5} \, \rm cm.
\label{minimum_source}
\end{equation}

We estimated the source emission radius ($R_{\rm c}$) $\simeq$ 4.9 $\times$ $10^{14}$ cm for \thisgrb. 

For most of the LGRBs, time lags, also known as spectral lags, often appear in light curves among different energy bands \citep{2017ApJ...834L..13W}. The spectral lag is generally interpreted due to the spectral evolution \citep{2011AN....332...92P, 2017ApJ...844..126S}. We calculated the time lags of light curves among different energy bands using the cross-correlation function \citep[(CCF);][]{2000ApJ...534..248N, 2010ApJ...711.1073U} from \fermiT to \fermiT + 5.0 sec, following the method described in \cite {2012ApJ...748..132Z}. We found negative lag values for \thisgrb, which indicates that higher energy photons arrived later than the lower energy photons. Negative lag could be connected with superposition effect, described in \cite{2014AstL...40..235M}. The analysis of spectral lag for \thisgrb is shown in Figure \ref{hr_t90} (c).

\begin{figure}
\centering
\includegraphics[scale=0.33]{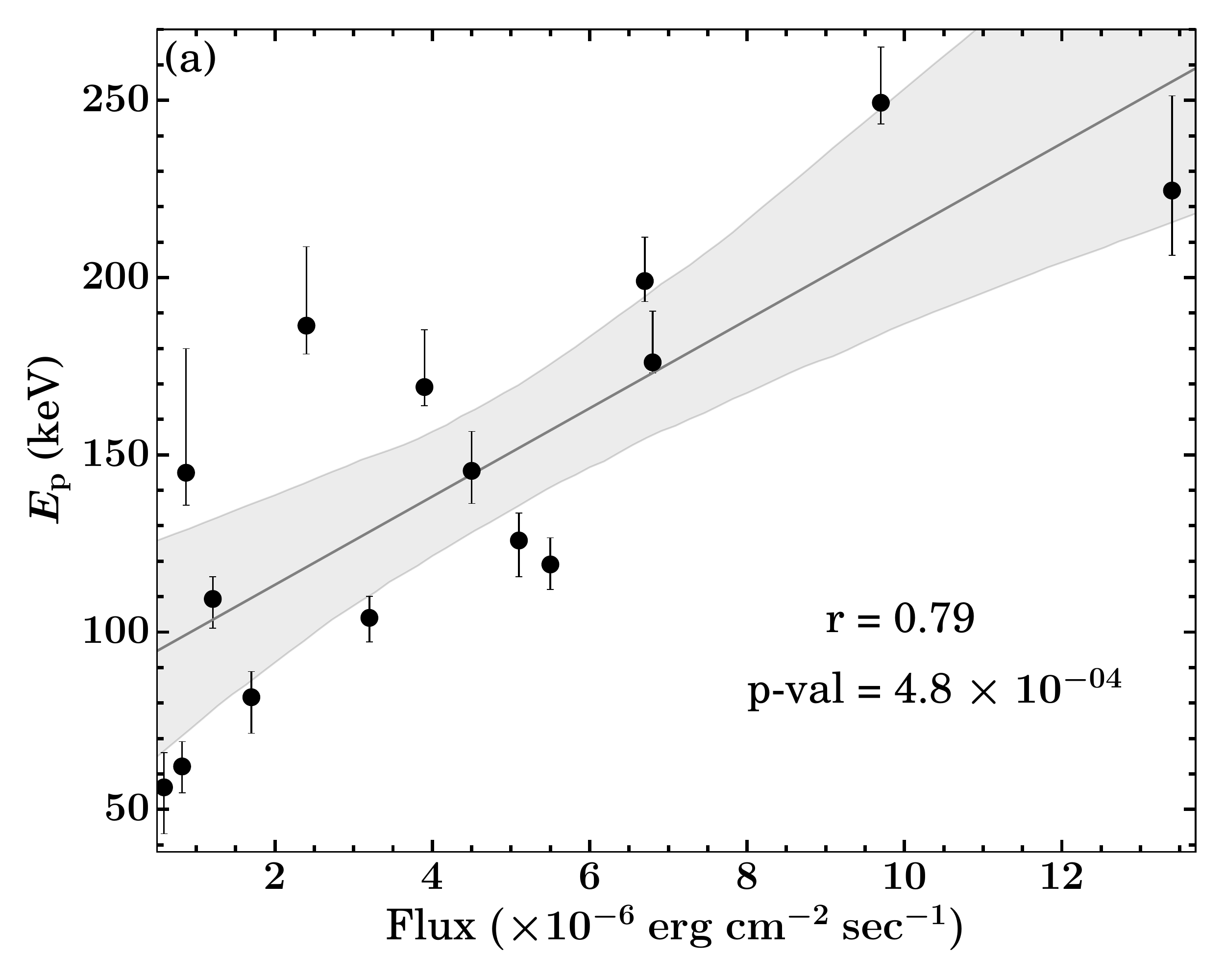}
\includegraphics[scale=0.33]{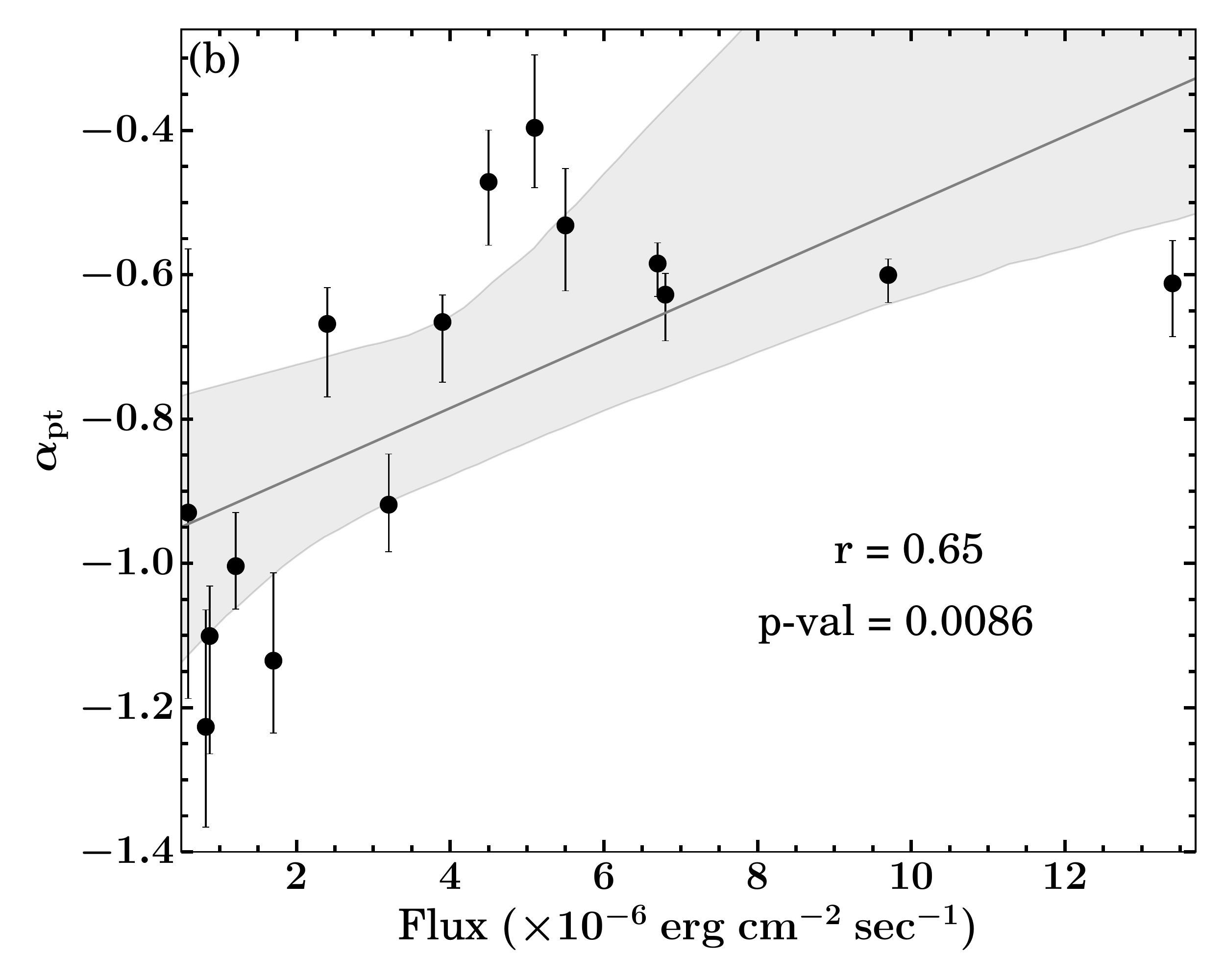}
\includegraphics[scale=0.33]{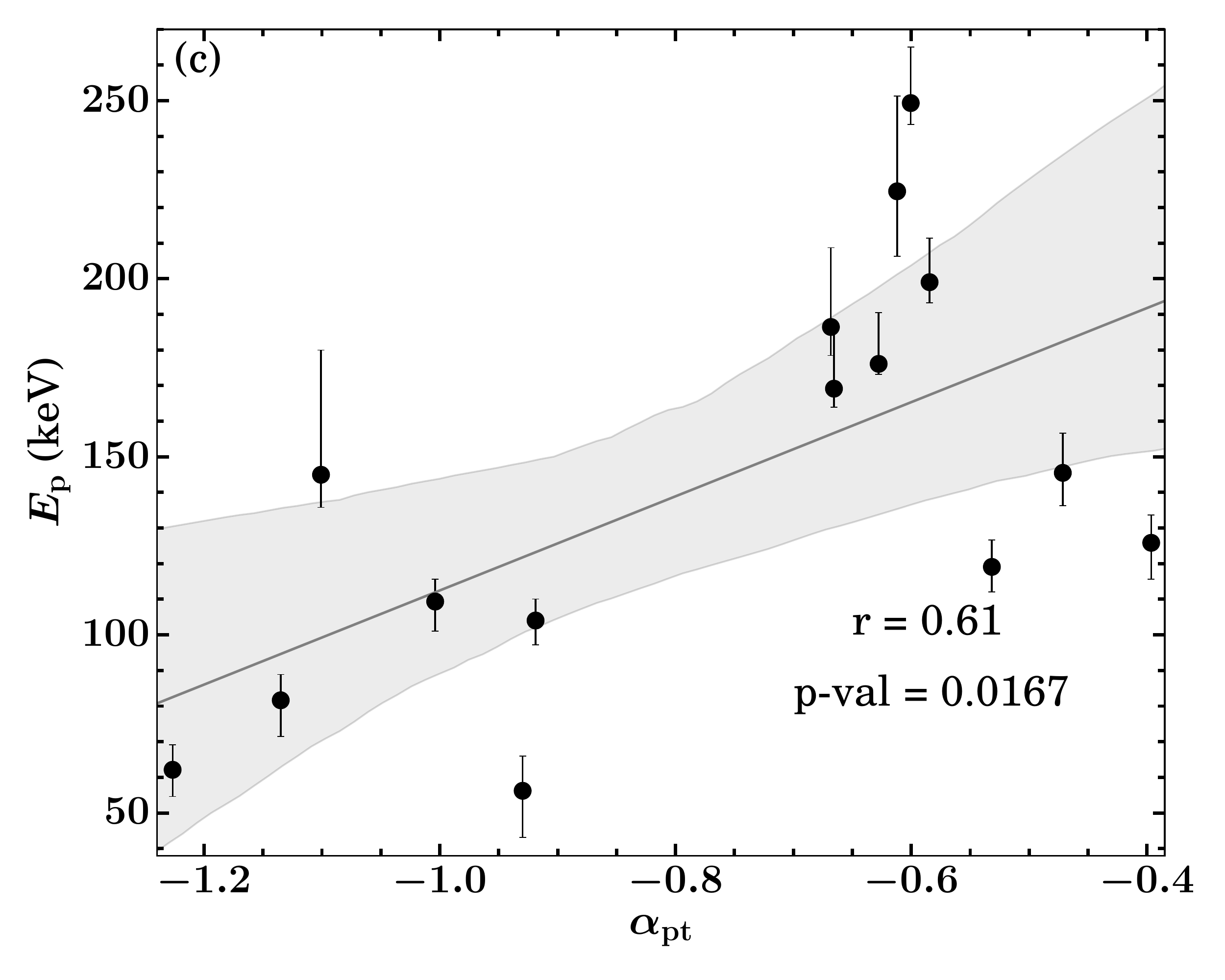}
\caption{{\bf Correlations between spectral model parameters :} (a) Peak energy versus and flux, (b) low-energy spectral index versus flux, (c) Peak energy versus low-energy spectral index. Correlation shown in (a), (b), and (c) are obtained using joint \fermi GBM and \swift BAT observations and modelling with \sw{Band} function. The best-fit lines are shown with solid grey lines, and the shaded grey region shows the 2 $\sigma$ confidence interval of the correlations.}
\label{spc}
\end{figure}

\subsubsection{\bf Correlations between Spectral parameters}
In the present section, we discussed the correlations between spectral parameters obtained from time-resolved spectral analysis. These correlations play an important role in probing the emission mechanisms of GRBs during the prompt emission phase.
Initially, we examine the correlation between \Ep - flux, $\it \alpha_{\rm pt}$-flux, and \Ep-$\it \alpha_{\rm pt}$ using the \sw{Band} function based on GBM data only for each bin obtained from the SNR binning method. We found a strong correlation between the \Ep of the \sw{Band} function and the flux in 8-900 \keV energy ranges with a Pearson coefficient (r) and \sw{p-value} of 0.86 and 1.88 $\times$ $10^{-5}$. We also found a strong correlation between $\it \alpha_{\rm pt}$ of \sw{Band} function and flux with r and \sw{p-value} of 0.75 and 7.4 $\times$ $10^{-4}$. As \Ep and $\it \alpha_{\rm pt}$ show a strong correlation with flux, we investigated the correlation between \Ep and $\it \alpha_{\rm pt}$. A moderate correlation with r and \sw{p-value} of 0.67 and 4.3 $\times$ $10^{-3}$ respectively. We also performed the time-resolved spectral analysis of \swift BAT data and find a strong correlation between BAT photon indices ($\it \Gamma_{\rm BAT}$) and the fluxes in 15-350 \keV energy ranges with r and \sw{p-value} of 0.86 and 1.02 $\times$ $10^{-13}$. 
 Furthermore, we jointly model the GBM and BAT spectrum of each bins obtained from Bayesian Block method to cross check the correlation results and we still find a strong correlation between the \Ep and flux with Pearson coefficient ($r$) and \sw{p-value} of 0.79 and 4.8 $\times$ $10^{-4}$. However, we notice moderate correlation between $\it \alpha_{\rm pt}$ and flux with $r$ and \sw{p-value} of 0.65 and 8.6 $\times$ $10^{-3}$. \Ep and $\it \alpha_{\rm pt}$ also show moderate correlation with r and \sw{p-value} of 0.61 and 1.7 $\times$ $10^{-2}$.
Therefore, \thisgrb has the characteristics of a `double-tracking' GRB (Both the low-energy spectral index and the peak energy follow the intensity-tracking pattern) similar to GRB 131231A \citep{2019ApJ...884..109L}. The spectral parameters $\it \alpha_{\rm pt}$, \Ep for \thisgrb are correlated with the observed flux, and we have shown this in Figure \ref{spc}. Our result is consistent with \cite{2020ApJ...890...90D}, where they performed the spectral analysis of Fermi LLE GRBs including \thisgrb using the Fermi GBM data only; they have used \sw{RMFIT} tool and \sw{$\chi^2$} statistics for measuring the goodness of fit. In contrast, we have performed the joint \fermi and \swift spectral analysis using the Bayesian spectral fitting package \sw{3ML}, which is a more appropriate method. \cite{2019MNRAS4841912R} also studied the correlation between the energy flux and $\it \alpha_{\rm pt}$, for a sample of the 38 \fermi GBM detected bursts having individual pulse structures. They found significant correlations in most pulses similar to the case of \thisgrb.


\subsubsection{\bf {Amati and Yonetoku correlations}}

\begin{figure}
\centering
\includegraphics[scale=0.34]{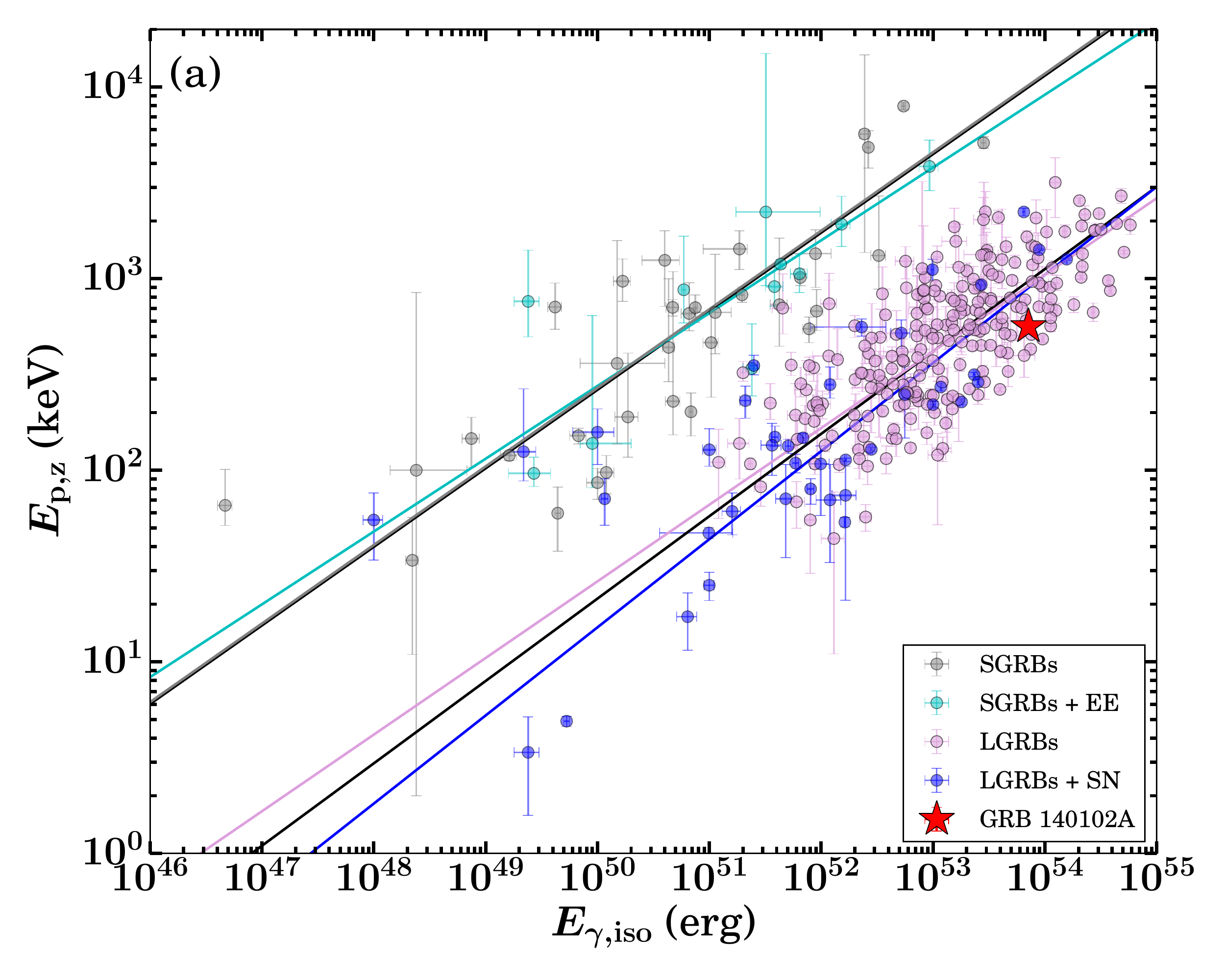}
\includegraphics[scale=0.34]{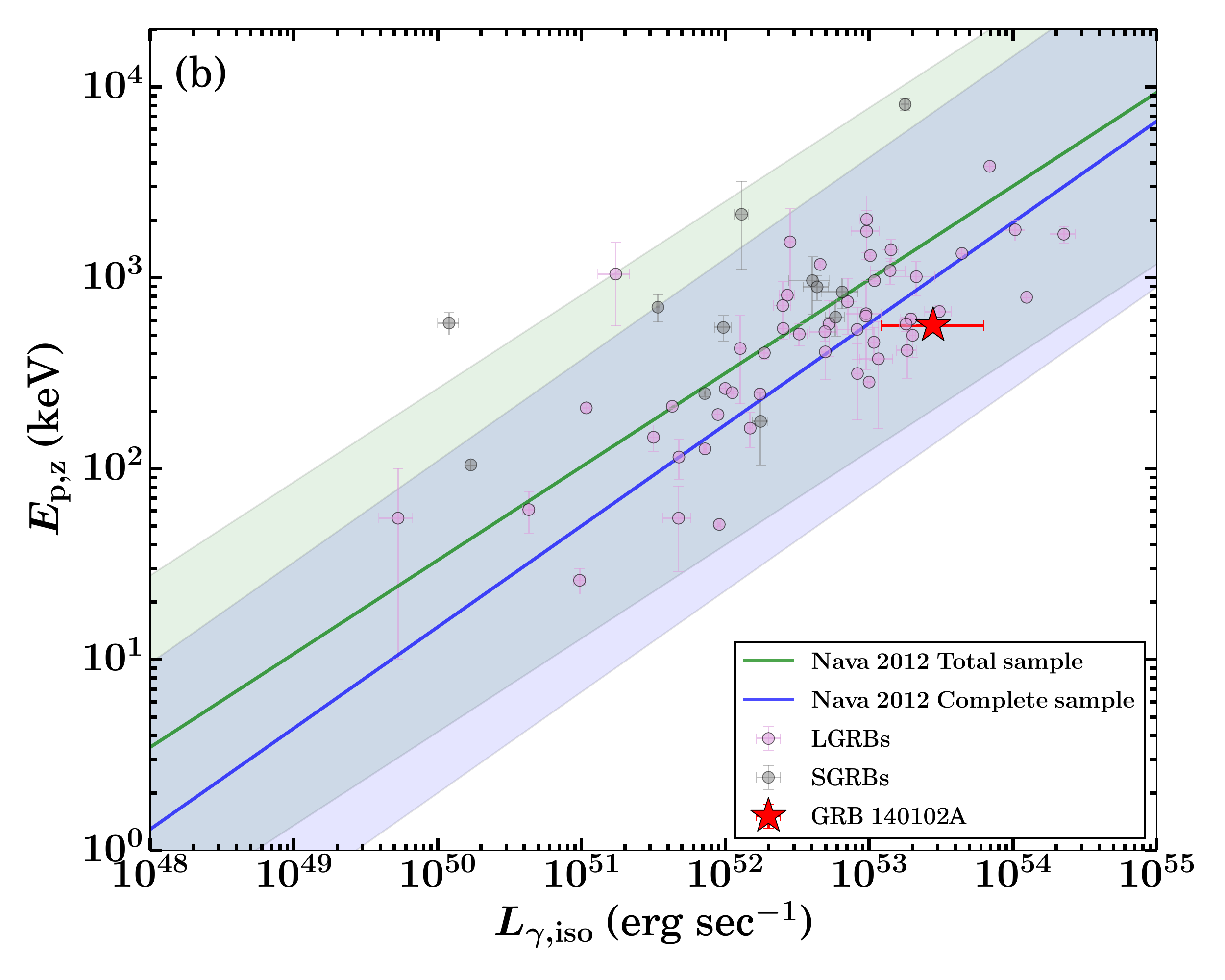}
\caption{{\bf Amati and Yonetoku correlation plane for \thisgrb :} (a) \thisgrb in the Amati correlation plane along with other the data points published in \protect\cite{2020MNRAS.492.1919M}. Various colored lines correspond to the best-fit lines for different classes mentioned in the legend. (b) \thisgrb in the Yonetoku correlation plane along with GRBs sample published in \protect\cite{2012MNRAS.421.1256N}. The colored lines show the best-fit and shaded region represents the 3 $\sigma$ scatter of the correlation.} 
\label{amati} 
\end{figure}

We study two well known prompt emission correlations for \thisgrb. The first one is Amati correlation \citep{2006MNRAS.372..233A, 2019ApJ...887...13F}, which is the correlation between spectral peak energy (see Table \ref{tab:prompt_properties}) corrected to the rest frame ($E_{\rm p,z}$) and isotropic equivalent $\gamma$-ray energy ($E_{\rm \gamma, iso}$), and the second one is Yonetoku correlation \citep{2004ApJ...609..935Y}, which is a correlation between $E_{\rm p,z}$ and isotropic peak luminosity ($L_{\rm \gamma, iso}$) in the cosmological rest frame ($z$ = 2.02). We estimated the time-integrated (\fermiT to \fermiT + 5.0 sec) total fluence in 1-$10^{4}$ \keV energy band for $E_{\rm \gamma, iso}$ calculation. We placed \thisgrb on the Amati correlation plane along with the data points for LGRBs and SGRBs published in \cite{2020MNRAS.492.1919M} and found that \thisgrb is consistent with the Amati correlation of LGRBs (see Figure \ref{amati} (a)). In Yonetoku correlation plane, \thisgrb lies within the 3 $\sigma$ scatter of total and complete samples of GRBs (see Figure \ref{amati} (b)) studied by \cite{2012MNRAS.421.1256N}.

\subsection{Multi-wavelength Afterglow}

 In this subsection, we present the afterglow properties of \thisgrb.
 
\subsubsection{\bf {Soft tail emission}}

\begin{figure}
\centering
\includegraphics[scale=0.31]{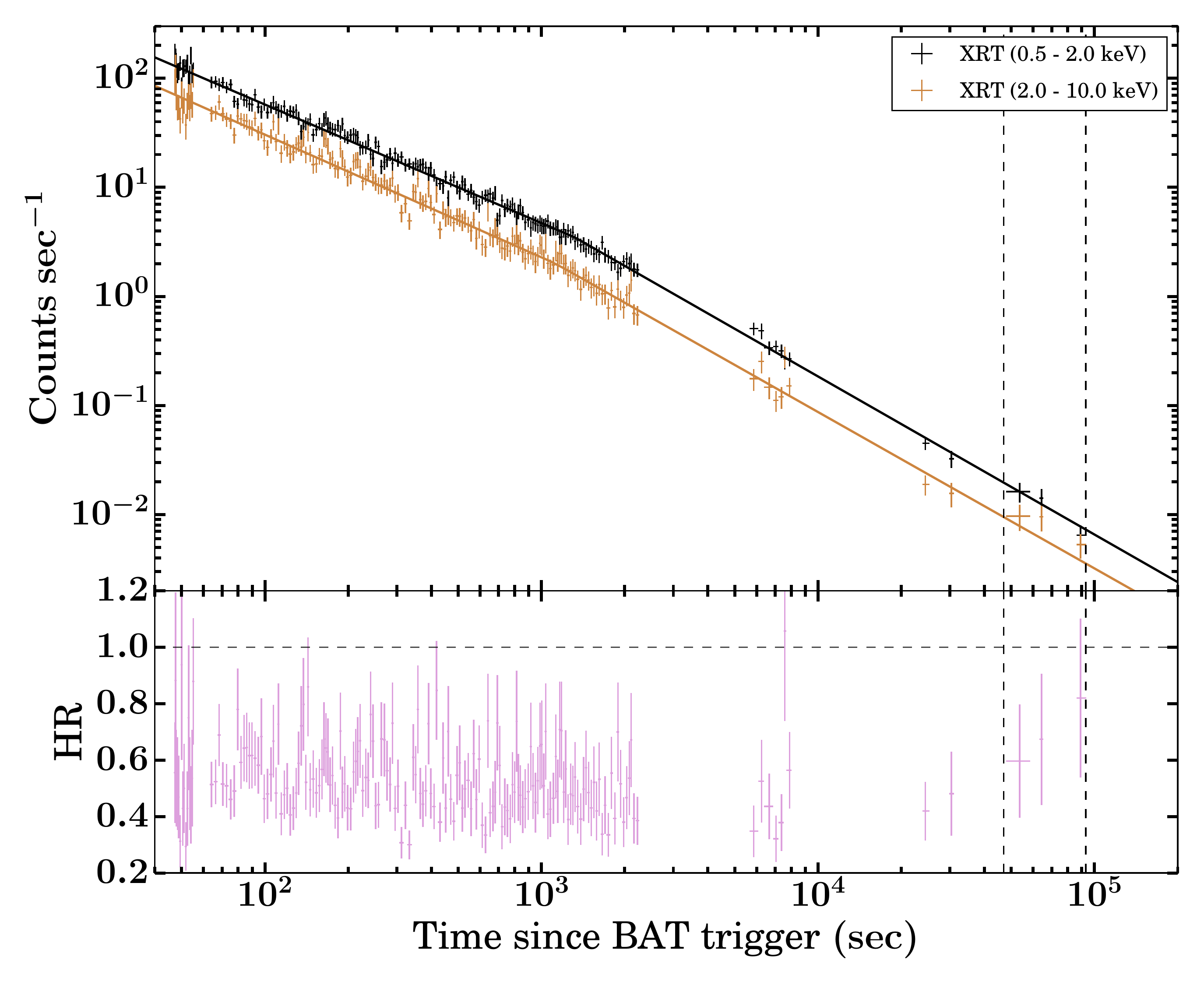}
\caption{The X-ray count rate light curves in soft (0.5-2.0 \keV) and hard (2.0-10.0 \keV) energy channels. The solid lines show the best-fit (broken power-law model) line to both the light curves. The bottom panel shows the evolution of HR in the same XRT energy bands. The vertical black dashed lines show the epoch of re-brightening activity in the X-ray light curve (at 10 \keV).}
\label{STE}
\end{figure}

The observed soft tail emission from GRBs is useful to constrain the transition time from prompt emission to afterglow phase \citep{2007ApJ...669.1115S}. \swift BAT observed the soft emission from \thisgrb until $\sim$ 200 sec. To understand the origin of this emission, we binned the BAT count-rate light curve based on Bayesian Blocks and performed the spectral analysis of each interval using \sw{pegpwrlw} model (a simple power-law model with pegged normalization \footnote{\url{https://heasarc.gsfc.nasa.gov/xanadu/xspec/manual/node207.html}}). The evolution of photon indices (black circle) calculated as mentioned above in BAT soft (15-25 \keV) energy channels are shown in Figure \ref{Xray_optical_afterglow} (c).  Also, spectral indices constrained using SEDs at 29-65 sec and 63-100 sec indicate agreement between BAT and optical emissions (after correcting for the RS contribution, in the case of SED at 29-65 sec, the optical data point is already within the BAT power-law uncertainty region due to large uncertainty on the index as the presence of a low signal in BAT). Furthermore, We notice that BAT photon indices do not show a rapid fluctuation and are consistent with those estimated at XRT frequencies during the soft emission phase. We also use early X-ray observations to investigate the underlying tail emission. We created the XRT light curve in soft (0.5-2.0 \keV) and hard (2.0-10.0 \keV) energy channels and examined the evolution of HR to identify spectral evolution at early epochs (see Figure \ref{STE}). We performed the temporal and spectral analysis for this early epoch ($\sim$ 47 sec to 200 sec post burst) XRT (WT mode) data. We find a temporal decay of 0.99$^{+0.03}_{-0.03}$ and spectral index equal to 0.59$^{+0.04}_{-0.04}$ during this epoch. These values are found to disagree with the expected closure relation for early X-ray observations to be prompt tail emission as described by \cite{2006ApJ...642..354Z} ($\alpha$ = 2 + $\beta$). Based on the above, we suggest that soft tail emission observed from \thisgrb had possibly afterglow origin.

\subsubsection{\bf {X-ray and Optical afterglow light curves}}
\label{xray_optical lc fitting}

The X-ray afterglow light curve, shown in Figure \ref{Xray_optical_afterglow}, declines from the beginning of observations and shows no features, such as steep and shallow decays phases or any flaring activity. A fit of a power-law to the light curve gives a $\chi^2$/dof = 1190/545, suggesting a more complex model is required to fit the data. A broken power-law model is a better fit with $\chi^2$/dof = 575/543;
according to the {\it F}-test, the broken power-law model gives an improvement at the $>5\sigma$ confidence level with respect to the power-law model. We also tested to see if a further break would provide an improvement. The double broken power-law model results in $\chi^2$/dof = 565/541. The {\it F}-test suggests the additional break is not required since the improvement is only at the 2.6 $\sigma$ level. The best-fit model is given in Table \ref{lcfits}.

The optical multi-band light curve is shown in Figure \ref{optical_multiband}. Observations in the UVOT $white$ and BOOTES-clear filters started before most of the other optical filters and have the best sampling and SNR during the first few thousand seconds. A power-law fit to both of these light curves results in $\chi^2/dof=1754/108$ and $\chi^2/dof=89/50$ for
BOOTES-clear and UVOT $white$, respectively. Both of which are statistically unacceptable at the $>3\sigma$ level.
Since the light curves display an early steep to shallow transition, suggesting the presence of a reverse dominated shock \citep{2015ApJ...810..160G} when fitting a more complex model such as a broken power-law, the sum of two power-law models, or their combinations. Both fits give slightly better values, with the broken power-law resulting in $\chi^2/dof=315/106$ and $\chi^2/dof=43/48$ for BOOTES-clear and UVOT $white$ respectively, and for the two power-law components, we achieved $\chi^2/dof=203/106$ and $\chi^2/dof=43/48$. While both models fit the UVOT $white$, for the higher sampled BOOTES-clear filter, both fits are still unacceptable at $>3\sigma$ level, although the two power-law model results in a smaller reduced $\chi^2$. The two power-law model appears to fit the data well, it is likely that the $\chi^2$ is large for BOOTES-clear due to the significant scatter in the data about the best-fit model.

The BOOTES photometry ends at 6ks and in UVOT, there is an observing gap between 2ks and 5ks, after which the S/N of the light curve is poor. Therefore we created a single filter light curve by combining the optical light curves from BOOTES and UVOT in order to create a light curve with better sampling and high SNR as described in \S~\ref{combLC}. A broken power-law fit to this light curve results in an unacceptable $\chi^2/dof=284/35$. We, therefore, fit the light curve with a double broken power-law. This resulted in a slightly improved $\chi^2/dof=256/33$; the {\it F}-test suggests the addition of a break is not required since the improvement is $<2\sigma$. However, on the basis that such a steep decay in the early optical light curve, may be due to the reverse shock, resulting in two-components producing the observed optical afterglow \citep{Wang_2015}, we also tested two-component models. Firstly we tested two power-laws, which results in $\chi^2/dof=192/35$. The fit is an improvement on the previous models but is still unacceptable. We therefore, also tried with a power-law plus broken power-law model. The fit results in a $\chi^2/dof=109/33$ with the {\it F}-test suggesting the addition of a power-law to the broken power-law model is an improvement at $3\sigma$ confidence.

We note that when constructing the combined light curve, color evolution was observed between the BOOTES and UVOT data,
with the UVOT $white$ and $V$ band data, during the initial steep decay, systematically lower than the BOOTES clear data.
The CCD of BOOTES is sensitive to much redder photons, covering the range 3000 {\AA} to 11000 {\AA} \footnote{\url{http://www.andor.com/scientific-cameras/ixon-emccd-camera-series/ixon-ultra-888}}, while the UVOT $white$ filter covers a wavelength range of between 1700-6500 {\AA} and the UVOT $V$ filter covers a range 5000-6000 {\AA}. This would suggest that the spectrum during the steep decay is redder in comparison to the rest of the afterglow. After normalization, the data were group together, weighted by the errors, since the early BOOTES and UVOT data do not align well, this may affect the measurement of temporal indices. However, we see that the initial steep decay
of the combined light curve is consistent with the tightly constrained steep decay measured with BOOTES-clear
and the less well constrained UVOT $white$. The steep decay of the combined light curve is most similar to that of the BOOTES clear filter, as the BOOTES light curve has better SNR than the UVOT $white$ filter and thus is the dominant component when creating the weighted binned light curve.

 We noticed a break in the X-ray light curve at $\sim$ 2 ks. The temporal break may be due to the change in spectral behavior or due to energy injection from the central engine, which we notice as a transition from shallow decay phase and followed by a steep decay ($1.50^{+0.02}_{-0.02}$). We sliced the X-ray light curve (count rate) into small temporal bins (see Figure \ref{Xray_optical_afterglow} (a)) to study the possible origin of this break. This resulted in 20 spectra, however, five bins do not have significant counts for the spectral modelling. The evolution of photon indices is shown in Figure \ref{Xray_optical_afterglow} (c, black circles). It indicates that there is no significant change in $\beta_{\rm x}$ (as it is expected for $\nu_{\rm c}$ passing through) close to the break in the X-ray light curve. When we fitted the photon index with a constant (\sw{CONS}) model, we get a very good fit, suggesting that there is no evolution. If we consider that $\nu_{\rm c}$ is between the BAT and LAT frequencies as suggested from SED 1 (29-65 sec, see \S~\ref{closure relation}), $\nu_{\rm c}$ could not pass through the X-ray band, inconsistent with expectations (i. e. t$^{-0.5}$), indicating that X-ray break is not due to the spectral break. We also tested the other possibilities of the origin of this break, such as energy injection. Considering adiabatic cooling with energy injection from the central engine, the inferred value of $\alpha_{\rm x}$ ($1.08 \pm 0.02 $) from observed $\beta_{\rm x}$ ($0.56 \pm 0.02$) matches with the observed value of $\alpha_{\rm x-ray}$ ($1.09 \pm 0.01$) for the spectral regime $\nu_m < \nu_{\rm x-ray} < \nu_c$ and for the slow cooling case in the ISM-like medium. We estimate the value of electron distribution index $ p = 2.12 \pm 0.04$ from $p= (2\beta + 1)$.  We notice that the inferred value of $p$ is close to what we obtained from modelling. We observed an early excess X-ray flux similar to those found in many other RS dominated bursts. It could be possible to explain early X-ray excess as an energy injection episode lasting up to 2 ks. Though we did not find an achromatic break in the X-ray and optical light curves as expected due to the end of the energy injection episode \citep{2015ApJ...814....1L}, however, we notice a break in the optical data $\sim$ 6 ks (see Table \ref{lcfits} and Figure \ref{Xray_optical_afterglow} (b)). 
 
 During the late time re-brightening phase in the unabsorbed X-ray flux light curve (at 10 keV), we notice the hardening in the photon index (see Figure \ref{Xray_optical_afterglow}) and in the hardness ratio (see Figure \ref{STE}). This could not be explained by the frequency crossing the X-ray band as the photon index reverses back to the original position after this phase. This unusual emission could be originating due to patchy shells or a refreshed shocks \citep{2020ApJ...898...42C, 2021A&A...646A..50H}.

\subsubsection{\bf {Optical and X-ray afterglow light curve modelling}}
\label{optical_afterglow_modelling}

The fireball synchrotron model for afterglows is currently the favoured scenario in terms of producing the observed multi-wavelength emission. In this model, the afterglow is a natural consequence of the beamed ejecta ploughing through the external medium and interacting with it, producing the observed synchrotron emission. The spectral and temporal behavior of the afterglow emission could be described by several closure relations \citep[][see the last reference for a comprehensive list]{sari98,mes97,sari99,dai01,racusin09}. This set of relations connect the values of $\alpha$ and $\beta$ to the power-law index of the electron energy distribution $p$, making it possible to estimate its value from the observations, which is typically found to be between 2 and 3 \citep{pan02,sta08,cur09}; the density profile of the external medium (constant or wind-like), and the relative positions of breaks in the synchrotron spectrum, primarily the synchrotron cooling frequency $\nu_{\rm c}$ and the synchrotron peak frequency $\nu_{\rm m}$. Another break frequency is called the synchrotron self-absorption frequency, though it does not influence the optical/UV or the X-rays during the timescales studied here. There are also closure relations describing temporal and spectral indices in the scenario that the afterglow is injected by an additional energy component \citep{2006ApJ...642..354Z}.

For jets having a single component, it is derived that the optical/UV and X-ray emissions are produced within the same region and therefore are explained by the same synchrotron spectrum, with the possibility that one or more of the break frequencies are between the two observing bands. This translates that the optical/UV and X-ray temporal indices, determined from an afterglow, should be described by closure relations that rely on the same assumptions about the ambient media, the electron energy index, $p$, and the energy injection parameter $q$.

The thin shell case ($T_{\rm dec} >$ \tninty) Type-II features  (also called flattening type, the peak of forward shock is below the reverse shock component) discussed by \cite{2015ApJ...810..160G} are used to interpret the optical emission. However, the forward shock emission dominates after $\sim 100$ sec. The optical emission in between $\sim 30-100$ sec follows the emission predicted by the reverse shock emission. This makes the early optical emission from this GRB a combination of RS+FS components. The RS model parameters for the early optical emission are useful to understand the magnetic energy available in the jet and the source environment \citep{2003ApJ...597..455K, 2000ApJ...542..819K}.  
 
The reverse shock crossing time is defined as $t_{x}$ = $\rm max (T_{\rm dec,}$ \tninty), which is important to estimate the peak emission for reverse shock. The break frequency evolution with time is $\nu_m^r \propto t^6$ for $ t< t_x$ and $ \nu_m^r \propto t^{-54/35}$ for $ t>t_x$, however $\nu_m^f$ is constant  before crossing time and decreases with time after this as $t^{-3/2}$, where $r$ denotes the reverse shock and $f$ is used for forward shock \citep{2015ApJ...810..160G}. The cooling frequency $\nu_c^r \propto t^{-2}$ for $t< t_x$ and $\nu_c^r \propto t^{-54/35}$ after $t_x$. The same for forward shock is $\nu_c^f \propto t^{-2}$ for $t< t_x$ and $\nu_c^f \propto t^{-1/2}$ after crossing time. The maximum synchrotron flux is defined as $f_{\rm max}^r \propto t^{3/2}$ for $t<t_x$ and after crossing time it decreases as $t^{-34/35}$. Similarly, $f_{\rm max}^f \propto t^{3}$ for $t<t_x$ and independent of time after crossing time \cite{2015ApJ...810..160G}.  We model the afterglow emission and the parameters obtained to explain \thisgrb are listed in Table \ref{sample_Modelling}. We have used Bayesian analysis \sw{PyMultiNest} software \citep{2014A&A...564A.125B} to estimate the afterglow modelling parameters and associated errors. A corner plot showing the analysis results is shown in appendix A. In the right panel of Figure \ref{Xray_optical_afterglow}, we have shown the optical and X-ray light curves based on afterglow modelling. The parameters electron energy index $p$ (2.00 for RS and 2.19 for FS, respectively), micro-physical parameters $\epsilon_e$ and $\epsilon_B$ for the RS and FS are constrained using optical and X-ray data. The model can explain the optical emission and produces a slightly lower X-ray flux than observed at early times (the excess can be explained in terms of energy injection, see \S~\ref{xray_optical lc fitting}). However, in a few cases of GRBs, this kind of feature (excess X-ray emission) has been seen, and possible scenarios like (i) wavelength-dependent origin and (ii) mass loss evolution dependence are discussed by \cite{xray_access, 2020ApJ...896....4X}.

\subsection{Host Galaxy SED modelling}

\begin{figure}
\centering
\includegraphics[height=8cm,angle=0]{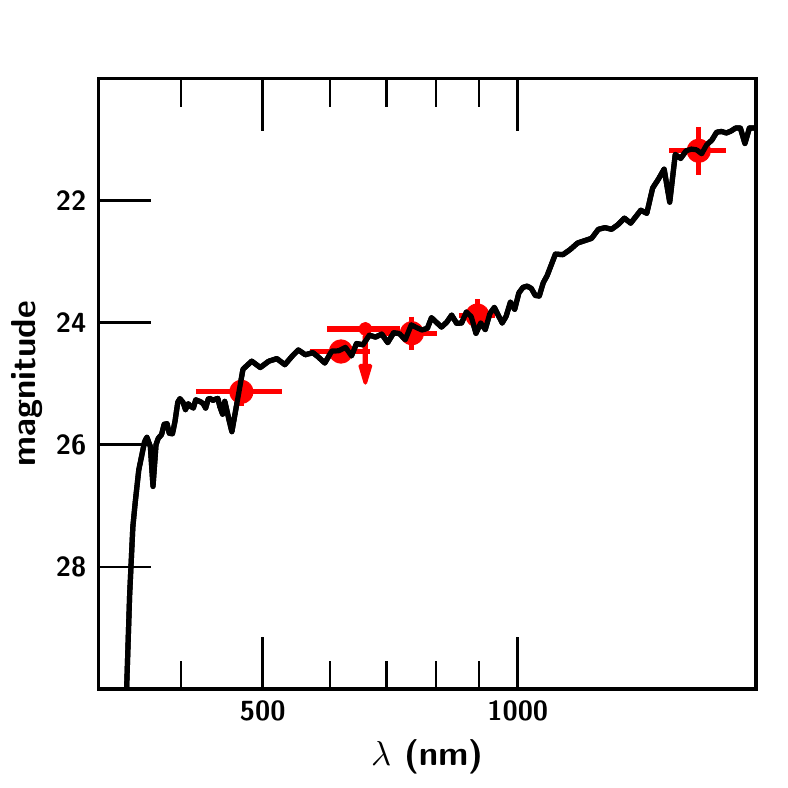}
\caption{The SED of the host galaxy of \thisgrb fitted by the LePhare with redshift $z$=$2.8^{+0.7}_{-0.9}$. Filled red circles depict respectively the data points in the filters g, r, R, i, z, H from original observations (see \S~\ref{Host Galaxy} and Table A7). All magnitudes are in the AB system.}
\label{host_sed}
\end{figure}

\begin{figure}
\centering
\includegraphics[height=8cm,width=8.7cm,angle=0]{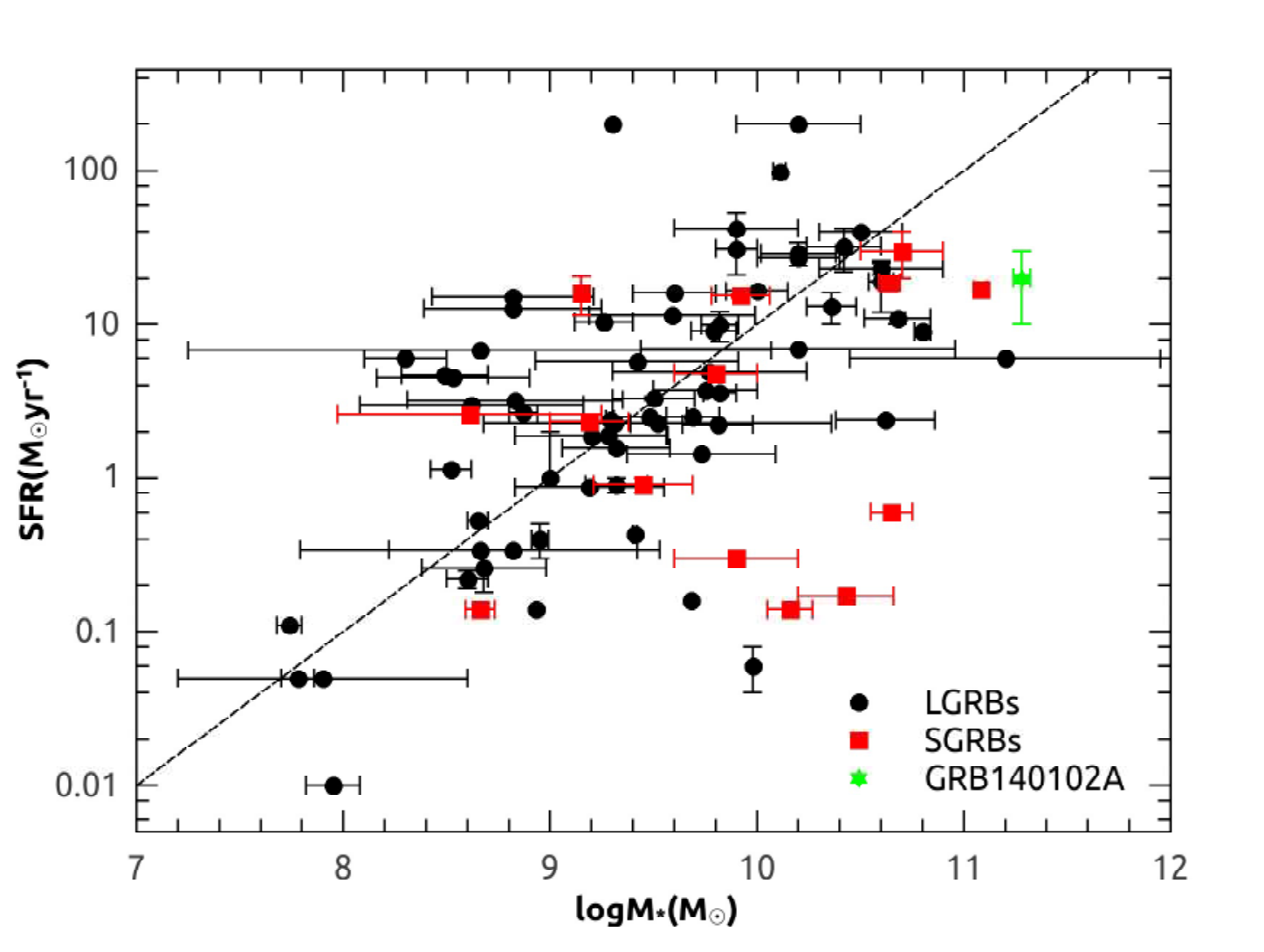}
\caption {The distribution of SFR a function of stellar mass for a sample of GRB hosts, inferred from template fitting to their photometric SEDs. A green star shows the host position for \thisgrb. Black dots and red squares indicate the LGRBs, and SGRBs hosts with SFRs measured from GHostS from 1997 to 2014 \citep{2006AIPC..836..540S, 2009ApJ...691..182S}. The dashed line shows a constant SSFR of 1 Gyr$^{-1}$.}
\label{host_sfr}
\end{figure}

\begin{figure}
\centering
\includegraphics[height=8cm,width=8.7cm,angle=0]{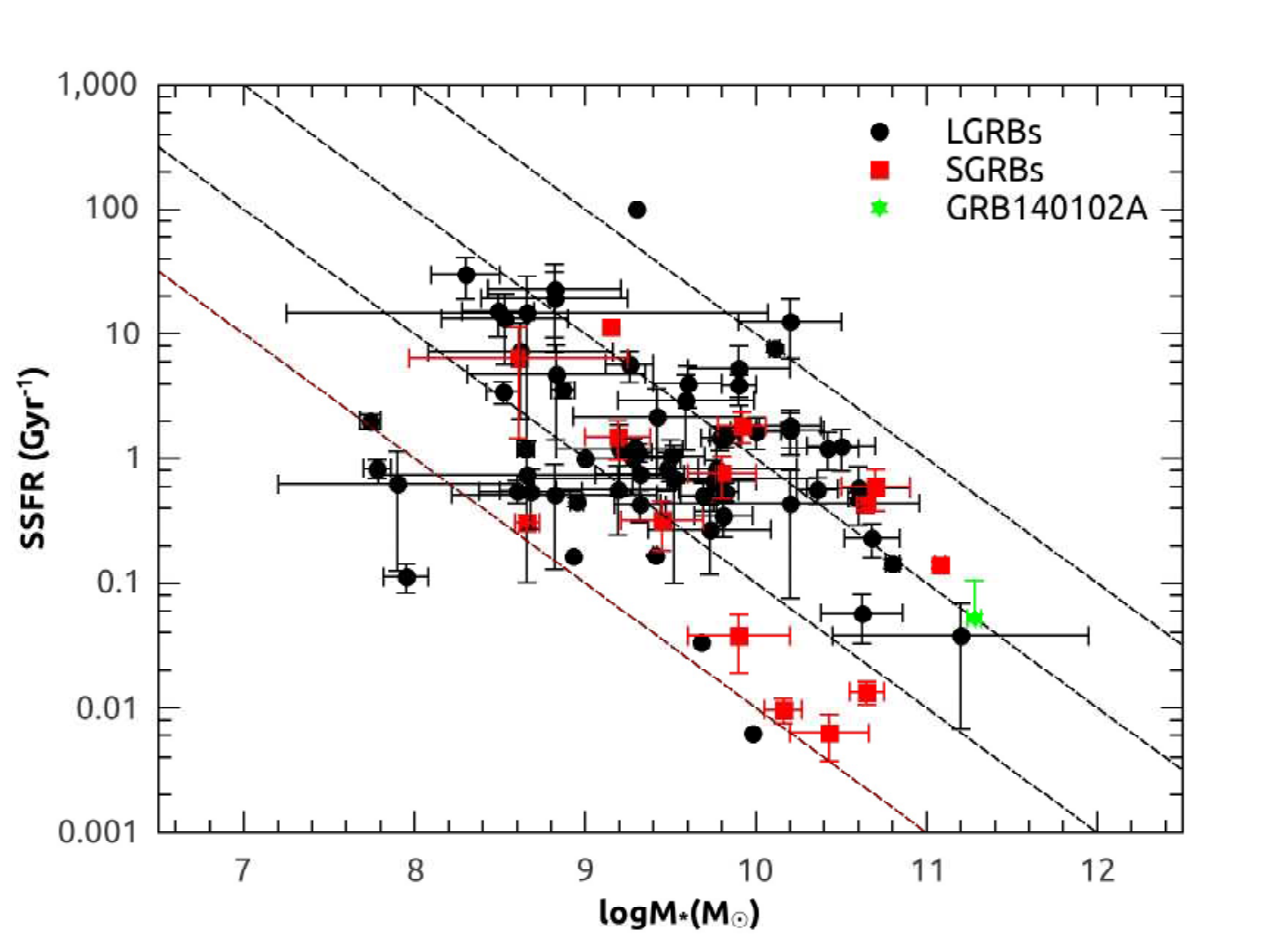}
\caption {The distribution of SSFR as a function of stellar mass for a sample of GRB hosts, inferred from template fitting to their photometric SEDs. A green star shows the host position for \thisgrb. Black dots and red squares show GRB hosts with SFRs measured from GHostS from 1997 to 2014 \citep{2006AIPC..836..540S, 2009ApJ...691..182S}. The dashed lines show the constant SSFRs of 0.1, 1, 10, and 100 M yr$^{-1}$ from left to right.}
\label{host_ssfr}
\end{figure}

The host characteristics, such as morphology, stellar population, offset, etc., of LGRBs and SGRBs, are different. These characteristics are likely associated with the physical conditions required to produce GRBs. LGRBs are largely localized in active star-forming and young stellar population dwarf galaxies. On the other hand, SGRBs belong to old stellar population galaxies \citep{2009ApJ...691..182S, 2009ApJ...690..231B}. Therefore, host parameters can be used to constrain about progenitors and their environment (useful in such cases where \tninty duration of burst lies in the brink regime as in the case of \thisgrb). We calculated the offset value of 0.52 $\pm$ 0.02 arcsec ($\sim$ 4.46 kpc) between the afterglow position and the center of the host galaxy of \thisgrb. This value is higher in comparison to the median offset value studied by \cite{2017MNRAS.467.1795L}. Recently \cite{2018JHEAp..18...21W} studied the possible correlations between rest-frame prompt emission (\tninty, $E_{\rm \gamma, iso}$, and $L_{\rm \gamma, iso}$) properties with host properties. They found that these parameters negatively correlate with offset (\thisgrb also follows it). As we could not constrain the redshift of \thisgrb using the spectroscopic observations of the host galaxy, we attempted the photometric observations of the host in six broad-band filters. These inspired us to perform a detailed SED modelling and compared the results with a well-studied GRBs host sample.

We used LePhare code \citep{1999MNRAS.310..540A:LePhare} with PEGASE2 stellar synthesis population models library \citep{1999astro.ph.12179F:Pegase2} for fitting (the redshift was left free to vary) the measured photometric magnitudes values of the host galaxy
 though with large errors associated with photometry. The best solution (see Figure \ref{host_sed}) was revealed for redshift $2.8^{+0.7}_{-0.9}$, E(B-V) = 0.05   $\pm$ 0.05, and $\chi^{2} = 0.1$ with parameters of host galaxy:  age of the stellar population in the galaxy = $9.1 \pm 0.1$ Gyr,  stellar mass = ($1.9 \pm 0.2) \times 10^{11} \msun$, and SFR (star formation rate) = $20\pm10 \msun$ $\rm yr^{-1}$. We find that LePhare host redshift is consistent with the redshift value obtained in \S~\ref{photoz} using joint X-ray and Optical afterglow SED. Furthermore, we compare these properties with other well studied samples of GRB host galaxies. We find that the SFR is higher than the median value of 2.5 $\msun$ $\rm yr^{-1}$ \citep{2009ApJ...691..182S}. The host galaxy of \thisgrb lies at the upper right position in the stellar mass and SFR correlation plane for star-forming galaxies \citep[see Figure \ref{host_sfr};][]{2007ApJ...670..156D}, indicating the higher mass of the host galaxy than other semi-SFR galaxies. We calculated the specific star formation rate (SSFR) of the host galaxies of \thisgrb and find that it is lower than the average value of 0.8 G $\rm yr^{-1}$, but it is consistent (lies towards the lower edge) with the correlation between the stellar mass and the SSFRs \citep[see Figure \ref{host_ssfr};][]{2009ApJ...691..182S}.

\section{Discussion}
\label{discussion}
Based on the above results, we extend discussions about important aspects of \thisgrb and possible implications towards our understanding about early time optical afterglows by comparing with a near-complete sample of LGRBs, listed in Table \ref{sample_Modelling}, exhibiting RS emission signatures.

\subsection{Afterglow behavior using Closure relationship}
\label{closure relation}

Comparison of the temporal slopes derived using power-law fits between the optical and X-ray light curves of \thisgrb throughout reveals an apparent mismatch. The optical emission has a steep to shallow transition, as predicted in the standard external shock model of RS and FS for a thin shell case. The optical light curve also displays a late time break. The X-ray emission exhibits a normal decay followed by a steeper decay. The optical and X-ray light curves are chromatic in behavior.



 To further investigate the chromatic nature of X-ray and optical afterglow, we performed a joint spectral analysis of the available simultaneous multiwavelength data. We created SEDs in five temporal segments, shown in Figure \ref{Xray_optical_afterglow}. The first two temporal segments are for the RS dominant phase (SED 1 and SED 2), and the last three segments are the FS dominant phase (SED 3, SED 4, and SED 5). We describe the joint spectral analysis method in \S~\ref{SED} and present the results in Figure \ref{SED_fig}. The SED 1 is produced from the data observed with \swift BAT and extrapolated the index towards lower (optical) and higher ($\gamma$-ray) energies, which covers from 1.98 $\times 10^{-3}$ \keV to 5.5 $\times 10^{5}$ \keV. We used $\alpha_{\rm opt}-\beta_{\rm opt}$, $\alpha_{\rm x-ray}-\beta_{\rm x-ray}$ closure relations of RS and FS \citep{2013NewAR..57..141G} to constrain the $p$ value and position of the cooling-break frequency ($\nu_c$) in the slow cooling case of synchrotron spectrum for an ISM-like medium. We notice that the observed LAT flux is lower than the extrapolated value during SED 1, indicating the presence of possible spectral break ($\nu_c$) between BAT and LAT frequencies. This is in agreement with spectral regime ($\nu~>~max \{ \nu_m,~\nu_c \}$) for the LAT photons discussed in \S~\ref{lat_TRS}.
 The optical and X-ray emission is consistent with $\nu_{\rm opt}$ $<$ $\nu_{\rm x-ray}$ $<$ $\nu_{\rm c}$ spectral regime. \cite{2018ApJ863138A} found that if the cooling break is either in between the XRT or LAT threshold energy or above this then the source can be detected in LAT band which can be modelled using synchrotron emission. We calculated the $p$ value using observed value of $\alpha_{\rm opt}-\beta_{\rm opt}$, $\alpha_{\rm x-ray}-\beta_{\rm x-ray}$ and find $p$ = 2.33 $\pm$ 0.21, this is consistent with that calculated from afterglow modelling (see \S~\ref{optical_afterglow_modelling}). The SED 2 is produced from the data observed with BOOTES optical telescope, \swift BAT, and \swift XRT from \fermiT+ 63 to \fermiT+ 100 sec. We consider a simple power-law model along with galactic and host X-ray absorption components for the joint BAT and XRT spectral analysis. We calculated a photon index of $\Gamma = -1.49 \pm 0.05$ (see Table \ref{SED_table}). Therefore, the soft and hard X-ray radiation during this temporal window should be originated from the same component. 
We calculated the $p$ value using observed value of $\alpha_{\rm opt}-\beta_{\rm opt}$, $\alpha_{\rm x-ray}-\beta_{\rm x-ray}$ during this temporal window, and find $p$ = 2.10 $\pm$ 0.21, this is consistent with that calculated from SED 1.  

The SED 3 is produced from the data observed with BOOTES optical telescope, \swift UVOT, and \swift XRT from \fermiT+ 700 to \fermiT+ 1000 sec. In this case, the X-ray and optical could be described with a simple power-law. The observed value of $\alpha_{\rm opt}-\beta_{\rm opt}$, $\alpha_{\rm x-ray}-\beta_{\rm x-ray}$  indicates that $\nu_{\rm c}$ was still beyond the $\nu_{\rm opt}$  and  $\nu_{\rm x-ray}$ spectral regime. We calculated the redshift and host extinction of the burst using this SED due to the availability of multi-band optical observations during this temporal window. Furthermore, we created two more SEDs (SED 4 and SED 5) at late epochs (23-31 ks and 51-65 ks) using optical and X-ray data. In these cases also, the X-ray and optical could be described with a simple power-law, and it appears that the X-ray and the optical are on the same spectral segment ($\nu_{\rm opt}$ $<$ $\nu_{\rm x-ray}$ $<$ $\nu_{\rm c}$), though their light curve decay slopes are different (1.50 $\pm $0.02 and 1.11 $\pm$ 0.15 for X-ray and optical, respectively). However, they are still consistent within 3 sigma.

\subsubsection{\bf Early Optical Afterglow and Reverse shock}
The optical afterglow of \thisgrb faded continuously after the first detection (\fermiT+ 29 sec, using BOOTES-4 robotic telescope) and showed a steep to shallow transition, not a simple power-law behavior as is more usually observed in the light curve of optical afterglow. It is likely that the early optical afterglow light curve is produced by the RS emission, and $\sim$ \fermiT+ 100 sec presents the beginning of the dominance of the FS emission. In the case of \thisgrb, we do not have optical observations before the peak time, but we constrain the decaying index after the peak as -1.72 $\pm$ 0.04 ($\sim$ \fermiT+100), and this decay index is consistent with the expected value for RS emission due to a thin shell expanding into the ISM-like medium. The closure relation for the RS II class i.e. $\alpha^{R}_{\rm dec, opt}$ = -(27p+7)/35 = - 1.74 considering the p = 2.00 (see Table \ref{sample_Modelling}), which is  consistent with the observed value. Furthermore, we also estimated the expected value of the rising index of RS component using the closure relation i.e. $\alpha^{R}_{\rm ris, opt}$ = (6p-3)/2 =  4.50 considering the same value of $p$.

\subsubsection{\bf Origin of the high energy LAT photons}

\label{lat_TRS}
\par
 For \thisgrb, the extended LAT emission becomes softer and marginally brighter (consistent with statistical fluctuation) after the end of GBM \keV-MeV emission. This indicates that the LAT GeV emission originated later than the GBM \keV-MeV emission and from a different spatial region. In this section, we study the possible external origin and radiation mechanism of detected LAT emission. 

 To investigate the radiation mechanism of high energy GeV LAT photons, we calculated the maximum photon energy emitted by the synchrotron radiation mechanism in an adiabatic external forward shock during the decelerating phase in the case of ISM external medium. We used the following expression (see equation \ref{maxsy_energy}) from \cite{2010ApJ...718L..63P}:
 \begin{equation}
 \label{maxsy_energy}
{\rm h} \nu_{max}=
9{\rm~GeV~} \left(\frac{E_{\rm iso,54}}{n_0}\right)^{1/8}
\left(\frac{1+z}{2}\right)^{-5/8} \left(\frac{t}{100}\right)^{-3/8}   
\end{equation}

 Where $E_{\rm iso,54}= E_{\rm iso}/10^{54}$ in ergs, $t$ is the arrival time of the event since \fermiT in sec, and $n_0$ is the ISM density. We consider $n_0$ =  0.70 {$\rm cm^{-3}$} for the present analysis (see Table \ref{sample_Modelling}). We noticed that some of the late time photons (even one photon with source association probability $>$ 90 $\%$) lies above the maximum synchrotron energy. It indicates the non-synchrotron origin of these photons. Photons above the maximum synchrotron energy from recent VHE detected GRBs confirms the Synchrotron self-Compton origin for these high energy photons \citep{2019Natur.575..455M, 2019Natur575464A, 2019Natur575448Z}. 
 
 Further, we fitted the LAT energy and photon flux light curves (see Figure \ref{fig:LAT_LCs}) obtained from the time-resolved analysis with a simple power-law decay model, which shows that emission could be continuously decreasing with time in both energy and photon fluxes during the temporal bins after the prompt phase (\fermiT + 5 sec):  $5 - 10$ sec, $10 - 100$ sec, and $100 - 1000$ sec. 
 The \fermi LAT photon flux light curve shows temporal variation as a power law with an index $-1.47 \pm 0.01$
 and the energy flux light curve shows temporal variation as a power law with an index $-1.29\pm 0.06$.  
 The LAT photon index 
($\it \Gamma_{\rm lat}$) is $-2.18\pm 0.08$ from a spectral fit obtained by fitting the first $10^{5}$ sec data. 
This gives spectral index $\beta_{\rm lat} = \it \Gamma_{\rm lat} + 1$ to be $-1.18 \pm 0.08$. The time-resolved spectra do not show strong temporal variation in the photon index in the first four temporal bins.

\begin{table*}
\centering
\scriptsize
\caption{Afterglow modelling results from our theoretical fits for \thisgrb using the combination of external reverse (RS) and forward shock (FS) model. We also present a sample of confirmed optical RS cases consistent with the thin shell in the ISM ambient medium to compare with the obtained parameters for \thisgrb. For the present analysis, we have calculated the values of $T_{\rm dec}$ and $R_{\rm dec}$ for all such events. In the case of GRB 161023A, we assume $n$ = 1 cm$^{-3}$.}
\label{sample_Modelling}
\begin{tabular}{|c|c|c|c|c|c|c|c|c|c|c|c|} 
\hline
\bf GRB/ &  {\boldmath$z$} & {\boldmath $T_{90}$} & \boldmath $p$ & \boldmath $\it \Gamma_{0}$& \boldmath $\epsilon_{\rm {e,f}}$  & \boldmath $\epsilon_{\rm {B,f}}$ & \boldmath $R_{\rm B}$  & \boldmath{$n_{0}$}  & \boldmath $E_{\rm K}$  & \boldmath$\eta_{\bf \gamma}$ & \boldmath $T_{\rm dec}/ R_{\rm dec}$  \\ 
 {\bf  (References)} &  & \bf  [sec]  &  & & $\bf [10^{-3}]$ & $\bf [10^{-5}]$ & \boldmath [$\epsilon_{\rm {B,r}}$ $/$ $\epsilon_{\rm {B,f}}$] & {$\bf [cm^{-3}]$} & $\bf [10^{52}  erg]$ &   & {\bf [sec] [$\bf \times ~10^{16}$ cm]}    \\ \hline
990123 & 1.60 & 63.3 &2.49 & 420 & 79.0 &  5 &  1156 & 0.3 & 108.0 & 0.2 $<$ $\eta_{\gamma}$ $<$ 0.9 &  36.36/ 14.80   \\ 
(1, 2, 3) &  &  & &  &  &   &   & &  &   &    \\ 
021211 & 1.006 & 2.41 &2.20 & 154 & 130.0  & 3  & 128 & 9.9 & 3.0 & $<$ 0.6&  38.45/ 2.73   \\
( 3, 4, 5 ) &  &  & &  &  &   &   & &  &   &   \\ 
060908 & 1.884 & 19.3 &2.24 & 107 & 14.0 &117 & 72 &  $190.0$ & 2.7 & 0.5 $<$ $\eta_{\gamma}$ $<$ 0.9 &   52.64/ 1.25  \\
(3, 6)&  &  & &  &  &   &   & &  &   &    \\ 
061126 &1.1588 & 191  &2.02 & 255 & 420.0 & 8 & 69 & 3.7  & 12.0 & 0.4 $<$ $\eta_{\gamma}$ $<$ 0.9 &  23.76/ 4.29  \\
(3,7)&  &  & &  &  &   &   & &  &   &    \\ 
080319B &0.937& $>$ 50 &2.57 & 286 & 68.0 &  4 & 16540 &  0.6  & 67.6 & $>$ 0.6 & 51.24/ 12.98 \\
(3, 8)&  &  & &  &  &   &   & &  &   &    \\ 
081007 & 0.5295 & 8.0 & 2.72 & 100 & - & -  & -  & 1 & 79  & 0.2 &  592.37/ 23.24  \\
(3, 9, 10, 11)&  &  & &  &  &   &   & &  &   &    \\ 
090102 & 1.547 & 27 &2.31 & 228 & 0.4 &  2 & 6666 &  359.0 &  816.0 & $<$ 0.4&  33.57/ 4.11 \\
(3, 12, 13)&  &  & &  &  &   &   & &  &   &    \\ 
090424 & 0.544 & 48 & 2.06 & 235  & 2.7 &  19 & 25 &  4.0 & 258 & $<$ 0.6 &  57.26/ 12.29 \\
(3, 10)&  &  & &  &  &   &   & &  &   &   \\ 
130427A & 0.34 & 162.83 &2.08 & 157& 3.3 & $22$ & 4 & $1.5$  & 521.0 & $<$ 0.8 & 255.34/ 28.18 5 \\
(3, 14, 15)&  &  & &  &  &   &   & &  &   &   \\ 
140512A & 0.725 & 154.8 &  2.25 & 112.3 & 290 &  0.00182 & 8187 & 9.7  & 765 & - &  490.05/ 21.50   \\ 
(16, 17, 18)&  &  & &  &  &   &   & &  &   &    \\ 
161023A &2.708& 80 & 2.32 & 140 & 4 & 1000 & 3.2 & - & 48 & - &  495.88/ 15.73  \\ 
(11, 19)&  &  & &  &  &   &   & &  &   &    \\ 
180418A & $<$ 1.31 & 1.5 &2.35 & 160 & 100 & 100 & 14 & 0.15 & 0.077 & -&  30.96/ 3.17 \\ 
(20, 21)&  &  & &  &  &   &   & &  &   &    \\ \hline
\bf 140102A & \bf 2.02$^{+0.05}_{-0.05}$ & \bf 3.58$^{+0.01}_{-0.01}$ & \bf 2.00$^{+0.01}_{-0.01}$ & \bf 218.98$^{+3.50}_{-3.67}$  & \bf 77.0$^{+6.7}_{-6.4}$ & \bf 420.0$^{+50.0}_{-40.0}$ & \bf 17.75 & \bf 0.70$^{+0.06}_{-0.05}$ & \bf 0.12 & \bf 0.99 & \bf 18.79/1.78  \\
(\bf Present work)&  &  & &  &  &   &   & &  &   &    \\ 
\hline
\end{tabular} \\
{(1) \cite{1999ApJ...518L...1B}; (2) \cite{1999GCN...224....1K}; (3) \cite{2014ApJ...785...84J}; (4) \cite{2006A&A...447..145V}; (5) \cite{2005ASPC..342..326P}; (6) \cite{2006GCN..5551....1P}; (7) \cite{2008ApJ...687..443G}; (8) \cite{2008GCN..7444....1V}; (9) \cite{2008GCN..8335....1B}; (10) \cite{2013ApJ...774..114J}; (11) \cite{2020ApJ...895...94Y}; (12) \cite{2009GCN..8766....1D}; (13) \cite{2010MNRAS.405.2372G}; (14) \cite{2013GCN.14455....1L}; (15) \cite{2013GCN.14470....1B}; (16) \cite{2014GCN.16310....1D}; (17) \cite{2014GCN.16258....1S}; (18) \cite{2016ApJ...833..100H}; (19) \cite{2018A&A...620A.119D}; (20) \cite{2019ApJ...881...12B}; (21) \cite{2020ApJ...905..112F}.}
\end{table*}

In the external shock model, for $\nu~>~max \{ \nu_{\rm m},~\nu_{\rm c} \}$ which is generally true for reasonable shock parameters we can derive
the power-law index of the shocked electrons by $f_\nu \propto \nu^{-p/2}$. We 
have synchrotron energy flux $\rm f_{\rm lat} \propto \nu^{-\beta_{\rm lat}} t^{-\alpha_{\rm lat}} $. We found $\alpha_{\rm lat} = -1.29 \pm 0.06$ (see the LAT light curve in Figure \ref{fig:LAT_LCs}) and $\beta_{\rm lat} = -1.18 \pm 0.08$. The value of $\beta_{\rm lat}$ gives us $p=2.36 \pm 0.16$. Thus, the power-law index for the energy flux decay can be predicted by using $f_{\rm lat}\propto t^{(2-3p)/4}$. 
The calculated value of $\alpha_{\rm lat}$ is $-1.27 \pm 0.16$, which agrees well with the observed value of $-1.29 \pm 0.06$. Hence, we can conclude that for \thisgrb, the extended LAT high energy afterglow is formed in an external forward shock.

\subsection{Derived physical parameters of \thisgrb and other thin shell cases}

\label{sample comparison}

\begin{figure}
\centering
\includegraphics[height=7.5cm,width=8.7cm]{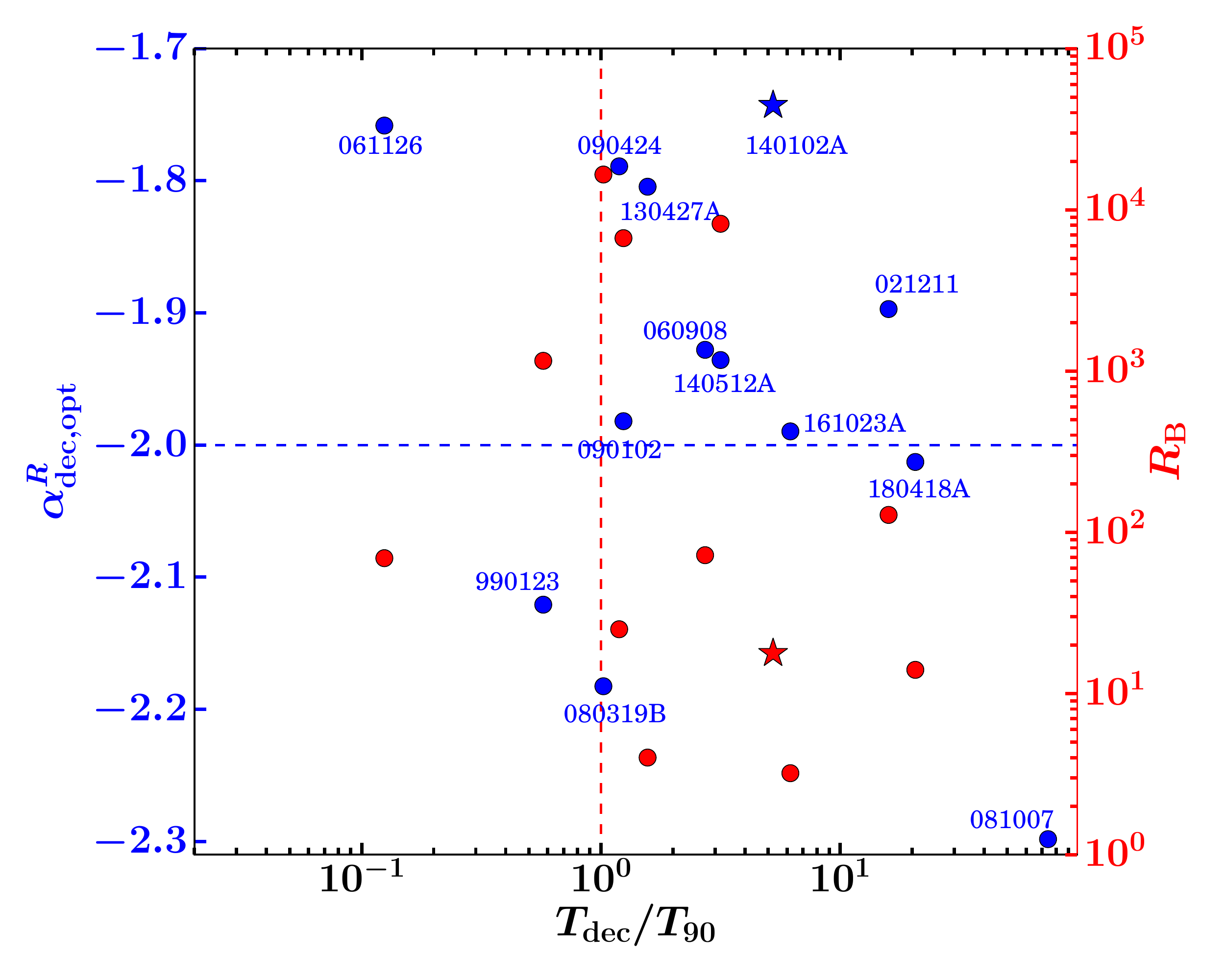}
\caption {A plot of decaying temporal indices of the RS emission ($\alpha^{\rm R}_{\rm dec, opt}$) versus ratio of deceleration time ($T_{\rm dec}$) and \tninty duration (left side Y scale) taken from the literature (see Table \protect\ref{sample_Modelling}). The right side Y-scale corresponds to the distribution of the magnetization parameter. Blue and red stars show the position of \thisgrb in this plane. The vertical red dashed line shows a ratio of $T_{\rm dec}$ and \tninty equal to one. The horizontal blue dashed line represents a line for $\alpha^{\rm R}_{\rm dec, opt}$ = -2.0.}
\label{thin_sample}
\end{figure}

In this section, derived parameters of \thisgrb ($T_{\rm dec}$ and $ R_{\rm B}$) are compared with other well-studied thin shell cases of optical RS emission.
We collected the sample of confirmed optical RS cases consistent with the thin shell in the ISM medium from the literature (see Table \ref{sample_Modelling}) to the completeness of the sample. We estimated the expected decay index ($\alpha^{R}_{\rm dec, opt}$) using the closure relation i.e. $\alpha^{R}_{\rm dec, opt}$ = -(27p+7)/35 for RS component, deceleration time ($T_{\rm dec}$) and radius ($R_{\rm dec}$) of the blast-wave for each of such events. We used blast-wave kinetic energy, Lorentz factor, circumburst medium density, and redshift parameters from the literature to calculate the $T_{\rm dec}$ and $R_{\rm dec}$ of these bursts. We show the distribution of $\alpha^{R}_{\rm dec, opt}$ and $\rm R_{B}$  with ratio of deceleration time ($T_{\rm dec}$), and \tninty duration. Interestingly, we find that GRB 990123 and GRB 061126 do not follow the criteria of the thin shell, i.e., $T_{\rm dec}$ $>$ \tninty. However, these events have a large value of magnetization parameter ($R_{\rm B}$ $>>$ 1), and ($\alpha^{R}_{\rm dec, opt}$) $\sim$ -2 as excepted from thin shell case of RS component in the ISM medium. This could be because of the dependency of \tninty duration on energy range and detector sensitivity. It could also be possible due to the prolonged central engine activity. In the sample of GRBs shown in Figure \ref{thin_sample}, bursts having RS emission ($T_{\rm dec} \ge$ \tninty) are considered. It is obvious from the figure that the observed $T_{\rm dec}$ values being spread over more than two magnitudes in time. We also notice that the deceleration radius of these events ranges from 1.25 $\times 10^{16}$ - 2.82 $\times 10^{17}$ cm, suggesting a diverse behavior of ejecta surrounding possible progenitors. A larger sample of GRBs with RS detection is fruitful to understand if a thin shell case leads to dominant RS emission. If the prompt emission is bright, then any sub-dominant early optical afterglow may not be observed. In the case of \thisgrb, the RS early optical afterglow is found with magnetization parameter $R_{\rm B} \sim$ 18, and this value lies {towards lower side} of the distribution of magnetization parameter for the sample of RS dominated bursts.

\section{Summary and Conclusions}
\label{conclusions}

We have reported a detailed prompt emission, early afterglow, and host properties of \thisgrb using the multiwavelength observations. We find a rare and exciting prompt emission spectral evolution where \Ep and $\it \alpha_{\rm pt}$ track the intensity of burst. The observed correlation between \Ep and intensity can be understood as the fireball cools and expands. The relativistic fluid loses energy, causing both the electron energy and magnetic field to reduce. The synchrotron energy scale will then naturally reduce, and both intensity and \Ep will fall. However, during the bursting phase, the opposite may be possible, and energy supply from the central engine would increase both of these parameters. Recently, \cite{2019MNRAS4841912R} explain the $\it \alpha_{\rm pt}$- intensity correlation in terms of subphotospheric heating in a flow with a varying entropy, all though, in the case of \thisgrb, we did not find a thermal signature in the time-resolved spectroscopy. For this burst, the spectral lag analysis reveals the presence of rarely observed negative lags \citep{2018JHEAp..18...15C}, and it is consistent with the prediction of \cite{2018ApJ...869..100U}, that the negative spectral lag is possible for \Ep-intensity correlation. It will be interesting to examine the spectral evolution-spectral lag correlation for a larger sample of bright bursts with negative spectral lags to understand the emission mechanism of prompt emission better.

Our afterglow modelling results suggest that the early ($\sim 30-100$ sec), bright optical emission of \thisgrb can be well described with the RS model, and the late emission can be explained with the FS model. The RS model parameters for the early optical emission are useful to understand the magnetic energy available in the jet and the source environment \citep{2003ApJ...597..455K, 2000ApJ...542..819K}. We find that the value of $\epsilon_{\rm \bf{B,r}}$ is larger than $\epsilon_{\rm \bf{B,f}}$, which provides the value of magnetization parameter $R_{\rm B} \approx 18$. It demands a moderately magnetized baryonic jet-dominated outflow for \thisgrb, similar to other cases of RS-dominated bursts \citep{2014ApJ...785...84J, 2015ApJ...810..160G}. We find a lower value of electron equipartition parameter for the reverse shock ($\epsilon_{\rm \bf{E,r}}$) than the commonly assumed value of $\epsilon_{\rm \bf{E,r}}$ =0.1.  \cite{2015ApJ...810..160G} suggest that a lower electron equipartition parameter in the external shock would lighten the `low-efficiency problem' of the internal shock model. We calculated the radiative efficiency ($\eta$= $E_{\rm \gamma, iso}$/($E_{\rm \gamma, iso}$+ $E_{\rm k}$)) value equal to 0.99 for \thisgrb. Our model predicts a slightly lower X-ray flux during the early phase, similar to other cases of RS-dominated bursts. It might be some intrinsic property of the source, either the central engine activity or wavelength-dependent origin \citep{xray_access}. It will be beneficial to investigate the possible origin of excess X-ray emission for a larger sample in the near future. The closure relations indicate that the optical and X-ray emission is consistent with $\nu_{\rm opt}$ $<$ $\nu_{\rm x-ray}$ $<$ $\nu_{\rm c}$ spectral regime for a slow cooling and ISM ambient medium. However, the observed LAT flux during the first SED lies below the extrapolated power-law decay slope, indicating the presence of a possible spectral break between BAT and LAT frequencies. We compare the physical parameters of \thisgrb with other well-known cases of RS thin shell in the ISM-like medium and find that GRB 990123 and GRB 061126 have $T_{\rm dec}$ $<$ \tninty even after a larger value of the magnetization parameter ($R_{\rm B}$ $>>$ 1, i.e., thin shell case) and observed $T_{\rm dec}$ values being spread over more than two magnitudes in time. We also notice that the deceleration radius of these events spread over more than a magnitude (Table \ref{thin_sample}), suggesting a diverse behavior of ejecta surrounding possible progenitors.

The host galaxy SED modelling suggests that the measured redshift value ($z = 2.8^{+0.7}_{-0.9}$) is consistent with the value obtained from SED. Furthermore, we find that the SFR and stellar mass of the host galaxy of this source is higher than the median value of the sample for star-forming galaxies studied by \cite{2009ApJ...691..182S}. Overall, \thisgrb provides a detailed insight into prompt spectral evolution and early optical afterglow along with GeV emission. In future, many more observations of such early optical afterglows and their multiwavelength modelling of RS-dominated GRBs using different RS parameters might help to resolve the open questions like low-efficiency problem, degree of magnetization, ejecta behavior, and environment, etc.

\section*{Acknowledgements}
 We thank the anonymous referee for providing constructive comments. RG, SBP, AA, DB, KM, and VB acknowledge BRICS grant {DST/IMRCD/BRICS/PilotCall1/ProFCheap/2017(G)} for the financial support. AA also acknowledges funds and assistance provided by the Council of Scientific \& Industrial Research (CSIR), India. SRO acknowledges the support of the Spanish Ministry, Project Number AYA2012-39727-C03-01. VAF was supported by RFBR 19-02-00311A grant. RG thanks Dr. V. Chand and Dr. P. S. Pal for helping in high energy data analysis and Bayesian Block algorithm. We are thankful to Dr. P. Veres for sharing data files related to Figure \ref{hr_t90} (a). RG and SBP are also thankful to Dr. Judith Racusin, Prof. Gudlaugur Johannesson, and Prof. Nicola Omodei from the LAT team for their valuable comments and suggestions on the manuscript. This research has made use of data obtained from the High Energy Astrophysics Science Archive Research Center (HEASARC) and the Leicester Database and Archive Service (LEDAS), provided by NASA's Goddard Space Flight Center 
and the Department of Physics and Astronomy, Leicester University, UK, respectively. This work is based on observations made with the Gran Telescopio Canarias (GTC), installed at the Spanish Observatorio del Roque de los Muchachos of the Instituto de Astrofísica de Canarias, in the island of La Palma. Also based on observations collected at the Centro Astronómico Hispano-Alemán (CAHA) at Calar Alto, operated jointly by Junta de Andalucía and Consejo Superior de Investigaciones Científicas (IAA-CSIC). The \textit{Fermi} LAT Collaboration acknowledges generous ongoing support from a number of agencies and institutes that have supported both the development and the operation of the LAT as well as scientific data analysis. These include the National Aeronautics and Space Administration and the Department of Energy in the United States, the Commissariat \`a l'Energie Atomique and the Centre National de la Recherche Scientifique / Institut National de Physique Nucl\'eaire et de Physique des Particules in France, the Agenzia Spaziale Italiana and the Istituto Nazionale di Fisica Nucleare in Italy, the Ministry of Education, Culture, Sports, Science and Technology (MEXT), High Energy Accelerator Research Organization (KEK) and Japan Aerospace Exploration Agency (JAXA) in Japan, and the K.~A.~Wallenberg Foundation, the Swedish Research Council and the
Swedish National Space Board in Sweden.
Additional support for science analysis during the operations phase is gratefully acknowledged from the Istituto Nazionale di Astrofisica in Italy and the Centre National d'\'Etudes Spatiales in France. This work performed in part under DOE Contract DE-AC02-76SF00515. The part of the work was performed as part of the government contract of the SAO RAS approved by the Ministry of Science and Higher Education of the Russian Federation.

\section*{Data Availability}
The data used in the present work can be made available based on the individual request to the corresponding authors.



\bibliographystyle{mnras}   
\bibliography{GRB140102A} 

\IfFileExists{\jobname.bbl}{}
 {\typeout{}
  \typeout{******************************************}
  \typeout{** Please run "bibtex \jobname" to optain}
  \typeout{** the bibliography and then re-run LaTeX}
  \typeout{** twice to fix the references!}
  \typeout{******************************************}
  \typeout{}
 }

\section*{Affiliations}
\small{$^{1}$ Aryabhatta Research Institute of Observational Sciences (ARIES), Manora Peak, Nainital-263002, India \\
 $^{2}$ Department of Physics, Deen Dayal Upadhyaya Gorakhpur University, Gorakhpur-273009, India \\
 $^{3}$ School of Physics and Astronomy, University of Birmingham, B15 2TT, UK \\
 $^{4}$ Instituto de Astrofisica de Andalucia (IAA-CSIC), Glorieta de la Astronomia s/n, E-18008, Granada, Spain.\\
$^{5}$ Departamento de Ingenier\'ia de Sistemas y Autom\'atica, Escuela de Ingenier\'ias, Universidad de M\'alaga, C\/. Dr. Ortiz Ramos s\/n, 29071 M\'alaga, Spain \\
$^{6}$ School of Astronomy and Space Science, Nanjing
University, Nanjing 210093, China \\
$^{7}$  Key Laboratory of Modern Astronomy and Astrophysics (Nanjing University), Ministry of Education, China \\
$^{8}$ Universidad de Granada, Facultad de Ciencias Campus Fuentenueva s/n E-18071 Granada, Spain \\
$^{9}$ Special Astrophysical Observatory of Russian Academy of Sciences, Nizhniy Arkhyz, Russia \\
$^{10}$ Crimean Astrophysical Observatory, Russian Academy of Sciences, Nauchnyi, 298409 Russia  \\
$^{11}$ School of Studies in Physics and Astrophysics, Pandit Ravishankar Shukla University, Chattisgarh 492 010, India \\
$^{12}$ Center for Research and Exploration in Space Science and Technology (CRESST) and NASA Goddard Space Flight Center, Greenbelt, MD 20771, USA \\
$^{13}$ Department of Physics, University of Maryland, Baltimore County, 1000 Hilltop Circle, Baltimore, MD 21250, USA \\
$^{14}$ National Astronomical Observatories, Chinese Academy of Sciences, 20A Datun Road, Chaoyang District, Beijing 100101, China.\\
$^{15}$ Yunnan National Astronomical Observatory, Chinese Academy of Sciences, Phoenix Hill, 650011 Kunming, Yunnan, China. \\
$^{16}$ Inter-University Center for Astronomy and Astrophysics, Pune, Maharashtra 411007, India\\
$^{17}$ Department of Physics, Adiyaman University, 02040 Adiyaman, Turkey.\\
$^{18}$ Unidad Asociada Grupo Ciencias Planetarias UPV/EHU-IAA/CSIC, Departamento de Fisica Aplicada I, E.T.S., Universidad del Pais Vasco UPV/EHU, Bilbao, Spain.\\
$^{19}$ Ikerbasque, Basque Foundation for Science, Bilbao, Spain.\\
$^{20}$ European Space Astronomy Centre (ESA-ESAC), Camino bajo del Castillo, s/n, Villafranca del Castillo, E-28.692 Villanueva dela Ca\~{n}ada (Madrid), Spain \\
$^{21}$ Universidad Internacional Valenciana, Valencia, Spain \\
$^{22}$ Mullard Space Science Laboratory - University College London, Holmbury Rd, Dorking RH5 6NT, UK \\
$^{23}$ Istituto Astrofisica Spaziale e Fisica Cosmica Palermo (INAF), Palermo, Via U. La Malfa 153, I-90146 Palermo, Italy \\
$^{24}$ Astronomical Institute of the Czech Academy of Sciences (ASU-CAS), Fri\v{c}ova 298, 251 65 Ond\v{r}ejov, CZ \\
$^{25}$ Fyzik\'{a}ln\'{\i} \'{u}stav AV \v{C}R, Na Slovance 2, CZ-182 21 Praha 8, Czech Republic  \\
$^{26}$ Space Research Institute, Moscow, Russia \\
$^{27}$ INAF, Istituto di Astrofisica e Planetologia Spaziali, via Fosso del Cavaliere 100, I-00133 Rome, Italy \\
$^{28}$ Nikolaev National University, Nikolska 24, Nikolaev 54030, Ukraine.\\
$^{29}$ Nikolaev Astronomical Observatory, Nikolaev, Ukraine. \\
$^{30}$ Centre for Astro-Particle Physics (CAPP) and Department of Physics, University of Johannesburg, PO Box 524, Auckland Park 2006, South Africa \\
$^{31}$ Indian Institute of Technology Bombay, Powai, Mumbai 400076, India \\
$^{32}$ Instituto de Educaci\'on Secundaria Beniaj\'an, Departamento de Matem\'aticas, Avda. Monteazahar, 17, 30570 Murcia, Spain.}

\appendix
\label{appendix}

\renewcommand{\thefigure}{A\arabic{figure}}

\setcounter{figure}{0}

\begin{figure*}
\centering
\includegraphics[height=19.5cm,width=18.5cm]{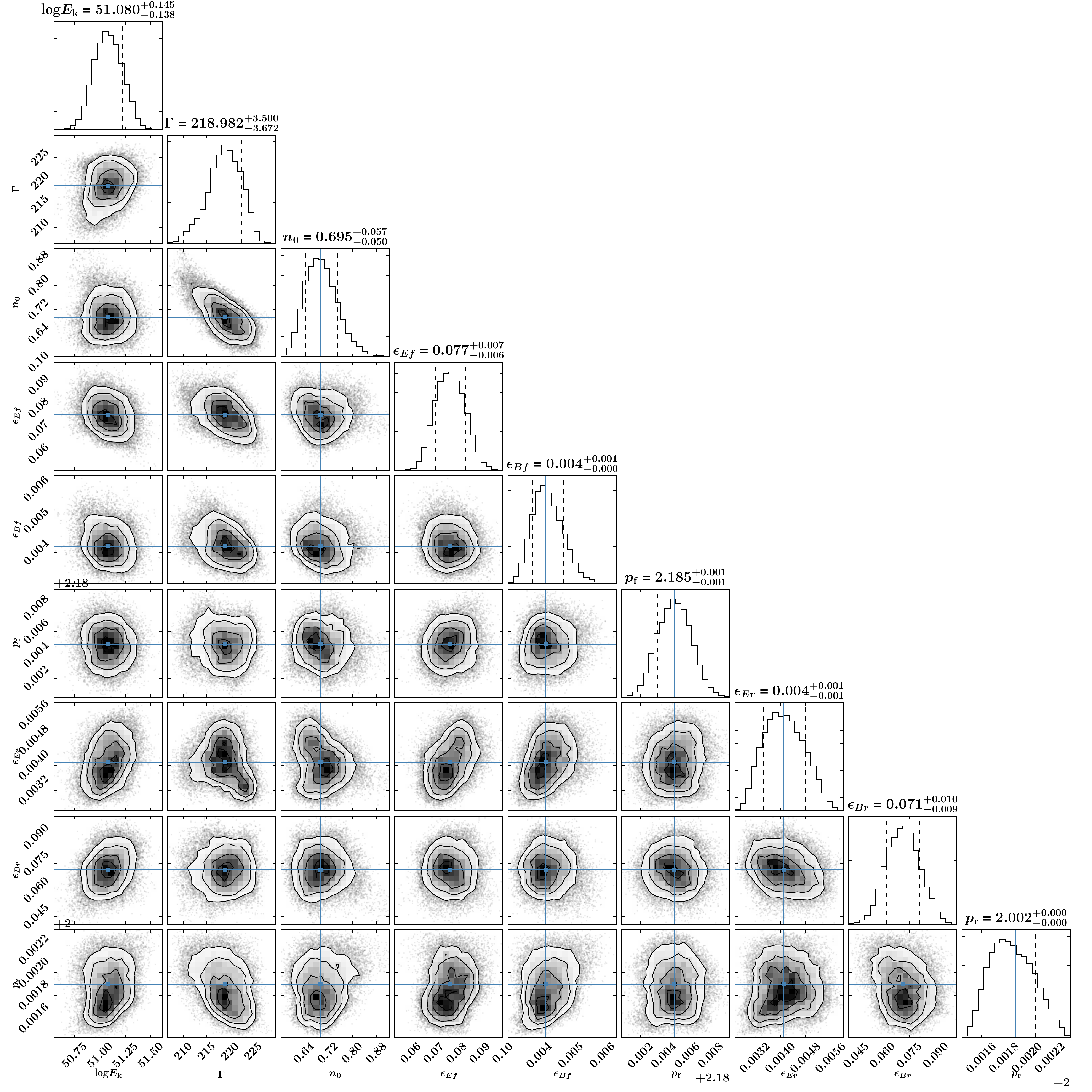}
\caption{The corner plot has been obtained from the \sw{PyMultiNest} simulation for each parameter using the thin shell RS and FS afterglow model. The best-fit parameter values in this figure are shown in blue.} 
\label{corner}
\end{figure*}

\section{Tables}
{ These tables belong to the analyses presented in \S~\ref{multiwaveength observations and data analyisis}.}
\begin{table*}
\begin{small}
\caption{\fermi LAT high energy emission ($>100$ MeV) in different  temporal bins for a fit with a power-law model for \thisgrb. All the photon and energy fluxes has been calculated in the 100.0 - 10000.0 MeV energy range.}
\label{tab:lat_sed}
\begin{center}
\begin{tabular}{|c|c|c|c|c|c|c|c|}
\hline
\bf Sr. no. & \bf Time   & \bf Index           & \bf Energy flux                        & \bf  Photon flux              & \bf Test Statistic & \bf Log(likelihood) \\
& \bf (sec)    &                &( $\rm  \bf 10^{-8} ~ergs$  $\rm \bf cm^{-2}$ $\rm \bf sec^{-1}$)   & ($\rm \bf \times 10^{-5} ~photons$   $\rm \bf cm^{-2}$ $\rm \bf sec^{-1}$) & \bf (TS)&\\ \hline
(0)& 0 - 5   & $-1.95 \pm 0.49$ & $ 13.4 \pm 9.85$& $  16.8 \pm 9.45$& 27& 23.98\\
(1)& 5 - 10 & $-2.47 \pm 0.61 $ & $ 9.99 \pm 6.39$ &    $22.4 \pm 10.2$  &33 & 17.96\\ 
(2)& 10 - 100 & $-2.14 \pm 0.50$ &$0.75 \pm 0.50 $&$1.20  \pm 0.60$ &23& 40.87\\
(3) &100 - 1000  &$-1.33 \pm 0.32 $ & $0.13 \pm 0.08$& $0.06 \pm 0.05$ &32&  208.10\\
(4) &2905 - 7460 &$-2$ (fixed)& $ <0.02$& $ < 0.03$& 1 & 847.28 \\
\hline 
\end{tabular}
\end{center}
\end{small}
\end{table*}

\begin{table*}
\tiny
\caption{The best fit model (shown with boldface) between different models used for the time-averaged joint spectral analysis of \fermi LAT standard, LAT- LLE, \fermi GBM (NaI +  BGO), and \swift BAT data of GRB 140102A. BB correspond to the \sw{Black Body} model.}
\begin{center}
\label{tab:TAS}
\begin{tabular}{|c|c|c|c|c|c|c|c|c|}
\hline
\textbf{Time (sec)/ Detectors} & \textbf{Model} & \multicolumn{4}{c|}{\textbf{Parameters}} & \textbf{-Log(Likelihood)} & \textbf{AIC} & \textbf{BIC} \\ \hline
\multirow{6}{*}{\begin{tabular}[c]{@{}c@{}}0-5\\ \\ \\ (BAT+ GBM +LAT)\end{tabular}} & \sw{bkn2pow} & $\it \Gamma_{1,2,3}$= 0.69$^{+0.06}_{-0.04}$, 1.26$^{+0.01}_{-0.01}$, 2.47$^{+0.03}_{-0.03}$ & \multicolumn{3}{c|}{$E_{\rm b1,b2}$= 26.92$^{+1.18}_{-1.12}$, 146.31$^{+3.24}_{-3.94}$} & 2407.72 & 4844.22 & 4903.75 \\ \cline{2-9} 
 & \sw{SBPL} & $\alpha_{1}$=  -0.92$^{+0.02}_{-0.01}$ & \multicolumn{2}{c|}{$\alpha_{2}$= -2.85$^{+0.06}_{-0.03}$} & $E_{0}$= 163.28$^{+5.94}_{-8.15}$ & 2435.64 & 4895.86 & 4946.97 \\ \cline{2-9} 
 & Band & $\it \alpha_{\rm pt}$= -0.85$^{+0.02}_{-0.01}$ & \multicolumn{2}{c|}{$\it \beta_{\rm pt}$= -2.72$^{+0.03}_{-0.05}$} & \Ep = 190.21$^{+2.91}_{-5.55}$ & 2426.26 & 4877.11 & 4928.22 \\ \cline{2-9} 
 & \sw{bkn2pow}+BB & $\it \Gamma_{1,2,3}$= 0.73$^{+0.01}_{-0.01}$, 1.49$^{+0.03}_{-0.03}$, 2.60$^{+0.02}_{-0.06}$ & $E_{\it b1,b2}$= 26.39$^{+0.16}_{-0.15}$, 254.72$^{+15.42}_{-21.30}$ & \multicolumn{2}{c|}{$\rm k{\it T}_{\rm BB}$ = 32.91$^{+1.19}_{-1.18}$} & 2403.39 & 4839.80 & 4907.71 \\ \cline{2-9} 
 & \sw{SBPL}+BB & $\alpha_{1}$= -0.58$^{+0.10}_{-0.09}$ & $\alpha_{2}$= -2.93$^{+0.07}_{-0.04}$ & $E_{0}$= 143.99$^{+10.33}_{-12.83}$ & $\rm k{\it T}_{\rm BB}$= 7.48$^{+0.25}_{-0.38}$ & 2402.90 & 4834.59 & 4894.11 \\ \cline{2-9} 
 & \textbf{Band+BB} & $\alpha_{\rm pt}$= -0.58$^{+0.04}_{-0.04}$ & $\rm \beta_{\rm pt}$ = -2.72$^{+0.02}_{-0.07}$ & \Ep= 186.57$^{+4.51}_{-4.16}$ & $\rm k{\it T}_{\rm BB}$= 7.90$^{+0.25}_{-0.28}$ & 2398.90 & 4826.58 & 4886.10 \\ \cline{2-9}
 & Band + PL & $\alpha_{1}$= -0.85$^{+0.01}_{-0.01}$ & $\alpha_{2}$= -2.72$^{+0.13}_{-0.03}$ & \Ep= 190.14$^{+1.87}_{-4.27}$ & $\it \Gamma_{\rm PL}$= -1.59$^{+0.75}_{-0.38}$ & 2425.75 & 4880.28 & 4939.80  \\ \hline
\end{tabular}

\end{center}
\end{table*}

\begin{table*}
\caption{Time-resolved spectral fitting of \thisgrb for \sw{Band} model based on Bayesian Block using \fermi GBM and \swift BAT data.}
\label{TRS_Table_band}
\begin{center}
\begin{tabular}{|c|c|c|c|c|c|c|}
\hline
\bf Sr. no. & $\rm \bf t_1$,$\rm \bf t_2$ \bf (sec) & \boldmath $\it \alpha_{\rm pt}$ & \boldmath $\it \beta_{\rm pt}$ & \boldmath \Ep (\keV) &  \bf (Flux $\times 10^{-06}$)\footnotemark  &\bf \sw{-Log(likelihood)/BIC} \\
\hline \hline

1& 0.238, 0.412&$-1.10_{-0.16}^{+0.07}$&$-5.00_{\infty}^{+0.47}$&$144.95_{-9.17}^{+35.00}$& $0.87_{-0.07}^{+0.06}$&$-205.74/-361.18$\\
2& 0.479, 0.633&$-0.67_{-0.10}^{+0.05}$&$-3.04_{-0.53}^{+0.53}$&$186.44_{-8.00}^{+22.28}$&$2.47_{-0.50}^{+0.70}$ &$-90.20/-130.09$\\
3& 0.633, 0.833&$-0.63_{-0.06}^{+0.03}$&$-2.93_{-0.25}^{+0.28}$&$176.08_{-2.95}^{+14.40}$&$6.83_{-0.80}^{+1.10}$ &$79.54/209.38$\\
4& 0.833, 0.929&$-0.53_{-0.09}^{+0.08}$&$-3.42_{-0.55}^{+0.65}$&$119.08_{-7.02}^{+7.53}$&$5.52_{-0.40}^{+0.80}$ &$-525.58/-1000.87$\\
5& 0.929, 1.118&$-0.92_{-0.06}^{+0.07}$&$\rm unconstrained$&$104.02_{-6.82}^{+6.02}$&$3.22_{-0.50}^{+0.60}$ &$-141.75/-233.19$\\
6& 1.118, 1.435&$-1.13_{-0.10}^{+0.12}$&$-3.01_{-0.92}^{+0.79}$&$81.63_{-10.17}^{+7.26}$&$1.71_{-0.50}^{+1.20}$ &$182.80/415.90$ \\
7& 1.435, 1.758 &$-1.23_{-0.14}^{+0.16}$&$-2.65_{-0.38}^{+0.40}$&$62.11_{-7.48}^{+7.01}$&$0.82_{-0.03}^{+0.50}$ &$184.68/419.66$\\
8& 1.758, 2.002&$-0.61_{-0.07}^{+0.06}$&$-2.28_{-0.15}^{+0.17}$&$224.54_{-18.21}^{+26.74}$&$13.44_{-2.70}^{+4.00}$ &$-175.92/-301.55$\\
9& 2.002, 2.420&$-0.58_{-0.05}^{+0.03}$&$-2.71_{-0.16}^{+0.18}$&$199.02_{-5.80}^{+12.37}$&$6.73_{-0.80}^{+0.90}$ &$665.37/1381.05$ \\
10& 2.420, 2.968&$-0.60_{-0.04}^{+0.02}$&$-2.81_{-0.21}^{+0.21}$&$249.31_{-5.96}^{+15.71}$&$9.66_{-0.90}^{+1.20}$ &$837.34/1724.99$\\
11& 2.968, 3.138&$-0.47_{-0.09}^{+0.07}$&$-3.51_{-0.74}^{+0.97}$&$145.48_{-9.21}^{+11.10}$&$4.54_{-0.90}^{+1.90}$ &$-133.23/-216.16$\\
12& 3.138, 3.329&$-0.67_{-0.08}^{+0.04}$&$-3.05_{-0.34}^{+0.45}$&$169.12_{-5.22}^{+16.12}$& $3.87_{-0.60}^{+0.90}$ &$36.92/124.15$\\
13& 3.329, 3.515 &$-0.40_{-0.08}^{+0.10}$&$-2.83_{-0.37}^{+0.34}$&$125.84_{-10.19}^{+7.79}$& $5.12_{-1.30}^{+1.7}$ &$-68.46/-86.62$\\
14& 3.515, 4.265&$-1.00_{-0.06}^{+0.07}$&$\rm unconstrained$&$109.35_{-8.29}^{+6.27}$  &$1.21_{-0.06}^{+0.05}$&$826.85/1704.01$ \\
15& 4.265, 4.999&$-0.93_{-0.26}^{+0.36}$&$-2.19_{-0.19}^{+0.17}$&$56.21_{-13.10}^{+9.74}$ &$0.59_{-0.22}^{+0.29}$&$763.89/1578.09$\\
\hline
\end{tabular}
\end{center}
\end{table*}
\footnotetext{erg $\rm cm^{-2}$ $\rm sec^{-1}$ in 8 \keV-30 MeV energy channel}

\begin{table*}
\caption{Time-resolved spectral fitting of \thisgrb for \sw{Black Body} + \sw{Band} and \sw{Cutoff-power law} model based on Bayesian block using \fermi GBM and \swift BAT data.}
\label{TRS_Table_bb_cpl}
\begin{center}
\begin{tabular}{|c|c|c|c|c|c|c|}
\hline
\bf Sr. no. & $\rm \bf t_1$,$\rm \bf t_2$ \bf (sec) & \boldmath ${\rm k}{\it T} ~(keV)$ & \bf \sw{-Log(likelihood)/BIC} & \boldmath $E_{\rm c}$ (\keV) &  \boldmath $\it \Gamma_{\rm Cutoff}$  &\bf \sw{-Log(likelihood)/BIC} \\
\hline \hline

1& 0.238, 0.412&$0.98_{-0.68}^{-0.04}$&$-205.88/-348.88$&$160.86_{-23.22}^{+82.05}$& $-1.10_{-0.17}^{+0.08}$&$-205.74/-367.46$\\
2& 0.479, 0.633&$0.38_{-0.15}^{+0.10}$&$-90.21/-117.54$&$156.30_{-7.32}^{+27.09}$&$-0.72_{-0.08}^{+0.04}$ &$-89.51/-135.00$\\
3& 0.633, 0.833&$0.31_{-0.07}^{+0.65}$&$79.54/221.96$&$145.64_{-6.69}^{+16.51}$&$-0.68_{-0.05}^{+0.04}$ &$81.23/206.47$\\
4& 0.833, 0.929&$1.86_{-1.23}^{+0.08}$&$-525.99/-989.11$&$86.51_{-9.95}^{+13.23}$&$-0.57_{-0.11}^{+0.10}$ &$-525.16/-1006.31$\\
5& 0.929, 1.118&$0.85_{-0.42}^{+1.37}$&$-141.75/-220.61$&$96.07_{-10.81}^{+12.46}$&$-0.92_{-0.07}^{+0.07}$ &$-141.90/-239.78$\\
6& 1.118, 1.435&$0.29_{-0.06}^{+0.004}$&$182.28/427.43$&$95.85_{-22.05}^{+20.67}$&$-1.14_{-0.10}^{+0.13}$ &$183.06/410.13$ \\
7& 1.435, 1.758 & unconstrained &$184.68/432.24$&$89.76_{-27.21}^{+40.63}$&$-1.27_{-0.20}^{+0.19}$ &$185.54/415.10$\\
8& 1.758, 2.002&$0.50_{-0.20}^{+0.002}$&$18.45/99.78$&$53.88_{-5.21}^{+9.97}$&$-0.59_{-0.16}^{+0.11}$ &$19.51/83.03$\\
9& 2.002, 2.420&$0.24_{-0.05}^{+1.96}$&$665.37/1393.62$&$164.71_{-4.43}^{+13.48}$&$-0.65_{-0.04}^{+0.02}$ &$675.42/1394.85$ \\
10& 2.420, 2.968&$0.98_{-0.31}^{+0.29}$&$837.34/1737.56$&$200.80_{-7.11}^{+12.70}$&$-0.64_{-0.03}^{+0.02}$ &$839.88/1723.77$\\
11& 2.968, 3.138&$0.80_{-0.62}^{-0.04}$ & $-133.81/-204.75$&$100.57_{-8.79}^{+11.97}$&$-0.50_{-0.08}^{+0.06}$ &$-133.27/-222.52$\\
12& 3.138, 3.329&$0.77_{-0.04}^{+1.32}$&$36.92/136.73$&$136.84_{-4.50}^{+22.57}$& $-0.70_{-0.08}^{+0.03}$ &$38.39/120.80$\\
13& 3.329, 3.515 &$\rm unconstrained$ &$-68.46/-74.04$&$88.64_{-7.38}^{+8.53}$& $-0.47_{-0.07}^{+0.07}$ &$-66.57/-89.12$\\
14& 3.515, 4.265&$0.61_{-0.33}^{-0.01}$&$823.85/1710.58$&$109.93_{-17.51}^{+15.35}$  &$-1.00_{-0.07}^{+0.09}$&$826.83/1697.68$ \\
15& 4.265, 4.999&$0.47_{-0.09}^{+0.008}$&$762.21/1587.29$&$115.84_{-35.62}^{+50.14}$ &$-1.31_{-0.16}^{+0.16}$&$765.86/1575.73$\\
\hline
\end{tabular}
\end{center}
\end{table*}

\begin{table*}
\caption{Log of \swift UVOT observations $\&$ photometry of \thisgrb afterglow. No correction for Galactic extinction is applied.}
\label{table:UVOT}
\begin{center}
\begin{tabular}{|c|c|c|c||c|c|c|c||}
\hline
\bf $\rm \bf  T_{\rm \bf  mid}$  &\bf Exp.& {\bf  Magnitude}  &\bf Filter  & \bf $\rm \bf T_{\rm \bf mid}$ &\bf Exp. & {\bf  Magnitude} & \bf Filter   \\
\bf  (sec) &   \bf  (sec)&  &  &\bf  (sec) &  \bf  (sec)&  &      \\
\hline \hline


67.82& 5.0 & $15.67^{+0.07}_{-0.07}$& $White$ & 49.90& 5.0& $13.97^{+0.11}_{-0.10}$ & $V$    \\
70.32& 10.0 & $15.70^{+0.05}_{-0.05}$& $White$ & 54.47& 4.1& $13.98^{+0.12}_{-0.11}$ & $V$   \\
72.82&5.0 & $15.74^{+0.07}_{-0.07}$ & $White$ & 617.37& 19.8& $17.21^{+0.36}_{-0.27}$ &$V$    \\
77.82& 5.0& $15.83^{+0.07}_{-0.07}$ & $White$ & 791.14& 19.8& $17.61^{+0.52}_{-0.35}$&$V$   \\
82.82&5.0 & $15.99^{+0.08}_{-0.07}$ & $White$ & 1132.96& 192.42&$ 17.80^{+0.41}_{-0.29}$& $V$   \\
82.82&15.0 & $15.95^{+0.04}_{-0.04}$& $White$ & 1479.24& 192.11& $>$18.58 & $V$   \\
87.82&5.0 & $16.02^{+0.08}_{-0.07}$ & $White$ & 1910.55& 364.92& $>$ 18.89& $V$  \\
92.82&5.0 &$16.12^{+0.08}_{-0.07}$ & $White$ &2255.54 &19.8&$>$18.36 &$V$  \\
97.82& 5.0& $16.42^{+0.09}_{-0.08}$ & $White$ & 6733.22& 199.8& $19.63^{+0.64}_{-0.40}$& $V$    \\
100.32& 20.0& $16.35^{+0.04}_{-0.04}$& $White$ &11470.66 &412.95 & $19.66^{+0.41}_{-0.30}$ & $V$   \\
102.82& 5.0&$16.29^{+0.08}_{-0.08}$ & $White$ & 53979.81& 199.8& $19.60^{+0.60}_{-0.39}$& $V$    \\
107.82& 5.0& $16.50^{+0.09}_{-0.08}$ & $White$ & 11470.66& 412.9 & $19.67^{+0.41}_{-0.30}$& $V$   \\
112.82 &5.0 & $16.42^{+0.09}_{-0.08}$ & $White$ & 53979.81&339.7& $>$ 20.61& $V$   \\
117.82& 5.0& $16.43^{+0.09}_{-0.08}$  & $White$ & 543.32&19.8 &$17.93^{+0.31}_{-0.24}$ & $B$  \\
120.32& 20.0& $16.56^{+0.04}_{-0.04}$& $White$ & 716.08 & 19.8 & $18.34^{+0.43}_{-0.31}$ & $B$   \\
122.82& 5.0& $16.58^{+0.09}_{-0.08}$ & $White$ &  1231.49 & 192.2& $18.17^{+0.25}_{-0.21}$&$B$  \\
127.82& 5.0 &$16.75^{+0.10}_{-0.09}$ & $White$ & 1577.50 & 191.8& $18.38^{+0.35}_{-0.26}$& $B$   \\
132.82& 5.0& $16.74^{+0.10}_{-0.09}$ & $White$ &  2008.91 &364.92 & $19.12^{+0.75}_{-0.44}$& $B$  \\
137.82& 5.0& $16.76^{+0.10}_{-0.09}$& $White$ & 6117.32  &199.8 &$19.89^{+0.34}_{-0.26}$ &$B$\\
142.82& 5.0&$16.81^{+0.10}_{-0.09}$ & $White$ & 7553.81 &199.8 & $20.26^{+0.73}_{-0.43}$& $B$  \\
142.82&25.0 & $16.83^{+0.04}_{-0.04}$& $White$& 36270.45 & 907.0& $>$21.50& $B$  \\
147.82& 5.0 &$16.86^{+0.10}_{-0.09}$ & $White$ & 61293.68& 7811.2& $>$ 22.19& $B$  \\
152.82& 5.0& $16.98^{+0.11}_{-0.10}$& $White$ & 288.13 &  20.0& $17.63^{+0.15}_{-0.13}$ & $U$ \\
157.82& 5.0& $16.84^{+0.10}_{-0.09}$& $White$ & 308.13  & 20.0 & $17.66^{+0.15}_{-0.13}$& $U$ \\
162.82& 5.0& $16.93^{+0.10}_{-0.09}$& $White$ &  328.13 & 20.0 & $17.61^{+0.15}_{-0.13}$& $U$  \\
167.82& 5.0 &$17.17^{+0.12}_{-0.11}$& $White$ & 348.13 & 20.0 & $17.67^{+0.15}_{-0.13}$& $U$  \\
170.32& 30.0& $17.06^{+0.04}_{-0.04}$& $White$ &368.13 & 20.0 & $17.78^{+0.0.16}_{-0.14}$ & $U$  \\
172.82&5.0& $17.12^{+0.11}_{-0.10}$& $White$ & 388.13 &20.0  &$18.10^{+0.20}_{-0.17}$ & $U$  \\
177.82&5.0 &$17.17^{+0.12}_{-0.10}$ & $White$ & 408.13 & 20.0 & $18.22^{+0.21}_{-0.18}$ &$U$ \\
182.82&5.0& $17.07^{+0.11}_{-0.10}$& $White$ & 428.13 & 20.0  & $17.71^{+0.15}_{-0.13}$& $U$ \\
187.82& 5.0&$17.29^{+0.12}_{-0.11}$ & $White$ & 448.13 & 20.0 & $18.10^{+0.20}_{-0.17}$& $U$    \\
192.82& 5.0&$17.20^{+0.12}_{-0.11}$ & $White$ & 468.13 & 20.0& $18.12^{+0.21}_{-0.17}$&$U$  \\
197.82&5.0 &$17.18^{+0.12}_{-0.11}$ & $White$ & 488.13 & 20.0 &$18.24^{+0.22}_{-0.18}$ &$U$   \\
200.20&29.76 & $17.31^{+0.05}_{-0.05}$& $White$ & 508.13& 20.0& $18.11^{+0.20}_{-0.17}$ & $U$    \\
202.82& 5.0& $17.24^{+0.12}_{-0.11}$& $White$ & 523.02 & 9.8 &$18.43^{+0.37}_{-0.28}$ &$U$  \\
207.82& 5.0& $17.44^{+0.13}_{-0.12}$& $White$ & 308.13 & 60.0  & $17.63^{+0.08}_{-0.08}$&$U$  \\
212.70& 4.8&$17.47^{+0.14}_{-0.12}$ & $White$ & 368.13& 60.0 &$17.86^{+0.09}_{-0.09}$ &$U$  \\
567.83&  19.75& $18.48^{+0.12}_{-0.11}$& $White$ & 438.13&80.0 & $18.04^{+0.09}_{-0.08}$ & $U$    \\
740.69& 19.75& $18.82^{+0.16}_{-0.14}$& $White$ &503.02 & 49.8& $18.23^{+0.13}_{-0.12}$ & $U$    \\
931.12& 149.76& $19.03^{+0.07}_{-0.06}$& $White$ &  691.33& 19.8& $18.34^{+0.23}_{-0.19}$ & $U$    \\
1255.78&192.09 & $19.23^{+0.15}_{-0.14}$& $White$ & 1206.62& 192.0& $19.20^{+0.31}_{-0.24}$ & $U$   \\
1602.07&192.42 & $19.46^{+0.21}_{-0.18}$& $White$ & 1552.79&191.9 & $18.90^{+0.27}_{-0.22}$ & $U$    \\
2033.22& 364.84& $19.53^{+0.21}_{-0.18}$& $White$ &1984.04 &364.7 & $19.19^{+0.35}_{-0.27}$ & $U$   \\
6321.93& 199.73& $20.68^{+0.23}_{-0.19}$& $White$ & 5911.80& 199.8& $20.16^{+0.22}_{-0.18}$ & $U$    \\
7758.75& 199.77& $21.15^{+0.57}_{-0.37}$& $White$ &7348.47 & 99.89200& $20.85^{+0.47}_{-0.32}$ & $U$    \\
58406.17& 211.27& $>$ 23.89 & $White$  &24964.78&725.6 &$21.39^{+0.48}_{-0.33}$ &$U$ \\
178945.23& 12160.06 & $>$ 23.94 & $White$  & 64529.61&906.8&$>$ 24.15&$U$ \\
&  &  &  & 89761.87&7192.6 &$>$ 23.59&$U$\\
\hline

\hline
\end{tabular}
\end{center}
\end{table*}

\onecolumn
\LTcapwidth=\linewidth
\centering

\begin{longtable}{|c|c|c|c|c||c|c|c|c|c|}
\caption{Log of optical photometry observations of \thisgrb afterglow. The magnitudes have not been corrected for Galactic Extinction E(B-V)= 0.0295. Magnitudes are in AB based system. Clear filter was calibrated to r filter.}
\label{optical_data}  \\ 
\hline 
 \bf $\rm \bf T_{start}$ (sec) & $\rm \bf T_{stop}$\bf (sec) & \bf Exposure (sec)  & \bf Magnitude  &\bf Filter &\bf $\rm \bf T_{start}$ (sec) & $\rm \bf T_{stop}$\bf (sec) & \bf Exposure (sec)  & \bf Magnitude  &\bf Filter\\
\hline 
\endfirsthead
\caption{continued.}\\
\hline\hline
 \bf $\rm \bf T_{start}$ (sec) & $\rm \bf T_{stop}$\bf (sec) & \bf Exposure (sec)  & \bf Magnitude  &\bf Filter &\bf $\rm \bf T_{start}$ (sec) & $\rm \bf T_{stop}$\bf (sec) & \bf Exposure (sec)  & \bf Magnitude  &\bf Filter\\
\hline 
\endhead
\hline
\endfoot
29.2     &   29.7     &   0.5     & $   12.68 \pm   0.23 $  &    C& 144.3    &   144.8    &   0.5     & $   15.24 \pm   0.06 $  &    C   \\ 
 34.3     &   34.8     &   0.5     & $   12.97 \pm   0.08 $  &    C&  146.0 &   146.5    &   0.5     & $   15.43 \pm   0.07 $  &    C   \\ 
38.9     &   39.4     &   0.5     & $   13.20 \pm   0.12 $  &    C&  147.8  &   148.3    &   0.5     & $   15.57 \pm   0.09 $  &    C   \\ 

     40.7     &   41.2     &   0.5     & $   13.35 \pm   0.27 $  &    C&  149.5    &   150.0    &   0.5     & $   15.59 \pm   0.25 $  &    C   \\ 

     42.5     &   43.0     &   0.5     & $   13.37 \pm   0.19 $  &    C&  151.3    &   151.8    &   0.5     & $   15.47 \pm   0.15 $  &    C   \\ 

     44.2     &   44.7     &   0.5     & $   13.45 \pm   0.14 $  &    C&  153.0    &   153.5    &   0.5     & $   15.63 \pm   0.20 $  &    C   \\ 

     45.9     &   46.4     &   0.5     & $   13.45 \pm   0.34 $  &    C&  154.8    &   155.3    &   0.5     & $   15.66 \pm   0.08 $  &    C   \\ 

     47.7     &   48.2     &   0.5     & $   13.63 \pm   0.19 $  &    C&  156.5    &   157.0    &   0.5     & $   15.75 \pm   0.15 $  &    C   \\ 

     49.4     &   49.9     &   0.5     & $   13.54 \pm   0.10 $  &    C&  158.2    &   158.7    &   0.5     & $   15.57 \pm   0.20 $  &    C   \\ 

     51.2     &   51.7     &   0.5     & $   13.73 \pm   0.14 $  &    C&  160.0    &   160.5    &   0.5     & $   15.59 \pm   0.19 $  &    C   \\ 

     52.9     &   53.4     &   0.5     & $   13.78 \pm   0.17 $  &    C&  161.7    &   162.2    &   0.5     & $   15.57 \pm   0.25 $  &    C   \\ 

  54.7     &   55.2     &   0.5     & $   13.87 \pm   0.14 $  &    C&  188.1    &   193.9    &   4x0.5   & $   16.04 \pm   0.08 $  &    C   \\ 

  56.4     &   56.9     &   0.5     & $   13.83 \pm   0.17 $  &    C&  195.2    &   200.9    &   4x0.5   & $   16.07 \pm   0.08 $  &    C   \\ 

  58.1     &   58.6     &   0.5     & $   13.97 \pm   0.09 $  &    C&  202.2    &   212.3    &   4x0.5   & $   16.09 \pm   0.12 $  &    C   \\

  59.9     &   60.4     &   0.5     & $   14.03 \pm   0.12 $  &    C&  213.6    &   219.3    &   4x0.5   & $   16.15 \pm   0.25 $  &    C   \\ 

  61.6     &   62.1     &   0.5     & $   13.99 \pm   0.11 $  &    C&  220.5    &   228.1    &   5x0.5   & $   16.01 \pm   0.13 $  &    C   \\

  63.4     &   63.9     &   0.5     & $   14.21 \pm   0.15 $  &    C&   240.6    &   246.4    &   4x0.5   & $   16.26 \pm   0.08 $  &    C   \\ 

  65.1     &   65.6     &   0.5     & $   14.16 \pm   0.09 $  &    C&  247.7    &   253.3    &   4x0.5   & $   16.36 \pm   0.11 $  &    C   \\ 

  66.8     &   67.3     &   0.5     & $   14.28 \pm   0.23 $  &    C&  255.5    &   261.3    &   4x0.5   & $   16.61 \pm   0.06 $  &    C   \\ 
  
  68.6     &   69.1     &   0.5     & $   14.17 \pm   0.05 $  &    C&  262.6    &   268.4    &   4x0.5   & $   16.57 \pm   0.06 $  &    C   \\
  
  70.4     &   70.9     &   0.5     & $   14.26 \pm   0.12 $  &    C&  269.6    &   277.4    &   4x0.5   & $   16.69 \pm   0.11 $  &    C   \\ 

  72.1     &   72.6     &   0.5     & $   14.40 \pm   0.16 $  &    C&  278.7    &   284.4    &   4x0.5   & $   16.45 \pm   0.10 $  &    C   \\ 

  73.8     &   74.3     &   0.5     & $   14.42 \pm   0.09 $  &    C&  285.6    &   291.4    &   4x0.5   & $   16.40 \pm   0.09 $  &    C   \\ 

  75.6     &   76.1     &   0.5     & $   14.49 \pm   0.19 $  &    C&  292.6    &   298.3    &   4x0.5   & $   16.75 \pm   0.06 $  &    C   \\ 

  77.3     &   77.8     &   0.5     & $   14.46 \pm   0.13 $  &    C&  299.6    &   307.1    &   4x0.5   & $   16.56 \pm   0.09 $  &    C   \\ 

  79.1     &   79.6     &   0.5     & $   14.56 \pm   0.12 $  &    C&  308.3    &   314.1    &   4x0.5   & $   16.45 \pm   0.10 $  &    C   \\ 

  80.8     &   81.3     &   0.5     & $   14.51 \pm   0.05 $  &    C&  315.3    &   321.1    &   4x0.5   & $   16.72 \pm   0.07 $  &    C   \\ 

  82.6     &   83.1     &   0.5     & $   14.49 \pm   0.13 $  &    C&  322.3    &   326.3    &   3x0.5   & $   16.49 \pm   0.14 $  &    C   \\ 

  84.3     &   84.8     &   0.5     & $   14.68 \pm   0.15 $  &    C&  327.8    &   339.1    &   2x5.0   & $   16.52 \pm   0.08 $  &    C   \\ 

  86.1     &   86.6     &   0.5     & $   14.73 \pm   0.12 $  &    C&  340.3    &   351.6    &   2x5.0   & $   17.03 \pm   0.06 $  &    C   \\ 

  87.8     &   88.3     &   0.5     & $   14.54 \pm   0.12 $  &    C&  352.8    &   364.1    &   2x5.0   & $   16.97 \pm   0.04 $  &    C   \\ 

  89.5     &   90.0     &   0.5     & $   14.63 \pm   0.11 $  &    C&  365.3    &   376.5    &   2x5.0   & $   16.76 \pm   0.04 $  &    C   \\ 

  96.9     &   97.4     &   0.5     & $   14.84 \pm   0.09 $  &    C&  377.8    &   389.0    &   2x5.0   & $   16.89 \pm   0.05 $  &    C   \\ 

  98.6     &   99.1     &   0.5     & $   15.02 \pm   0.12 $  &    C&  390.3    &   401.5    &   2x5.0   & $   16.95 \pm   0.05 $  &    C   \\

  100.3    &   100.8    &   0.5     & $   14.89 \pm   0.12 $  &    C&  485.3    &   496.6    &   2x5.0   & $   17.35 \pm   0.10 $  &    C   \\

  102.1    &   102.6    &   0.5     & $   15.10 \pm   0.15 $  &    C&  497.9    &   509.1    &   2x5.0   & $   17.34 \pm   0.06 $  &    C   \\

  111.1    &   111.6    &   0.5     & $   15.18 \pm   0.18 $  &    C&  510.4    &   527.2    &   2x5.0   & $   17.43 \pm   0.07 $  &    C   \\

  112.9    &   113.4    &   0.5     & $   15.10 \pm   0.12 $  &    C&  528.4    &   539.7    &   2x5.0   & $   17.36 \pm   0.09 $  &    C   \\

  114.7    &   115.2    &   0.5     & $   15.23 \pm   0.16 $  &    C&  540.9    &   552.2    &   2x5.0   & $   17.39 \pm   0.04 $  &    C   \\

  116.4    &   116.9    &   0.5     & $   15.34 \pm   0.08 $  &    C&  553.4    &   572.2    &   2x5.0   & $   17.35 \pm   0.06 $  &    C   \\ 

  118.2    &   118.7    &   0.5     & $   15.26 \pm   0.16 $  &    C&  727.6    &   757.6    &   30.0    & $   17.75 \pm   0.05 $  &    C   \\ 
  119.9    &   120.4    &   0.5     & $   15.28 \pm   0.21 $  &    C&  758.9    &   788.9    &   30.0    & $   17.78 \pm   0.03 $  &    C   \\ 

  121.6    &   122.1    &   0.5     & $   15.28 \pm   0.15 $  &    C&  1082.6   &   1112.6   &   30.0    & $   18.17 \pm   0.04 $  &    C   \\ 

  123.4    &   123.9    &   0.5     & $   15.32 \pm   0.19 $  &    C&  1113.9   &   1143.9   &   30.0    & $   18.11 \pm   0.05 $  &    C   \\ 

  125.1    &   125.6    &   0.5     & $   15.16 \pm   0.17 $  &    C&  1729.3   &   1849.3   &   120.0   & $   18.43 \pm   0.06 $  &    C   \\ 

  126.9    &   127.4    &   0.5     & $   15.28 \pm   0.10 $  &    C&  2037.1   &   2157.1   &   120.0   & $   18.62 \pm   0.06 $  &    C   \\ 

  128.6    &   129.1    &   0.5     & $   15.37 \pm   0.11 $  &    C&  2341.6   &   2461.6   &   120.0   & $   18.82 \pm   0.04 $  &    C   \\ 

  130.4    &   130.9    &   0.5     & $   15.22 \pm   0.24 $  &    C&  3163.0   &   3283.0   &   120.0   & $   18.79 \pm   0.05 $  &    C   \\ 

  132.1    &   132.6    &   0.5     & $   15.58 \pm   0.10 $  &    C&  3470.7   &   3590.7   &   120.0   & $   18.96 \pm   0.05 $  &    C   \\ 

  133.9    &   134.4    &   0.5     & $   15.50 \pm   0.08 $  &    C&  3775.2   &   3895.2   &   120.0   & $   19.02 \pm   0.05 $  &    C   \\ 

  135.6    &   136.1    &   0.5     & $   15.37 \pm   0.27 $  &    C&  4619.7   &   4739.7   &   120.0   & $   18.93 \pm   0.04 $  &    C   \\ 

  137.3    &   137.8    &   0.5     & $   15.54 \pm   0.18 $  &    C&  4978.0   &   5098.0   &   120.0   & $   19.14 \pm   0.04 $  &    C   \\ 

  139.1    &   139.6    &   0.5     & $   15.70 \pm   0.09 $  &    C&  5282.5   &   5402.5   &   120.0   & $   19.01 \pm   0.05 $  &    C   \\ 

  140.8    &   141.3    &   0.5     & $   15.56 \pm   0.20 $  &    C&  448.7    &   468.7    &   20.0    & $   17.66 \pm   0.08 $  &    g   \\ 

  142.6    &   143.1    &   0.5     & $   15.60 \pm   0.08 $  &    C&   622.3    &   642.3    &   20.0    & $   17.94 \pm   0.04 $  &    g  \\

  643.6    &   663.6    &   20.0    & $   18.04 \pm   0.08 $  &    g &   1853.2   &   2033.2   &   180.0   & $   18.56 \pm   0.07 $  &    r   \\ 

  854.9    &   914.9    &   60.0    & $   18.12 \pm   0.05 $  &    g   &   3286.8   &   3466.8   &   180.0   & $   19.29 \pm   0.08 $  &    r   \\ 

  1209.9   &   1269.9   &   60.0    & $   18.27 \pm   0.06 $  &    g   &  4768.9   &   4948.9   &   180.0   & $   19.32 \pm   0.07 $  &    r   \\ 

  2159.4   &   2339.4   &   180.0   & $   19.08 \pm   0.04 $  &    g&  918.6    &   978.6    &   60.0    & $   18.05 \pm   0.13 $  &    i   \\ 

  3592.9   &   3772.9   &   180.0   & $   19.47 \pm   0.05 $  &    g&  1273.6   &   1333.6   &   60.0    & $   18.20 \pm   0.11 $  &    i   \\ 

  5100.3   &   5280.3   &   180.0   & $   19.65 \pm   0.05 $  &    g&  2467.1   &   2647.1   &   180.0   & $   18.67 \pm   0.07 $  &    i   \\ 

  405.3    &   425.3    &   20.0    & $   17.35 \pm   0.10 $  &    r&  3900.6   &   4080.6   &   180.0   & $   18.60 \pm   0.05 $  &    i   \\ 

  426.6    &   446.6    &   20.0    & $   17.30 \pm   0.07 $  &    r&  5408.0   &   5588.0   &   180.0   & $   19.13 \pm   0.12 $  &    i   \\ 

  578.9    &   598.9    &   20.0    & $   17.75 \pm   0.16 $  &    r&  1335.9   &   1515.9   &   180.0   & $   17.96 \pm   0.15 $  &    Z   \\ 

  600.2    &   620.2    &   20.0    & $   17.71 \pm   0.10 $  &    r&  2649.4   &   2889.4   &   240.0   & $   18.81 \pm   0.23 $  &    Z   \\ 

  792.7    &   852.7    &   60.0    & $   18.17 \pm   0.08 $  &    r&  4083.0   &   4323.0   &   240.0   & $   19.27 \pm   0.21 $  &    Z   \\ 

  1147.7   &   1207.7   &   60.0    & $   18.57 \pm   0.08 $  &    r&
  
  5590.4   &   5830.4   &   240.0   & $   18.98 \pm   0.14 $  &    Z   \\
  
    5668   &   5768   &   100.0    & $   18.99 \pm   0.01 $  &    R&
  
  5794  &   5894    &   100.0   & $   19.05 \pm   0.05 $  &    R   \\
  
    5913  &  6013    &   100.0    & $   19.06 \pm   0.05 $  &    R&
  
   96259   &   98238   &   300.0 $\times$ 6  & $   22.21 \pm   0.27 $  &    R  \\
  \hline
  
\end{longtable}

\clearpage
\begin{table*}
  \caption{Log of the host galaxy observations of \thisgrb. The tabulated magnitudes are in AB magnitude system and have not been corrected for Galactic extinction.}
  \begin{tabular}{|c|c|c|c|c|}
  \hline
  \bf Date & \bf Exposure (sec) & \bf Magnitude & \bf Filter & \bf Telescope\\ \hline \hline
   13.05.2014   &   59x65.0   & $   21.18 \pm   0.26 $  &  H  & CAHA \\
  18.07.2017   &   7x120.0   & $   25.13 \pm   0.16 $  &  g  & GTC \\
  18.07.2017   &   7x90.0    & $   24.47 \pm   0.13 $  &  r  & GTC \\
  18.07.2017   &   7x90.0    & $   24.17 \pm   0.13 $  &  i  & GTC \\
  18.07.2017   &   6x60.0    & $   23.88 \pm   0.18 $  &  z  & GTC \\
 16.01.2021   &  3x300.0, 2x900.0   &   $\geq$ 24.10  &  R  &  3.6m DOT \\
  \hline
  \end{tabular}
  \label{host}
\end{table*}

\end{document}